\documentclass{aa}
\usepackage[varg]{txfonts}
\usepackage{graphicx}
\usepackage{threeparttable}
\usepackage{natbib,twoopt}
\usepackage[draft]{hyperref}
\usepackage{color} 
\usepackage{amsmath}
\bibpunct{(}{)}{;}{a}{}{,}
\makeatletter
  \newcommandtwoopt{\citeads}[3][][]{\href{http://adsabs.harvard.edu/abs/#3}%
    {\def\hyper@linkstart##1##2{}%
     \let\hyper@linkend\@empty\citealp[#1][#2]{#3}}}
  \newcommandtwoopt{\citepads}[3][][]{\href{http://adsabs.harvard.edu/abs/#3}%
    {\def\hyper@linkstart##1##2{}%
     \let\hyper@linkend\@empty\citep[#1][#2]{#3}}}
  \newcommandtwoopt{\citetads}[3][][]{\href{http://adsabs.harvard.edu/abs/#3}%
    {\def\hyper@linkstart##1##2{}%
    \let\hyper@linkend\@empty\citet[#1][#2]{#3}}}
  \newcommandtwoopt{\citeyearads}[3][][]%
    {\href{http://adsabs.harvard.edu/abs/#3}
    {\def\hyper@linkstart##1##2{}%
     \let\hyper@linkend\@empty\citeyear[#1][#2]{#3}}}
\makeatother

\newcommand{\degree}{^{\circ}}

\newcommand{\HI}{{\rm H\,\scriptstyle I}}
\newcommand{\HII}{{\rm H\,\scriptstyle II}}

\begin{document}

\title{Investigation of the cosmic ray population and magnetic field strength in the halo of NGC\,891}
\author{D.\,D.~Mulcahy\inst{1,2}
\and A.~Horneffer\inst{2}
\and R.~Beck\inst{2}\fnmsep\thanks{corresponding author, rbeck@mpifr-bonn.mpg.de}
\and M.~Krause\inst{2}
\and P.~Schmidt\inst{2}
\and A.~Basu\inst{2,3}
\and K.\,T.~Chy{\.z}y\inst{4}
\and R.-J.~Dettmar\inst{5}
\and M.~Haverkorn\inst{6}
\and G.~Heald\inst{7}
\and V.~Heesen\inst{8}
\and C.~Horellou\inst{9}
\and M.~Iacobelli\inst{10}
\and B.~Nikiel-Wroczy{\'n}ski\inst{4}
\and R.~Paladino\inst{11}
\and A.\,M.\,M.~Scaife \inst{1}
\and Sarrvesh~S.~Sridhar\inst{10,12}
\and R.\,G.~Strom\inst{10,13}
\and F.\,S.~Tabatabaei\inst{14,15}
\and T.~Cantwell\inst{1}
\and S.\,H.~Carey\inst{16}
\and K.~Grainge\inst{1}
\and J.~Hickish\inst{16}
\and Y.~Perrot\inst{16}
\and N.~Razavi-Ghods\inst{16}
\and P.~Scott\inst{16}
\and D.~Titterington\inst{16}
}

\institute{Jodrell Bank Centre for Astrophysics, Alan Turing Building, School of Physics and Astronomy, The University of Manchester, Oxford Road, Manchester, M13 9PL, United Kingdom
\and Max-Planck-Institut f{\"u}r Radioastronomie, Auf dem H{\"u}gel 69, 53121 Bonn, Germany
\and Fakult{\"a}t f{\"u}r Physik, Universit{\"a}t Bielefeld, Postfach 100131, 33501 Bielefeld, Germany
\and Astronomical Observatory, Jagiellonian University, ul. Orla 171, 30-244 Krakow, Poland
\and Astronomisches Institut der Ruhr-Universit{\"a}t Bochum, Universit{\"a}tsstr. 150, 44780 Bochum, Germany
\and Department of Astrophysics/IMAPP, Radboud University, PO Box 9010, 6500 GL Nijmegen, The Netherlands
\and CSIRO Astronomy and Space Science, PO Box 1130, Bentley, WA 6102, Australia
\and University of Hamburg, Hamburger Sternwarte, Gojenbergsweg 112, 21029 Hamburg, Germany
\and Department of Space, Earth and Environment, Chalmers University of Technology, Onsala Space Observatory, 439 92 Onsala, Sweden
\and Netherlands Institute for Radio Astronomy (ASTRON), Postbus 2, 7990 AA Dwingeloo, The Netherlands
\and INAF/Istituto di Radioastronomia, via Gobetti 101, 40129 Bologna, Italy
\and Kapteyn Astronomical Institute, University of Groningen, PO Box 800, 9700 AV, Groningen, The Netherlands
\and Astronomical Institute `Anton Pannekoek', Faculty of Science, University of Amsterdam, Science Park 904, 1098 XH Amsterdam, The Netherlands
\and Instituto de Astrof{\'i}sica de Canarias, V{\'i}a L{\'a}ctea S/N, 38205 La Laguna, Spain
\and Departamento de Astrof{\'i}sica, Universidad de La Laguna, 38206 La Laguna, Spain
\and Astrophysics Group, Cavendish Laboratory, 19 J.\,J. Thomson Avenue, Cambridge
CB3 0HE, United Kingdom
}

\date{Received 15 February 2018 /
Accepted 29 March 2018}

\abstract{Cosmic rays and magnetic fields play an important role for the formation and dynamics of gaseous halos of galaxies.}
{Low-frequency radio continuum observations of edge-on galaxies are ideal to study cosmic-ray electrons (CREs) in halos via radio synchrotron emission and to measure magnetic field strengths. Spectral information can be used to test models of CRE propagation. Free--free absorption by ionized gas at low frequencies allows us to investigate the properties of the warm ionized medium in the disk.}
{We obtained new observations of the edge-on spiral galaxy NGC\,891 at 129--163\,MHz with the LOw Frequency ARray (LOFAR) and at 13--18\,GHz with the Arcminute Microkelvin Imager (AMI) and combine them with recent high-resolution Very Large Array (VLA) observations at 1--2\,GHz, enabling us to study the radio continuum emission over two orders of magnitude in frequency.}
{The spectrum of the integrated nonthermal flux density can be fitted by a power law with a spectral steepening towards higher frequencies or by a curved polynomial. Spectral flattening at low frequencies due to free--free absorption is detected in star-forming regions of the disk.
The mean magnetic field strength in the halo is $7\pm2\,\mu$G. The scale heights of the nonthermal halo emission at 146\,MHz are larger than those at 1.5\,GHz everywhere, with a mean ratio of $1.7 \pm 0.3$, indicating that spectral ageing of CREs is important and that diffusive propagation dominates. The halo scale heights at 146\,MHz decrease with increasing magnetic field strengths which is a signature of dominating synchrotron losses of CREs. On the other hand, the spectral index between 146\,MHz and 1.5\,GHz linearly steepens from the disk to the halo, indicating that advection rather than diffusion is the dominating CRE transport process. This issue calls for refined modelling of CRE propagation.}
{Free--free absorption is probably important at and below about 150\,MHz in the disks of edge-on galaxies. To reliably separate the thermal and nonthermal emission components, to investigate spectral steepening due to CRE energy losses, and to measure magnetic field strengths in the disk and halo, wide frequency coverage and high spatial resolution are indispensable.}

\keywords{galaxies: halos -- galaxies: individual: NGC\,891 -- galaxies: ISM -- galaxies: magnetic fields -- ISM: cosmic rays -- radio continuum: galaxies}
\titlerunning{Cosmic rays and magnetic fields in the halo of NGC\,891}
\maketitle

\section{Introduction}

Magnetic fields and cosmic rays are dynamically relevant in the disks of spiral galaxies because the magnetic energy density is similar to the kinetic energy density of turbulence but larger than the thermal energy density \citep[e.g.][]{Beck2015}. Part of the energy input from supernova remnants goes into the acceleration of cosmic rays and into the amplification of turbulent magnetic fields. This scenario can explain the tight correlation between radio and far-infrared emission that holds for the integrated luminosities of galaxies as well as for the local intensities within galaxies \citep[e.g.][]{Tabatabaei2013,Tabatabaei2017,Heesen2014}.

The physical relationships between the various components of the interstellar medium (ISM) are less understood in gaseous galaxy halos. Continuum emission from thick disks or halos is observed from the radio to the X-ray spectral ranges, as well as $\HI$ radio line emission of neutral hydrogen and optical emission lines of ionized gas. As sources in the disk probably provide most of the energy, warm and hot gas, cosmic rays, and magnetic fields have to be transported from the disk into the halo. The required pressure could be thermal or nonthermal (cosmic rays and magnetic fields). Possible transport mechanisms are a ``galactic wind'' with a velocity sufficient for escape \citep{Uhlig2012}, a ``galactic fountain'' with cold gas returning to the galaxy \citep{Shapiro1976,Fraternali2017}, or ``chimney'' outbreaks driven by hot superbubbles \citep{Norman1989}. Cosmic rays can propagate relative to the gas with a velocity limited to the Alfv\'en speed by the streaming instability \citep{Kulsrud1969} or can diffuse along or across the magnetic field lines \citep{Buffie2013}. Especially unclear is the role of magnetic fields in outflows. A turbulent field may be transported together with the gas in an outflow while an ordered field can support or suppress the outflow, depending on its strength and orientation. This is complicated further due to the interplay of the magnetic field with an inhomogeneous outflow in which isotropic turbulent fields are converted into anisotropic turbulent fields due to shear and compression, thus creating an ordered field in the halo \citep{Elstner1995, Moss2017}.

Edge-on galaxies are ideal laboratories to study the disk--halo connection and to investigate the driving forces of outflows. The discovery of a huge radio halo around NGC\,4631 by \citet{Ekers1977} revealed the importance of nonthermal processes. The theory of cosmic-ray propagation into the halos was developed in great detail  \citep{Lerche1981,Lerche1982,Pohl1990}, but comparisons with observations were inconclusive due to the limited quality of the radio data at that time.

Recently, this topic has been revived with high-quality radio spectral index maps becoming readily available. This is driven both by the advent of broadband correlators at existing interferometric telescopes such as the Australia Telescope Compact Array (ATCA) and the Very Large Array (VLA), operating at GHz frequencies, as well as the arrival of entirely new facilities such as the LOw Frequency ARray (LOFAR) that opens a hitherto unexplored window on radio halos at MHz frequencies. As a promising step, \citet{Heesen2016}, using a 1D cosmic ray transport model {\tt SPINNAKER} (SPectral INdex Numerical Analysis of K(c)osmic-ray Electron Radio-emission), were able to extract properties of the cosmic-ray propagation in the halo of two edge-on galaxies.\footnote{\href{www.github.com/vheesen/Spinnaker}{www.github.com/vheesen/Spinnaker}} As a key result, they could show that the vertical profiles of the spectral index can be used to distinguish between advection- and diffusion-dominated halos, the latter representing the case of no significant outflows. Extending the sample to 12 edge-on spiral and Magellanic-type galaxies, \citet{Krause2017} and \citet{Heesen2018a} showed that many halos are advection dominated with outflow speeds similar to the escape velocity, raising the possibility of cosmic ray-driven winds in them.

A prime target for such studies is NGC\,891 that is a fairly nearby edge-on spiral galaxy. NGC\,891 is similar to our own Milky Way in terms of optical luminosity \citep{deVaucouleurs1991}, Hubble type \citep[Sb;][]{vanderKruit1981}, and rotational velocity \citep[225\,km\,s$^{-1}$;][]{Rupen1991}, but has a high star-formation rate (3.3\,M$_{\odot}$\,yr$^{-1}$, \citealt {Arshakian2011}) compared to the Milky Way 
\citep[the Galactic value is $1.66\pm0.20$\,M$_{\odot}$\,yr$^{-1}$;][]{Licquia2014}; this is in accordance with the presence of approximately twice the amount of molecular gas of the Milky Way, with the radial distribution of CO remarkably similar in both galaxies \citep{Scoville1993}. Due to its proximity and very high inclination, NGC\,891 is an observational testing ground for the study of disk and halo interactions and the galactic halo. The physical parameters of NGC\,891 are summarized in Table~\ref{tab:physicalpara}.

\begin{table}
\centering
\begin{threeparttable}
\caption{Observational data of NGC\,891}
\label{tab:physicalpara}
\begin{tabular}{l c}
\hline\hline
Morphology\tnote{a} & Sb \\
Position of the nucleus\tnote{b} & RA(J2000) = 02$^\mathrm{h}$ 22$^\mathrm{m}$ 33$^\mathrm{s}.4$\\
 & DEC(J2000) = +42$\degr$ 20$\arcmin$ 57$\arcsec$ \\
Position angle of major axis\tnote{c} &  23$\degr$ (0$\degr$ is north)\\
Inclination of disk\tnote{d} & $89.8\degr$ ($0\degr$ is face on)  \\ 
Distance\tnote{a} & 9.5\,Mpc ($1\arcsec \approx$ 46\,pc) \\
Star-formation rate\tnote{e} & 3.3 M$_{\odot}$\,yr$^{-1}$ \\
Total mass \tnote{c} & 1.4 $\times$ 10$^{11}$ M$_{\odot}$ \\ 
\hline
\end{tabular}
\begin{tablenotes}
\item[a] \citet{vanderKruit1981}
\item[b] \citet{Vigotti1989}
\item[c] \citet{Oosterloo2007}
\item[d] \citet{Xilouris1999}
\item[e] \citet{Arshakian2011}
\end{tablenotes}
\end{threeparttable}
\end{table}

NGC\,891 possesses a bright, well-studied halo and for which a plethora of ancillary data from various gas components is available. \citet{Rand1990} and \citet{Dettmar1990} independently detected diffuse H$\alpha$ emission from ionized gas up to 4\,kpc distance from the galaxy's plane with an exponential scale height of about 1\,kpc. A huge halo of neutral atomic $\HI$ gas with up to 22\,kpc extent was observed by \citet{Oosterloo2007}. \citet{Howk1997} detected prominent dust lanes emerging vertically into the halo of NGC\,891 which could partly be associated with energetic processes connected to massive star formation in the disk. \citet{Sofue1987} interpreted such dust lanes as tracers of vertical magnetic fields.
Diffuse line emission from CO molecules is observed up to about 1\,kpc distance from the plane \citep{Garcia1992}. Infrared emission from polycyclic aromatic hydrocarbons (PAH) particles and warm dust is detected to about 2.5\,kpc from the plane with similar scale heights of about 300\,pc \citep{Whaley2009}, excited by photons escaping from the disk. Emission from cold dust in the sub-millimetre range extends up to about 2\,kpc \citep{Alton1998} and has a larger scale height than the warm dust. Diffuse X-ray emission from hot gas was found in the halo up to 4\,kpc from the plane \citep{Bregman1994,Hodges2013}.

NGC\,891 has also been extensively observed in radio continuum emission throughout the past few decades. The first extensive interferometric investigation in radio continuum at 610, 1412 and 4995\,MHz with the Westerbork Synthesis Radio Telescope (WSRT) by \citet{Allen1978} revealed a strong steepening of the spectral index in the halo, but this result was affected by missing large-scale emission in the 4995\,MHz image. \citet{Hummel1991} observed NGC\,891 at 327\,MHz and 610\,MHz with the WSRT and at 1490\,MHz with the VLA. The spectral index between 610\,MHz and 1.49\,GHz with a resolution of $40\arcsec$ showed that the inner and outer radio disks (at small and large distances from the centre) have significantly different spectra, partly due to the larger thermal fraction in the inner disk. The spectral steepening towards the halo is mild on the eastern side but steep on the western side. The radio disk at 1490\,MHz was described by an exponential scale height of 1.2\,kpc (scaled to the distance adopted in this paper, see Table~\ref{tab:physicalpara}). \citet{Beck1979} and \citet{Klein1984b} observed NGC\,891 at 8.7\,GHz and 10.7\,GHz with the 100-m Effelsberg telescope. When comparing their data to those at 610\,MHz, they found only a mild steepening of the spectral index~\footnote{$I_\nu \propto \nu^{\alpha}$ where $I_\nu$ is the intensity at frequency $\nu$} from $\alpha \approx -0.75$ in the disk to about $-0.9$ in the halo and no further steepening until up to 6\,kpc above the galaxy's plane (scaled to the distance to NGC\,891 adopted in this paper).
The spectral index in the halo of NGC\,891 allowed for the first time a comparison with CRE propagation models. The models of diffusion and advection by \citet{Strong1978} predicted almost linear gradients of spectral index which were in conflict with the observations by \citet{Allen1978}, but consistent with those by \citet{Beck1979} and \citet{Klein1984b}.
\citet{Dumke1995} observed a sample of edge-on galaxies, including NGC\,891, with the Effelsberg telescope at 10.55\,GHz.
The vertical profile was described by two exponential scale heights of 270\,pc and 1.8\,kpc for the disk and halo, respectively \citep{Krause2012}.



The structure of the magnetic field in the halo of NGC\,891 has been investigated through measurement of linearly polarized radio synchrotron emission. The aforementioned observations of \citet{Dumke1995} with the Effelsberg telescope at 10.55\,GHz revealed diffuse polarized emission from the disk with an orientation predominately parallel to the plane. Such a field structure is to be expected from magnetic field amplification by the action of a mean-field $\alpha \Omega$--dynamo in the disk \citep[e.g.][]{Beck1996} or from a small-scale dynamo in a differentially rotating disk \citep{Pakmor2014} or from shearing of an initially vertical field \citep{Nixon2018}. From Effelsberg observations of NGC\,891 at 8.35\,GHz, \citet{Krause2009} showed that the large-scale orientation of the halo magnetic field in the sky plane appears to be ``X-shaped''. Such a field structure could arise from a mean-field dynamo including a galactic wind \citep{Moss2010} or from a helical field generated by a velocity lag of the rotating halo gas \citep{Henriksen2016}. With the aid of Faraday rotation measures, \citet{Krause2009} also found indication of a large-scale regular magnetic field within the disk of NGC\,891, likely part of a spiral magnetic field.
 
\citet{Israel1990} found that the integrated flux densities of 68 galaxies at 57.5\,MHz are systematically below the extrapolation from measurements at 1.4\,GHz if one assumes a power-law spectrum with a constant slope. They also reported that the 57.5\,MHz flux density is a function of inclination angle of the disk with respect to the sky plane, with lower flux densities observed for galaxies with larger inclination angles which they interpreted as increasing free--free absorption caused by a clumpy medium with an electron temperature of $T_\mathrm{e} \approx 1000$\,K and an electron density of order 1\,cm$^{-3}$. However, no direct observational evidence of such a medium exists, not even in our own Galaxy. \cite{Hummel1991b}, re-analyzing the data of \citet{Israel1990}, and \cite{Marvil2015} observed a flattening in the integrated spectra of nearby galaxies, but found no dependence on inclination; this makes it less likely that free-free absorption is the cause of this flattening. Therefore, low-frequency observations of nearby galaxies, specifically of edge-on galaxies, need to be performed to clear up some of these contradictions.

The LOw Frequency ARray \citep[LOFAR;][]{vanhaarlem2013} opened a new era of studying the diffuse, extended radio continuum emission in halos of nearby galaxies and their magnetic fields -- the study of which has so far been hampered at GHz frequencies by spectral ageing of CREs. The range of short baselines of the High Band Antenna (HBA) Array allow the detection of  extended emission from nearby galaxies. To date, results on two few star-forming low-inclination galaxies with LOFAR were published \citep{mulcahy2014,Heesen2018b}, but the nature of halos around edge-on spiral galaxies still needs to be investigated at low radio frequencies.

In this paper, we present and analyse the first LOFAR observations of NGC\,891 with the HBA array at a central frequency of 146\,MHz. These data are complemented by observations from the Arcminute MicroKelvin Imager (AMI) at 15.5\,GHz, thus providing us with two decades of radio frequency coverage. This paper is organized as follows: In Sections~\ref{sec:observation} and \ref{sec:amiobs}, we describe the observational setup of the LOFAR and AMI observations along with the data reduction and imaging process. In Section~\ref{sec:morphology}, we present the maps of NGC\,891 at both 146\,MHz and 15.5\,GHz; this is followed by a discussion in Section~\ref{sec:compareotherwavelengths} on the morphology of the galaxy in comparison with other wavelengths; here, we pay particular attention to broadband observations with the VLA at central frequencies of 1.5 and 6\,GHz from the CHANG-ES survey \citep[Continuum Halos in Nearby Galaxies: An EVLA Survey;][]{Irwin2012}. In Section~\ref{sec:spectrum}, we present the separation of thermal and nonthermal emission components, discuss the spectrum of the integrated nonthermal emission, and present maps of the total and nonthermal radio spectral indices. Estimates of the magnetic field strength are given in Section~\ref{sec:magneticfield}. In Section~\ref{sec:radioemissionprofile}, we measure the scale heights of the nonthermal emission in the halo at different distances along the projected major axis of the galaxy. We discuss the implications of our findings in Section~\ref{sec:discussion}: free-free absorption, energy loss processes of CREs, and CRE propagation. A summary is given in Section~\ref{sec:conclusion}.

\section{LOFAR observations and data processing}
\label{sec:observation}

\subsection{Observational setup and data selection}
\label{subsec:lofarobs}

The observations of NGC\,891 were done in interleaved mode, switching between scans on the calibrator 3C48 
(at RA(J2000) = $01^\mathrm{h} 34^\mathrm{m} 41^\mathrm{s}3$, DEC(J2000) = $+33^\circ 09^\prime 35^{\prime \prime}$)
and the target NGC\,891. A total of 44 stations were used for this observation, of which 32 were core stations and 12 were remote stations.
Full details of the observational setup are shown in Table~\ref{tab:obs_para_lofar}.

\begin{table}[h!]
\caption{Parameters of the NGC\,891 LOFAR HBA observations}
\label{tab:obs_para_lofar}
\centering
\begin{tabular}{l c}
\hline\hline
Start date (UTC) & 31 March 2013 / 09:00 \\
End date (UTC) & 31 March 2013 / 16:59 \\
Interleaved calibrator & 3C48 \\
Scan length on calibrator & 2 min \\
Scan length on target & 15.8 min \\
Duration of observations & 8 hours \\
Final time on target & 5.56 hours (21 scans) \\
\hline
Frequency range & 129.2--176.8\,MHz  \\ 
Final frequency range & 129.2--163.6\,MHz \\ 
Total bandwidth on target & 47.6\,MHz  \\
Final bandwidth on target & 34.4\,MHz \\ 
Reference frequency & 146.38\,MHz \\
\hline
\end{tabular}
\end{table}

Two of the scans on the target were not recorded properly and one had excessive
radio-frequency interference (RFI), so that these scans were discarded after pre-processing. After subtracting overheads for the calibrator observations and for the switching of the beam positions, the remaining on-source integration time was $5.6$~h. During facet calibration (see Section~\ref{subsec:facet-cal}), the self-calibration solutions of the calibrator for the target facet, containing NGC\,891, did not converge for frequencies $>$163.6~MHz, so those data were excluded as well. Hence, the resulting bandwidth that was used for the imaging was 34.4\,MHz. Finally, the automatic flagging in the pipelines flagged a further 15\% of the visibilities, mostly data affected by RFI that were identified as spikes when plotting amplitudes as function of time or frequency.

Most of the short baselines between the ``ears'' of the core stations were flagged to avoid cross-talk. The remaining shortest baselines between the core stations of about 50\,m ensure the detection of large diffuse emission on scales of up to $2\degr$.


\subsection{Calibration}

\begin{figure*}
\centering
\includegraphics[width=8cm]{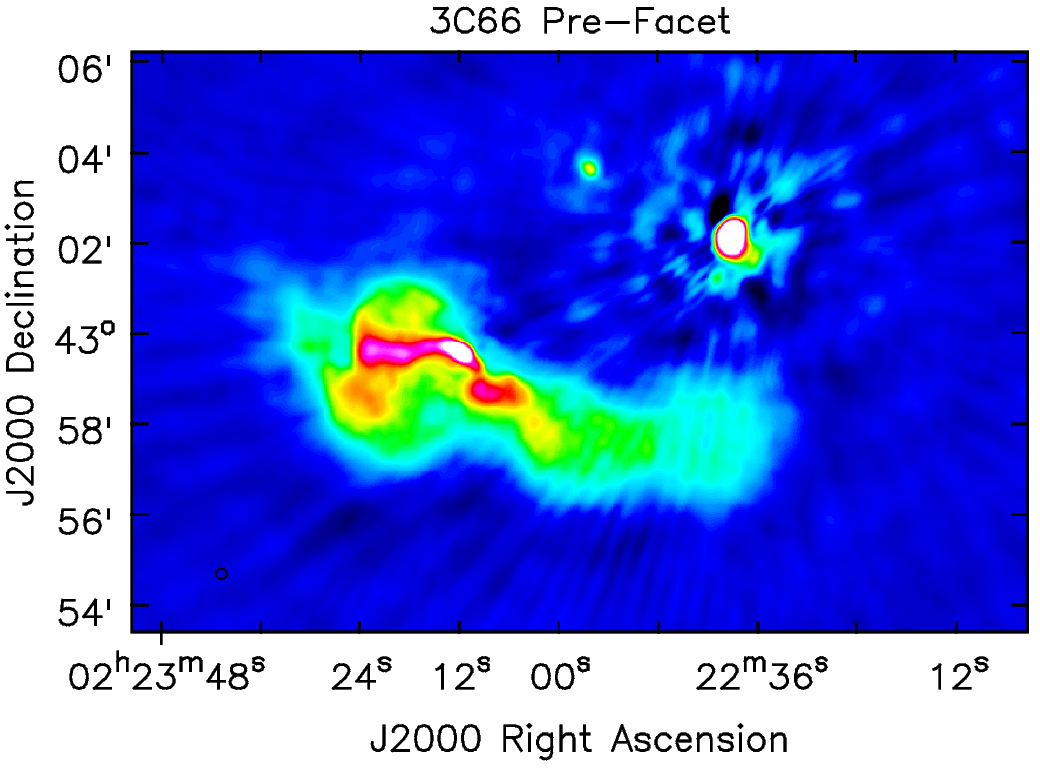} \hskip 0.7cm
\includegraphics[width=8cm]{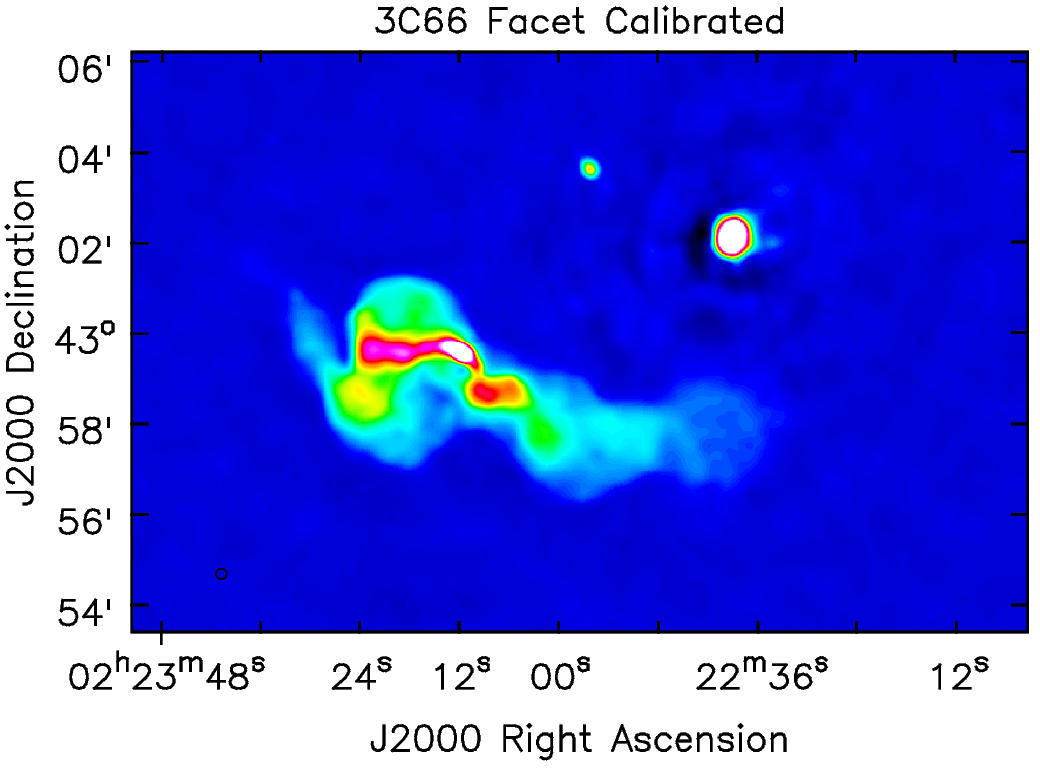}
\caption{Image of the compact source 3C66\,A and the extended source 3C66\,B after the initial direction-independent calibration (left) and after the direction-dependent facet calibration (right). Both images are smoothed to the same resolution of $14\arcsec$ and are displayed on the same colour scale. The strong artefacts around 3C66\,A after the direction-independent calibration in the left map are caused by the calibration not being adequate for this direction. After calibrating specifically for this direction these artifacts are mostly gone.}
\label{fig:3c66}
\end{figure*}

\begin{figure*}
\centering
\includegraphics[width=13cm]{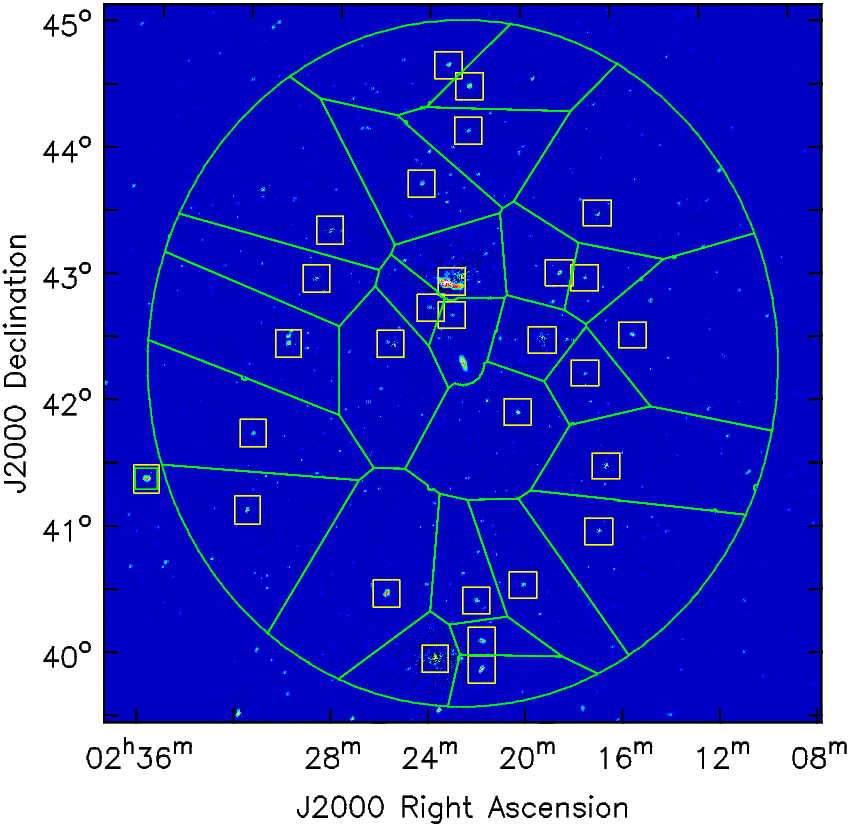}
\caption{Image of the full FoV of the LOFAR observations after initial calibration, with NGC\,891 in the centre, 3C66\,A/B above the centre, and 3C65 at the lower edge of the image. Overlaid is the faceting scheme that was used for the facet calibration: in orange the squares of the calibrator regions that were used to calibrate the facets and in green the resulting facets that are defined by {\tt Voronoi tessellation} around the calibrators.}
\label{fig:facetimage}
\end{figure*}

Low-frequency, wide-field imaging data are subject to several effects that are usually negligible at higher frequencies, most notably distortions of the Earth's ionosphere and artifacts caused by the large number of bright background sources. In order to overcome these difficulties, we applied the novel technique of \emph{facet calibration}. This method uses three main steps. First, a direction-independent calibration of the data with the {\tt prefactor} pipeline that applies the flux density scale, corrects the Faraday rotation of the Earth's ionosphere, corrects for instrumental effects such as clock offsets between different LOFAR stations, and performs an initial round of phase calibration; second, the subtraction of all sources that are present in the supplied sky model with the {\tt Initial} {\tt Subtract} pipeline;  last, the application of the {\tt Factor} pipeline. This pipeline divides the field of view into a mosaic of smaller facets using {\tt Voronoi} {\tt tesselation} \citep[e.g.][]{Okabe2000},
each of them located around a bright source which serves as its calibrator. Then, for each of these patches the calibrator source is added back in and several loops of phase and amplitude calibration are applied. Once the final solutions are calculated, all other sources are added back in and corrected using the gains derived from the corresponding facet calibrator. The reduction process of these pipelines is described in more detail by \citet{vanWeeren2016} and \citet{Williams2016}. In the next two subsections, we provide a brief overview of each of these steps.

\subsubsection{Initial calibration}
\label{subsec:prefactor}

To calibrate our LOFAR dataset of NGC\,891, we used a prototype version of the {\tt prefactor}~\footnote{https://github.com/lofar-astron/prefactor} pipeline for the initial, direction-independent calibration and {\tt Version 1.0pre} of {\tt Factor}~\footnote{https://github.com/lofar-astron/factor}
for the direction-dependent facet calibration.

The data of the calibrator and target were first pre-processed with the New Default Pre-Processing Pipeline ({\tt NDPPP}) that includes RFI excision with {\tt aoflagger}~\citep{Offringa2010,Offringa2012}, removal of edge-channels, and averaging to 4\,s time- and 49\,kHz (4 channels per subband) frequency-resolution.

The data of the calibrator observations were then calibrated against a known model characterizing the detailed frequency dependence of the flux density of 3C48  \citep{Scaife2012}. This model thus sets the flux density scale of our data. The gain solutions from this calibration were then used to extract instrumental calibration parameters: gain amplitudes, station clock delays, and phase offsets between the X- and Y-dipoles within a station.

These solutions were copied to the target data in order to correct for instrumental effects. To deal with the effects of strong off-axis sources, we predicted the visibilities of the four strongest sources (Cas~A, Cyg~A, Tau~A, and Vir~A) and flagged all visibilities to which they contributed more than an apparent (not corrected for primary-beam attenuation) flux density of 5\,Jy. \footnote{This threshold was found to be the best compromise between flagging too much data and thus increasing the rms noise and flagging too few data and thus not removing the residual sidelobes of these sources (van Weeren, priv. comm.).}
Then the data were averaged to the final resolution of 8\,s and 98\,kHz (2 channels per subband) and concatenated into groups of 11 subbands to form 16 bands with 2.15\,MHz total bandwidth each. This resolution is sufficient to avoid decorrelation due to rapid changes in the ionospheric phase, while keeping the data volume manageable at 14.5\,GByte per band. An additional round of automatic RFI excision with {\tt aoflagger} was then performed on the concatenated data, because the algorithm is more effective in detecting RFI in data sets with wide bandwidths as compared with single subbands.
As a final step for the direction-independent calibration, the data were phase calibrated on a model generated from the LOFAR Global Sky Model \citep[GSM;][]{vanhaarlem2013}.

Finally, the 16 bands were imaged separately, first at a medium resolution (outer \textit{uv} cut at 7\,k$\lambda$, about $20 \arcsec$ resolution) then -- after subtracting the sources found in the medium resolution images -- at a lower resolution (outer \textit{uv} cut at 2\,k$\lambda$, about $1.5 \arcmin$ resolution). The field of view (FoV) of the medium resolution images was 2.5 times the full width to half maximum (FWHM) of the station beam (between $9.4\degree$ and $12.8\degree$ depending on the frequency of the band) and the FoV of the low-resolution images 6.5 times the FWHM of the station beam ($24.5\degree$ -- $33.3\degree$). The reason to create different kinds of images is to pick up low surface-brightness emission and to be able to image a larger FoV without prohibitive computing requirements. The final result is the combined list of sources from both imaging steps, the residual visibilities in which all detected sources have been subtracted, and the phase solutions from the last calibration step, for each of the 16 bands.


\subsubsection{Facet calibration}
\label{subsec:facet-cal}


The first step of the facet calibration is the selection of the calibration directions that contain the facet calibrators. For this step and for what follows we used the {\tt Factor} pipeline that automates most of the necessary steps. In the first step, {\tt Factor} uses the sky model for the highest frequency band and searches for strong and compact sources. This was done with the following selection parameters:
\begin{itemize}
\item maximum size of a single source: $2\arcmin$
\item minimum apparent flux density of a single source: 100\,mJy
\item maximum distance of single source to be grouped into one calibrator region: $6\arcmin$
\item minimum total apparent flux density of the sources of a calibrator region: 250~mJy.
\end{itemize}

The resulting list of calibration directions was then manually modified in order to tailor it to our specific requirements. Apart from removing some of the weaker calibrators that were too close to each other, we also experimented how to define best the calibrator region around 3C66 (Fig.~\ref{fig:3c66}) -- by far the brightest source in our FoV.
In the end we decided to fully include both 3C66\,A and 3C66\,B into the region: while the extended emission of 3C66~B is not suited for calibration, there is enough compact flux density in 3C66\,A to allow for good calibration solutions even on the long baselines, whereas excluding 3C66\,B made calibration worse, probably because the short baselines could not be calibrated well as the sky model did not include the extended emission.

The 29 calibration directions after our manual adjustments are shown in Fig.~\ref{fig:facetimage}. The facets are set up by {\tt Voronoi tesselation} around the calibration direction up to a maximum radius of 2.5$\degree$ in RA and 2.7$\degree$ in DEC~\footnote{The region is elliptical to account for the elongation of the LOFAR primary beam at lower elevations.} around the pointing centre. The facet boundaries are slightly deflected to avoid intersecting with detected sources in the sky model. We defined a region with a radius of 10.2\arcmin\ around NGC\,891 as the target facet, so that sources, as well as our target, are not split between two facets. The resulting facet boundaries are shown as lines in Fig.~\ref{fig:facetimage}. The one direction outside the faceting radius (see the orange box outside of the green ellipse in Fig.~\ref{fig:facetimage}) has only a calibration region but no facet associated with it which means that only a small area around the calibrator is imaged.

The core of {\tt Factor} is the direction-dependent calibration. For each direction the residual visibilities are phase shifted to the direction of the calibration region and the sources within that region are added back to the visibilities. Then these data are self-calibrated, starting with the direction-independent calibration.  The self-calibration has two loops: in the first loop, {\tt Factor} only solves for a fast phase term to track the ionospheric delay; in the second loop, {\tt Factor} solves again for this fast phase term and simultaneously for a slowly varying gain (phase + amplitude) term to also correct for residual effects from discrepancies in the beam model and from any other causes. After the calibration region has been calibrated, the full facet is imaged again: the visibilities are prepared in a similar fashion as they were for self-calibration, except that all sources within the facet are added back in and that the data are corrected with the self-calibration solutions. From this image an updated sky model for the facet is created which, together with the calibration solutions for this direction, is used to update the residual visibilities by subtracting the difference between the new and the original model. The following directions are then processed with the improved residual visibilities. This way the strong sources can be subtracted first and relatively weak sources can be used as calibrators.

We only processed one direction at a time, starting with 3C66 which has the largest apparent flux density and thus the highest signal-to-noise ratio, proceeding down in apparent flux density. Ordered in this way, the region that contains NGC\,891 is in the $12^{\rm th}$ facet. Before processing the target facet, the directly adjacent facets were processed, too. The remaining facets contribute only little noise to the region around NGC\,891, so we stopped processing there. The visibilities we used for the final imaging were the ones that {\tt Factor} generated for the imaging of the full target facet; this  means that all detected sources outside the target facet were subtracted, so that it was sufficient to image only the relatively small area of the target facet.

\subsection{Final imaging}
\label{subsec:finalimage}

We used the {\tt Common Astronomy Software Applications}~\footnote{http://casa.nrao.edu} \citep[{\tt CASA};][]{MCMULLIN2007}, Version 4.7, to image the facet containing NGC\,891. Whilst CASA does not implement the LOFAR primary beam, NGC\,891 is much smaller than the size of the LOFAR primary beam (3.8$^\circ$ FWHM at 146\,MHz) and is located at the phase centre of our observation. Thus, systematic flux density errors due to the missing primary beam correction are minimal.

\begin{figure*}
\centering
\includegraphics[width=9cm]{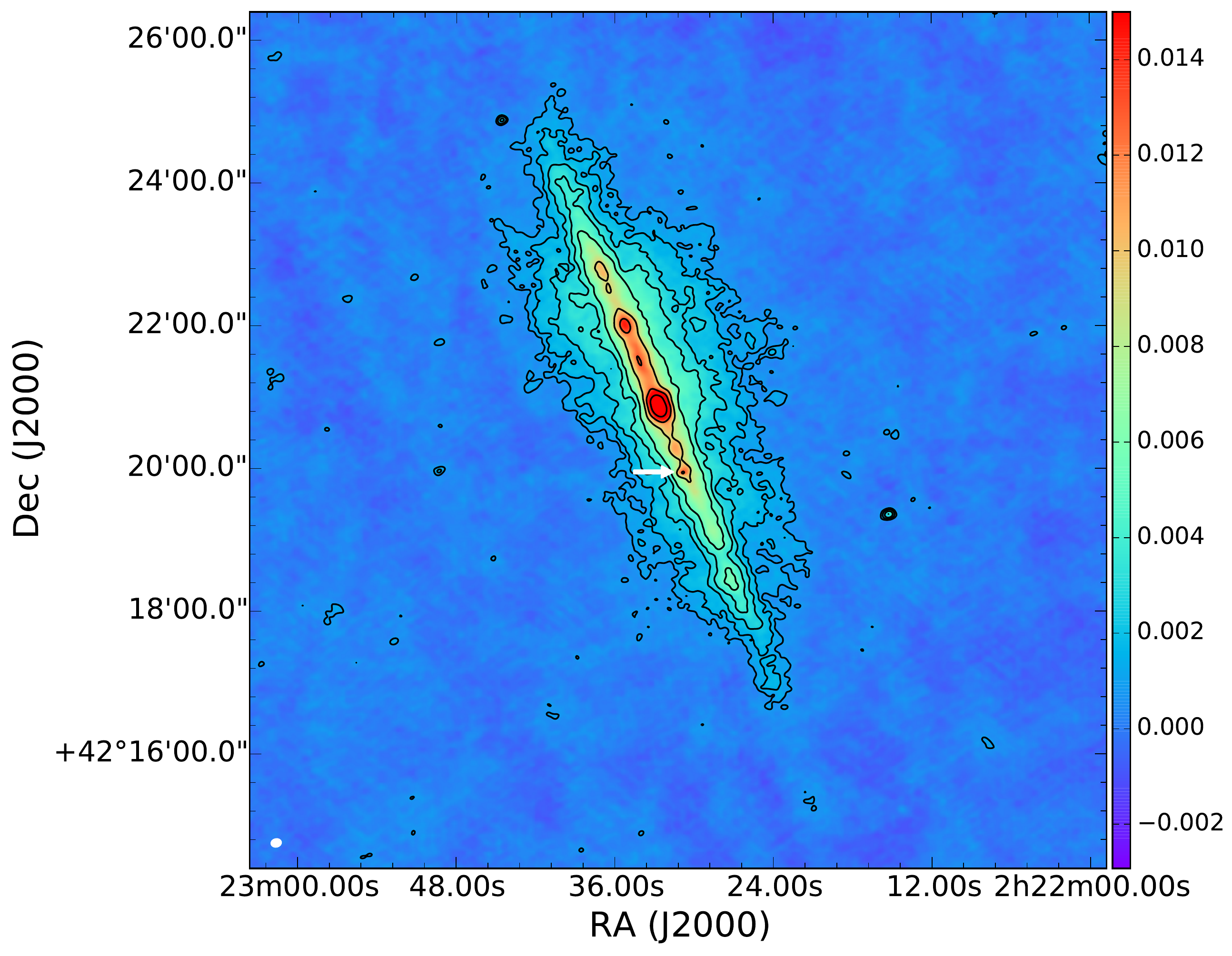}\hfill
\includegraphics[width=9cm]{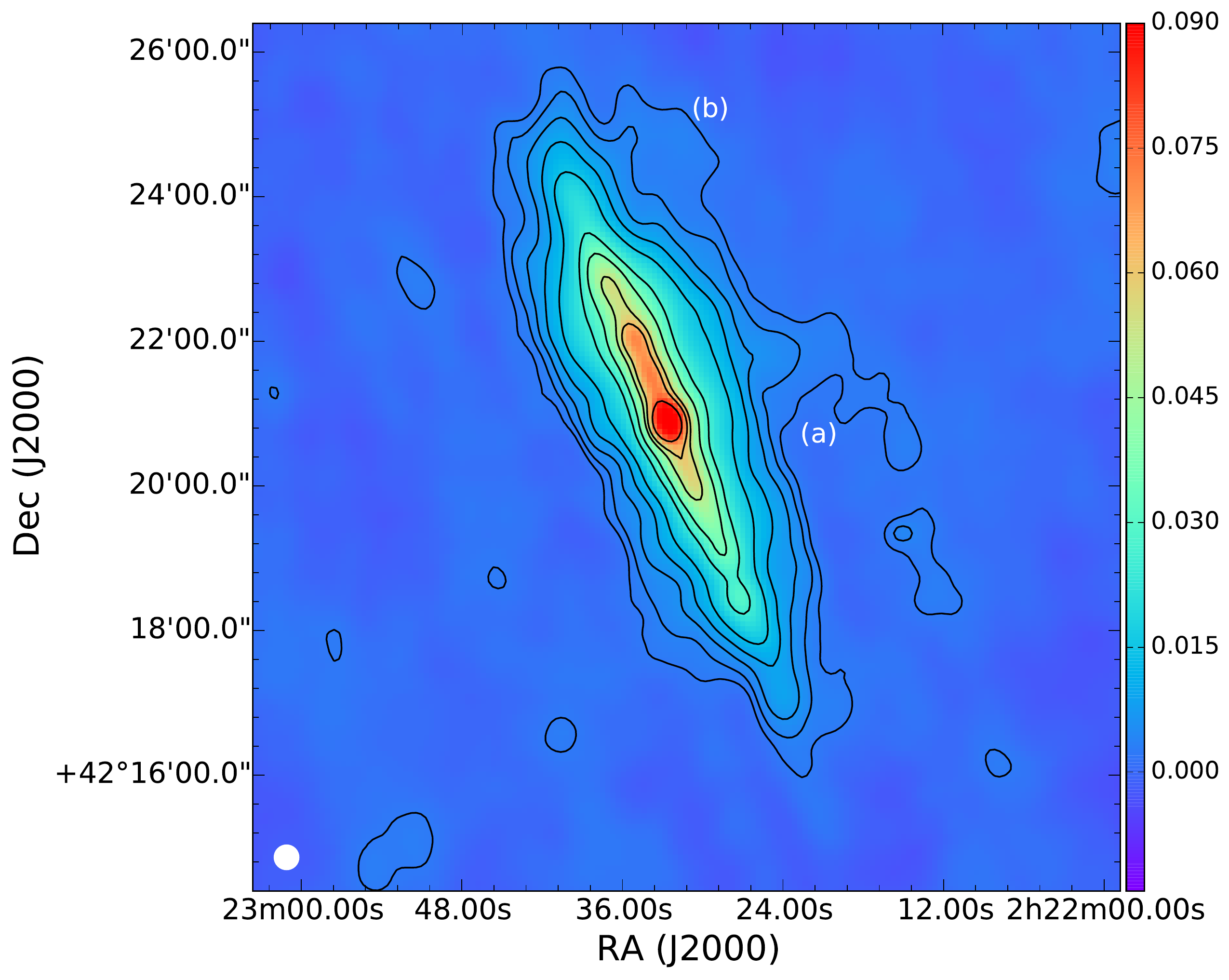} 
\caption{LOFAR maps of NGC\,891 observed at a central frequency of 146.4\,MHz with a bandwidth of 34.4\,MHz at two different resolutions. Left: Resolution of $8.3 \arcsec \times 6.5 \arcsec$ (shown by the filled ellipse in the bottom left corner). Contours are at 3, 5, 8, 12, 18, 32, 44, 64 $\times \sigma$ where $\sigma = 0.29$\,mJy/beam is the rms noise level. 
The location of SN1986J is shown by the white arrow. Right: Resolution of $20 \arcsec \times 20 \arcsec$ (shown by the filled ellipse in the bottom left corner). Contours are at 3, 5, 8, 12, 18, 32, 44, 64, 76 $\times \sigma$ 
where $\sigma = 0.8$\,mJy/beam is the rms noise level. 
The features denoted by `(a)' and `(b)' are discussed in Section~\ref{sec:morphology}. The colour scale is in units of Jy/beam.}
\label{fig:NGC891lofarimage}
\end{figure*}

We created four different images. The first image, with no \textit{uv} taper applied and a robust weighting of $-1.0$, achieves a high resolution of $8.3 \arcsec \times 6.5 \arcsec$, sufficient to resolve various features in the disk and the inner halo. Another version of this image was created by convolution to $12 \arcsec \times 12 \arcsec$. The third image has a moderate resolution, with an outer \textit{uv} taper of 7k$\lambda$, a robust weighting of 0, and convolved to $20 \arcsec \times 20 \arcsec$. The fourth, our low-resolution image, with an outer \textit{uv} taper of 2k$\lambda$ and a robust weighting of +1.0, is best suited to detect the low surface brightness of the extended halo.

For all images, we performed multi-scale and multi-frequency synthesis {\tt CLEAN} \citep{Hogbom1974,Cornwell2008,Rau2011} on the facet containing NGC\,891, for which we used {\tt CLEAN} manually drawn masks. The imaging parameters are given in Table~\ref{table:imagingpara}.

\section{AMI observations and data processing}
\label{sec:amiobs}

\subsection{Observational setup}

The Arcminute Microkelvin Imager (AMI) telescope \citep{Zwart2008} consists of two radio arrays, the Small Array (SA) and the Large Array (LA), located at the Mullard Radio Astronomy Observatory (Cambridge, UK). The SA is a compact array of ten $3.7$-m paraboloid dishes and is sensitive to structures on angular scales between $2\arcmin$ and $10\arcmin$. The LA is an array of eight $12.8$-m dishes and is sensitive to scales between $0.5\arcmin$ and $3\arcmin$. Each array observes at a frequency range of $13$--$18$\,GHz with 4096 channels split into two bands \citep{Hickish2018}. 

The SA data were taken as a single pointing with interleaved observations of the phase calibrator J0222+4302 and 3C286 used as a flux density and bandpass calibrator. The LA data were taken as a mosaic, consisting of seven pointings arranged on a hexagonal raster centred on NGC\,891. Individual pointings were cycled between the pointing centers every 60 seconds, switching to the phase calibrator J0222+4302 every 10 minutes for 2 minutes. Details of these observations are given in Table~\ref{tab:obs_para_ami}.

\begin{table}[h!]
\caption{Parameters of the NGC\,891 AMI observations}
\label{tab:obs_para_ami}
\centering
\begin{tabular}{l c}
\hline\hline
1$^{\rm st}$ Small Array time (UTC) & 7-8 Dec 2016 / 20:07-01:50 \\
2$^{\rm nd}$ Small Array time (UTC) & 9-10 Dec 2016 / 17:23-01:42 \\
Large Array time (UTC) & 7 Dec 2016 / 19:01-22:51 \\
Flux calibrators & 3C48 (LA); 3C286 (SA) \\
\hline
Frequency range & 13-18\,GHz  \\ 
Reference frequency & 14.5\,GHz \\
\hline
\end{tabular}
\end{table}

\subsection{Calibration and imaging}

Calibration and imaging of the visibilities from both arrays were carried out using {\tt CASA}. The first step, in calibrating AMI data, is the generation of the rain gauge correction which is a correction for the system temperature dependent on the weather during the observation. An initial round of flagging was carried out on the full resolution data, using the {\tt rflag} option in {\tt CASA's} {\tt flagdata} task, in order to remove strong narrow-band RFI. The data were then averaged from 2048 to 64 channels for each band. We used the flux density scale of \cite{Perley2013} for calibrators 3C48 and 3C286. A tailored, in-house version of the {\tt CASA} task {\tt setjy} was used to correct for the fact that AMI measures single polarization $I$+$Q$. An initial round of phase only calibration was performed on the flux calibrator which was then used for delay and bandpass calibration. This was followed by another phase and amplitude calibration, applying the delay and bandpass solutions on the fly. The calibrated data were used for a second round of flagging, before performing a phase and amplitude calibration on the phase calibrator, applying again the delay and bandpass solutions on the fly. After the flux density of the phase calibrator was bootstrapped \citep[e.g.][]{Lepage1992}
the calibration tables were applied to the target data which were then averaged to 8 channels for imaging. 

The data from both arrays were imaged with multi-scale {\tt CLEAN} \citep{Cornwell2008} and each image was cleaned interactively. The SA image with its single pointing was primary beam corrected using the task {\tt PBCOR} in {\tt AIPS} with the defined SA primary beam.\footnote{{\tt AIPS}, the Astronomical Image Processing System, is free software available from NRAO.} Each of the seven LA pointings were imaged separately and converted into FITS files. These images were combined into a mosaic using the {\tt AIPS} task {\tt FLATN} which was then corrected for attenuation by the LA primary beam.

\section{Results: NGC\,891 at 146\,MHz and 15.5\,GHz}
\label{sec:morphology}

\subsection{The LOFAR images}
\label{subsec:LOFARimage}

\begin{figure}[h!]
\includegraphics[scale=0.35]{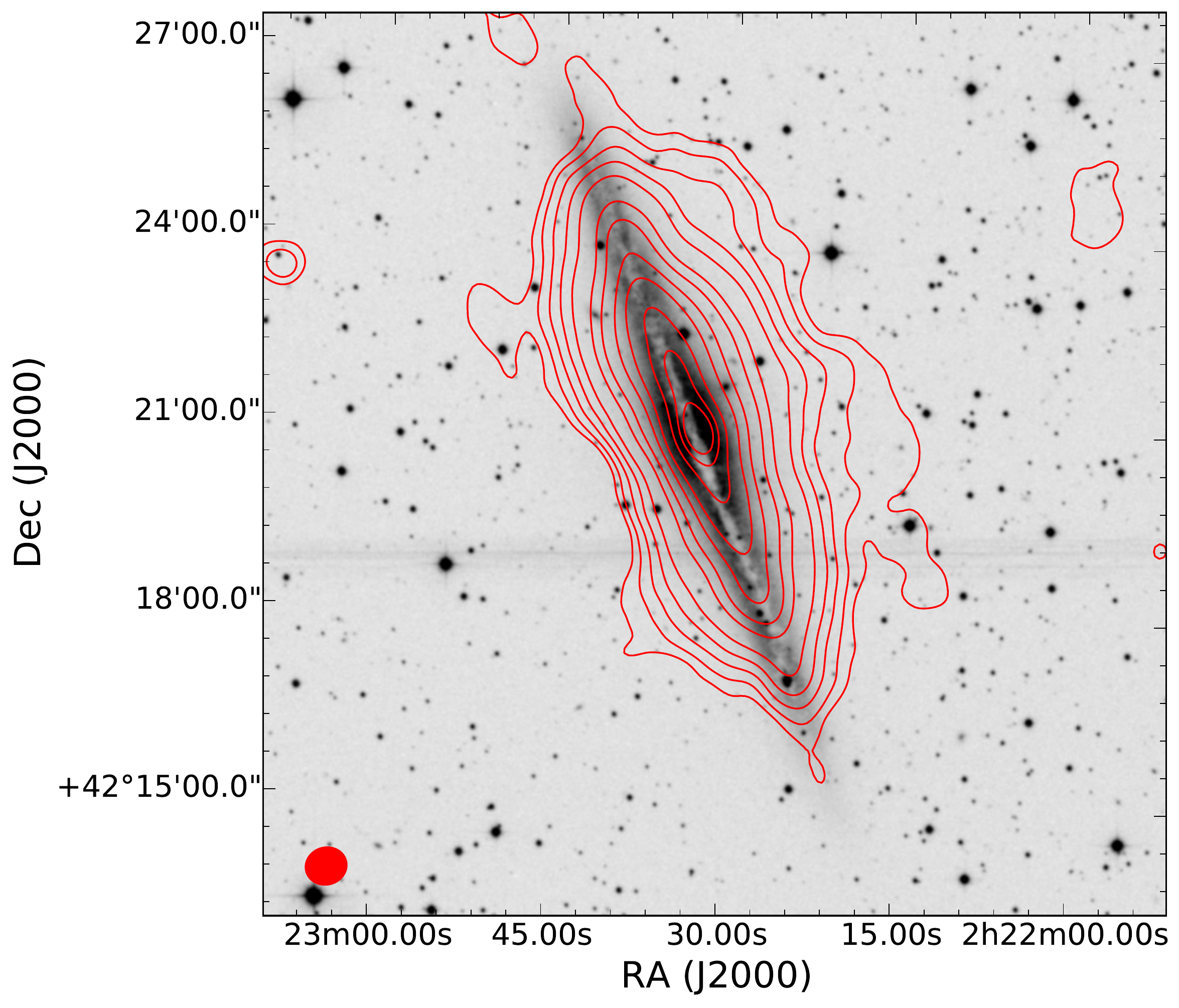}
\caption{LOFAR map of NGC\,891 at a resolution of $40.0 \arcsec \times 35.7 \arcsec$ shown by the filled ellipse in the bottom left, overlaid onto an optical image from the Digitized Sky Survey. Contours are at $3, 5, 8, 12, 18, 32, 44, 88, 164, 200 \times\sigma$ where $\sigma = 1.1$\,mJy/beam is the rms noise level.} 
\label{low_res_image}
\end{figure}

The images of NGC\,891 at a central frequency of 146\,MHz are shown at the highest resolution ($8.3 \arcsec \times 6.5 \arcsec$) in Fig.~\ref{fig:NGC891lofarimage} (left), at medium resolution ($20 \arcsec \times 20 \arcsec$) in Fig.~\ref{fig:NGC891lofarimage} (right), and at low resolution ($40.0 \arcsec \times 35.7 \arcsec$) in Fig.~\ref{low_res_image}. The image characteristics are given in Table~\ref{table:imagingpara}. The image at $12 \arcsec \times 12 \arcsec$ resolution is shown in the composite of Fig.~\ref{fig:composite}.

\begin{table*}
\caption{Imaging parameters for the LOFAR images}
\centering
\begin{tabular}{l c c c c}
\hline\hline
& High-resolution image & Medium-resolution image & Low-resolution image &\\
\hline \hline
\textit{uv} taper & -- & 7k$\lambda$ & 2k$\lambda$ \\
Weighting & -1 & 0 & +1 \\
Angular resolution & $8.3 \arcsec \times 6.5 \arcsec$ ($381 \times 299$\,pc$^{2}$) & $20\arcsec \times 20\arcsec$ ($920 \times 920$\,pc$^{2}$)  & $40.0 \arcsec \times 35.7 \arcsec$ ($1.84 \times 1.64$\,kpc$^{2}$) \\
Cell size & $1.0\arcsec$ & $3.0\arcsec$ & $4.0\arcsec$ & \\
\hline
\end{tabular}
\label{table:imagingpara}
\end{table*}

The mean root-mean-square (rms) noise at high resolution is approximately $\sigma\simeq 0.29$\,mJy/beam next to NGC\,891 and about 0.26\,mJy/beam in a quiet region. The noise at medium resolution is approximately 1.0\,mJy/beam next to NGC\,891 and 0.8\,mJy/beam in a quiet region.

The morphology of NGC\,891 at 146\,MHz at low resolution is quite similar to that at higher frequencies (327\,MHz and 610\,MHz) as seen in \citet{Hummel1991}. The radio halo bulges out in the northern sector of the galaxy. This is especially evident in the high-resolution image (Fig.~\ref{fig:NGC891lofarimage} left) where diffuse emission extends to the north-east and north-west. It is known that 
the northern part of the disk has a larger star-formation rate than the southern part \citep{Strickland2004}.

\begin{figure*}
\centering
\includegraphics[width=9cm]{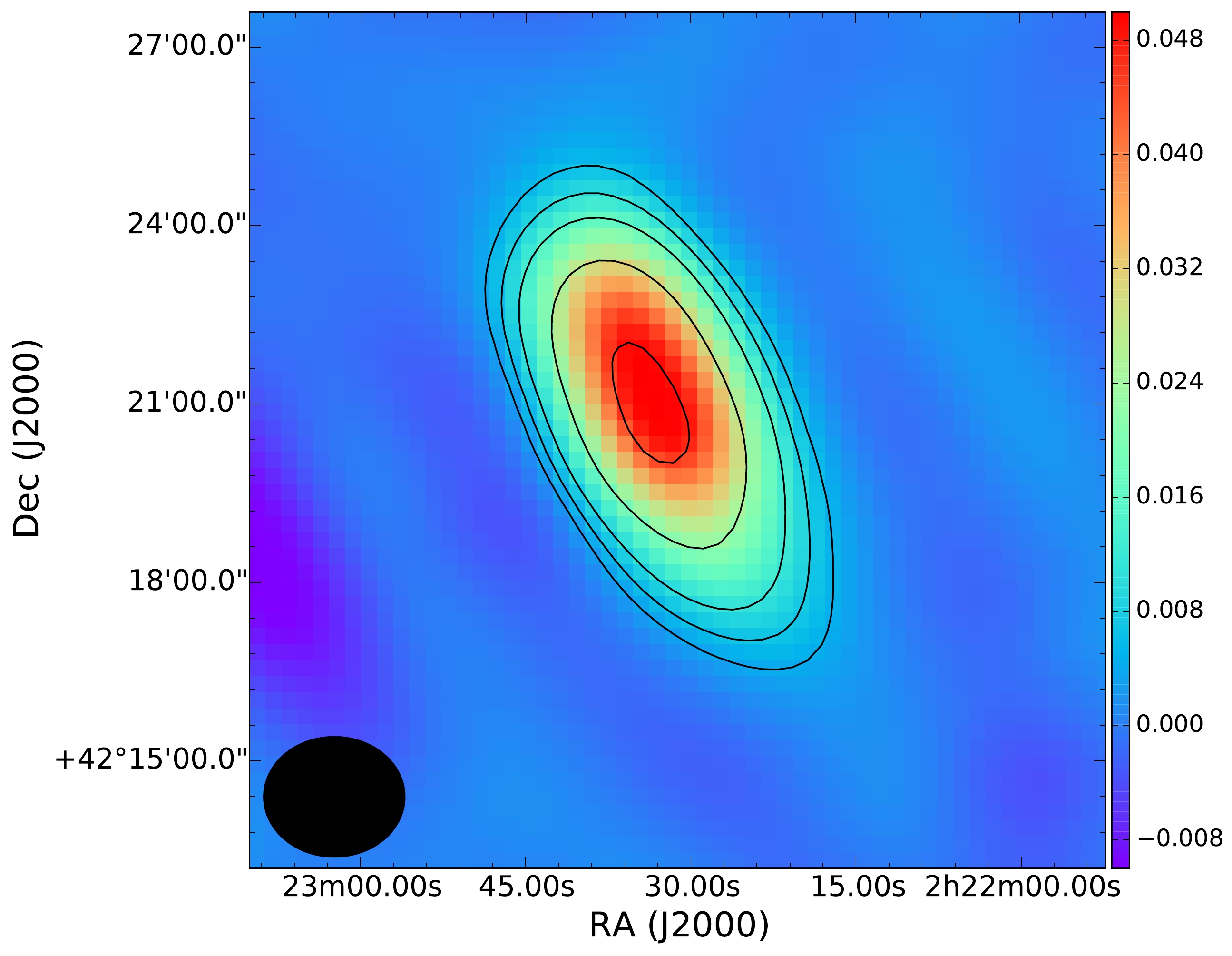} \hfill
\includegraphics[width=9cm]{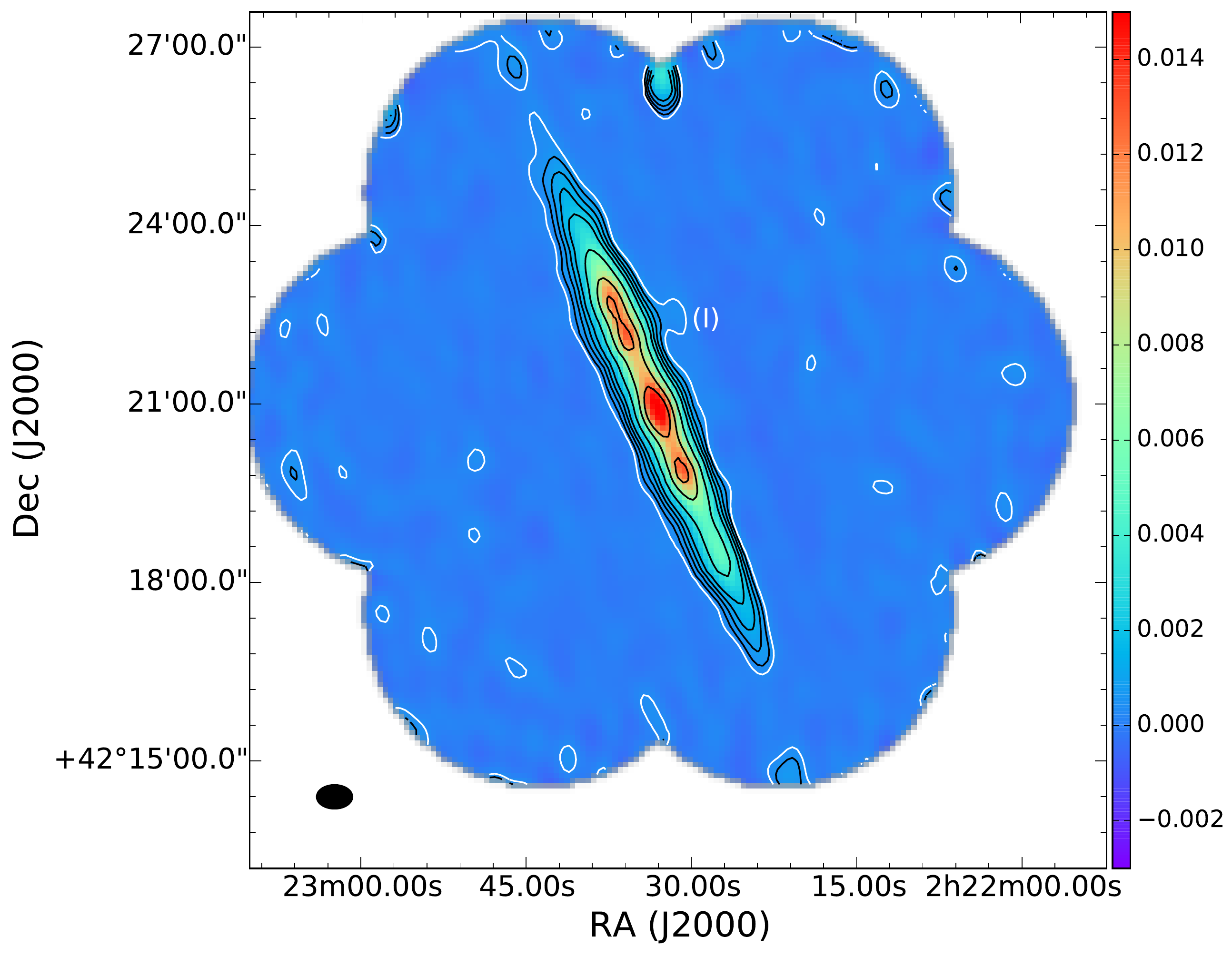}
\caption{AMI maps of NGC\,891 observed at a central frequency of 15.5\,GHz with the Small Array (SA) at $142 \arcsec \times 121 \arcsec$ resolution (left) and the Large Array (LA) at $36 \arcsec \times 24 \arcsec$ resolution (right). For the SA image, contours are at 3, 5, 8, 16, 32, 64 $\times \sigma$ where $\sigma = 1.5$\,mJy/beam is the rms noise level. 
For the LA image, contours are at 3 (white), 5, 8, 12, 18, 36, 72, 108 $\times \sigma$ where $\sigma = 0.11$\,mJy/beam is the rms noise level. 
The colour scales are in units of Jy/beam. The sizes of the synthesized beams are shown by the filled ellipses in the bottom left corners.}
\label{fig:AMIimages}
\end{figure*}

Interesting extensions (marked `(a)' and `(b)' in Fig.~\ref{fig:NGC891lofarimage}) are seen in the western halo which were also observed at higher frequencies \citep{Schmidt2016,Schmidt2018}, but are more prominent at lower frequencies. Signs of feature `(a)' are also seen as low brightness emission in Fig.~\ref{low_res_image}.
These features are revisited later with information on their spectrum in Section~\ref{sec:spectrum}.

The eastern radio halo of NGC\,891 displays a ``dumbbell'' shape in Fig.~\ref{low_res_image}, similar to what is observed on both sides of the halo of NGC\,253 \citep{Heesen2009}. Such a shape can be a signature of dominating synchrotron losses (see Section~\ref{subsec:lossprocesses}). The two extensions
on the eastern side of the galaxy out to approximately 5.5\,kpc from the major axis are similar to the ones found in NGC\,5775 \citep{Soida2011} and may be signs of outflows which eject gas from galaxies and enrich the local intergalactic medium. Between these two galactic spurs the radio emission extends out to about 4.5\,kpc from the plane. The maximum extent of the western halo of NGC\,891 (outermost contour in Fig.~\ref{low_res_image}) is about 9\,kpc from the major axis. The extents are measured using the $3\sigma$ level in Fig.~\ref{low_res_image}.

The radio disk extends to about 16\,kpc from the centre along the major axis in the north and south. With the full extents of disk and the halo of about 32\,kpc along the plane and 14.5\,kpc perpendicular to the plane, respectively, the halo-to-disk extent ratio at 146\,MHz is $\approx 0.45$. This is similar to the ratio for NGC\,253 at 200\,MHz \citep{Kapinska2017}.
However, we caution that the halo-to-disk extent ratio is strongly dependent on sensitivity and angular resolution; a better way to quantify the halo emission is via the scale height (see Section~\ref{sec:radioemissionprofile}).

Several features in the disk are resolved in the high-resolution image (Fig.~\ref{fig:NGC891lofarimage}, left panel). The most intense radio emission in the galaxy is seen in the central region of NGC\,891 and the north of the disk due to the larger star-formation rate in this region of the disk. To the south of the disk, less intense radio emission is observed compared to the north. 

The position of the radio supernova SN1986J is indicated by an arrow in Fig.~\ref{fig:NGC891lofarimage} (left panel).
SN1986J \citep{vangorkom1986} is one of the most luminous radio supernovas ever discovered \citep{Bietenholz2010} and has been studied extensively since its discovery. The date of its explosion is uncertain; the best estimate is $1983.2\pm1.1$ \citep{Bietenholz2002}.
We detect SN1986J in our LOFAR high-resolution image (Fig.~\ref{fig:NGC891lofarimage} left) as an unresolved point source located in the south-west of the disk. A Gaussian fit gives a flux density at 146\,MHz of $5.5\pm0.2$\,mJy/beam above the background disk emission at the position of RA(J2000) = 02$^\mathrm{h}$ 22$^\mathrm{m}$ 30$^\mathrm{s}.8$, DEC(J2000) = +42$\degr$ 19$\arcmin$ 57$\arcsec$.

\subsection{The AMI images}
\label{subsec:AMIimage}

The images of NGC\,891 at a central frequency of 15.5\,GHz observed with both the AMI arrays are shown in Fig.~\ref{fig:AMIimages}. For the SA image the final resolution is $142 \arcsec \times 121 \arcsec$ (6.5 $\times$ 5.5 \,kpc$^{2}$) and $36 \arcsec \times 24 \arcsec$ (1.6 $\times$ 1.1 \,kpc$^{2}$) for the LA image. For the SA and LA images the rms noise is 1.5\,mJy/beam and 0.11\,mJy/beam, respectively.

The AMI SA image shows no distinct features for NGC\,891 due to its low resolution.
The AMI LA image is able to resolve the star-forming disk of NGC\,891 and is similar to the image of the H$\alpha$ line emission from ionized hydrogen gas (Fig.~\ref{fig:haoverlay}) and to the infrared image of thermal emission of warm dust at $24\,\mu$m \citep[see Fig.~6 in][]{Whaley2009}. The northern star-forming region stands also out in the emission of cold dust at $850\,\mu$m \citep{Alton1998}, sub-millimetre line emission from rotational transitions of the CO molecule, a tracer for molecular hydrogen \citep{Garcia1992}, and in optical H$\alpha$ line emission \citep[e.g.][]{Dahlem1994}. 

We measure an integrated flux density that is only about 2\% lower for the LA image than for the SA image (avoiding the negative sidelobes), indicating that no significant flux density is lost in the LA image due to missing spacings. This seems surprising because the largest visible structure the LA is sensitive to is only about 3\arcmin. However, due to the highly elliptical shape of NGC\,891 the visibilities at short baselines that are aligned at the position angle of the major axis can detect the entire flux density.

Very few previous observations of NGC\,891 with a sufficiently small beam size to resolve the disk exist at similar frequencies. The closest in frequency are those of \cite{Gioia1982} and \cite{Dumke1995} who used the Effelsberg 100-m telescope at 10.7\,GHz and 10.55\,GHz, respectively, and found integrated flux densities consistent with the AMI value (see Table~\ref{tab:integratedflux}).

The extent of the disk above the major axis is approximately 2\,kpc which is similar to the beam size at the distance of the galaxy, so that the disk thickness cannot be properly measured. No extended halo emission is observed at the given sensitivity. One feature, observed emanating from the region to the north of the galaxy (marked by `I'), extends to $\approx$3\,kpc from the plane and coincides with X-ray emission of hot gas observed with XMM-Newton at 0.4--0.75\,keV \citep{Hodges2013}. At 13--18\,GHz, a substantial fraction of the emission we are observing is thermal, so this feature could be the result of outflow of warm gas (in addition to hot gas) from star-forming regions in the northern region of the galaxy where the star-formation rate is larger \citep{Dahlem1994}.

\section{Comparison with other wavelengths}
\label{sec:compareotherwavelengths}

\subsection{Radio continuum}
\label{subsec:compareradio}

The most recent radio continuum data for comparison are the 1.5\,GHz and 6\,GHz VLA observations from the CHANG-ES survey \citep{Wiegert2015,Schmidt2016,Schmidt2018}.
When the 146\,MHz image is smoothed to a larger beam (Fig.~\ref{low_res_image}) in order to detect the most extended emission, the halo extends about as far out as in the 1.5\,GHz D-array image (with a similar beam size), but not further out. This is due to the relatively limited sensitivity of our LOFAR image. The exponential scale heights will give us a better indication of the halo extent, as will be shown in Sec.~\ref{sec:radioemissionprofile}.

\subsection{H$\alpha$}
\label{subsec:compareHalpha}

\begin{figure}
\includegraphics[scale=0.35]{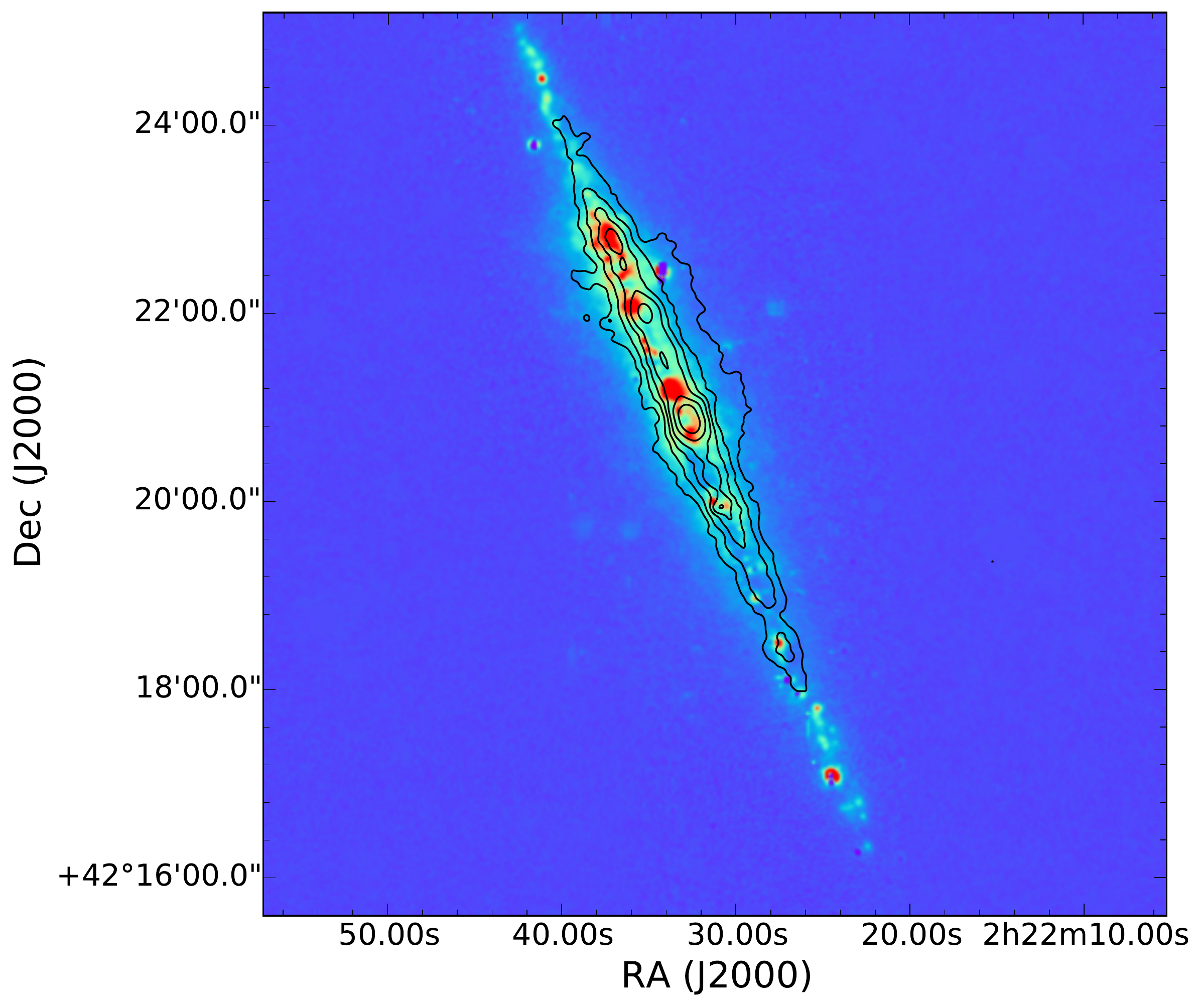}
\caption{H$\alpha$ image of NGC\,891 from \citet{Rand1990} (in colours) with the high resolution LOFAR image (Fig.\ref{fig:NGC891lofarimage} left) overlaid as contours. Contours are at 12, 18, 25, 32, 44, 64 $\times \sigma$
where $\sigma = 0.29$\,mJy/beam is the rms noise level.} 
\label{fig:haoverlay}
\end{figure}

The diffuse H$\alpha$ emission from the halo of NGC\,891 has an exponential scale height of about 1\,kpc \citep{Dettmar1990}. \citet{Rand1990} and \citet{Rossa2004} observed many vertical H$\alpha$ filaments or ``worms'' extending up to 2\,kpc off the plane of the galaxy. They interpreted these ``worms'' as providing evidence for a galactic ``chimney'' mode \citep{Norman1989}. \citet{Rossa2004} speculated that the very narrow H$\alpha$ filaments could be magnetically confined.

We observe a relation between the high-resolution 146\,MHz radio continuum and H$\alpha$ emission only in the H$\alpha$ complex in the northern disk (Fig.~\ref{fig:haoverlay}). At this low frequency, we do not expect significant thermal emission (Sec.~\ref{subsec:thermalprop}) while the 15.5\,GHz emission (Fig.~\ref{fig:AMIimages} right) has a much larger thermal fraction (Sec.~\ref{subsec:integratedspectrum}) and hence is more similar to the H$\alpha$ image. However, the observations at 15.5\,GHz are not sensitive enough to detect diffuse thermal emission from the halo.

Fig.~\ref{fig:composite} shows a composite of radio, H$\alpha$, and optical emission.

\subsection{Neutral gas}
\label{subsec:compareHI}

The galactic fountain model is invoked to explain the huge halo of neutral atomic $\HI$ gas of NGC\,891 \citep{Oosterloo2007}. The extent is up to 22\,kpc from the plane in the north-western quadrant. The exponential scale height increases from 1.25\,kpc in the central regions to about 2.5\,kpc in the outer parts beyond about 15\,kpc radius (``flaring''). The bulk of the cold CO-emitting molecular gas and the cold dust, on the other hand, are much more concentrated to the plane \citep{Scoville1993,Alton1998}, but some CO emission could be traced up to 1.4\,kpc height above the plane \citep{Garcia1992} and infrared emission up to 2.5\,kpc height \citep{Whaley2009}.

We do not observe such a large extension in the north-western quadrant as seen in $\HI$ by \citet{Oosterloo2007}. The size of the radio halo is limited by the synchrotron lifetime of the cosmic-ray electrons of about $2\times 10^8$\,yrs (Sec.~\ref{subsec:lossprocesses}), longer than the duty cycle of a typical galactic fountain of about $10^8$\,yrs \citep{Fraternali2017}. However, the north-western extension is much larger than a typical fountain because either its timescale is larger or the origin is different.

\begin{figure*}[h!]
 \begin{center}
 \includegraphics[scale=0.75,trim={1cm 6cm 1cm 2cm},clip]{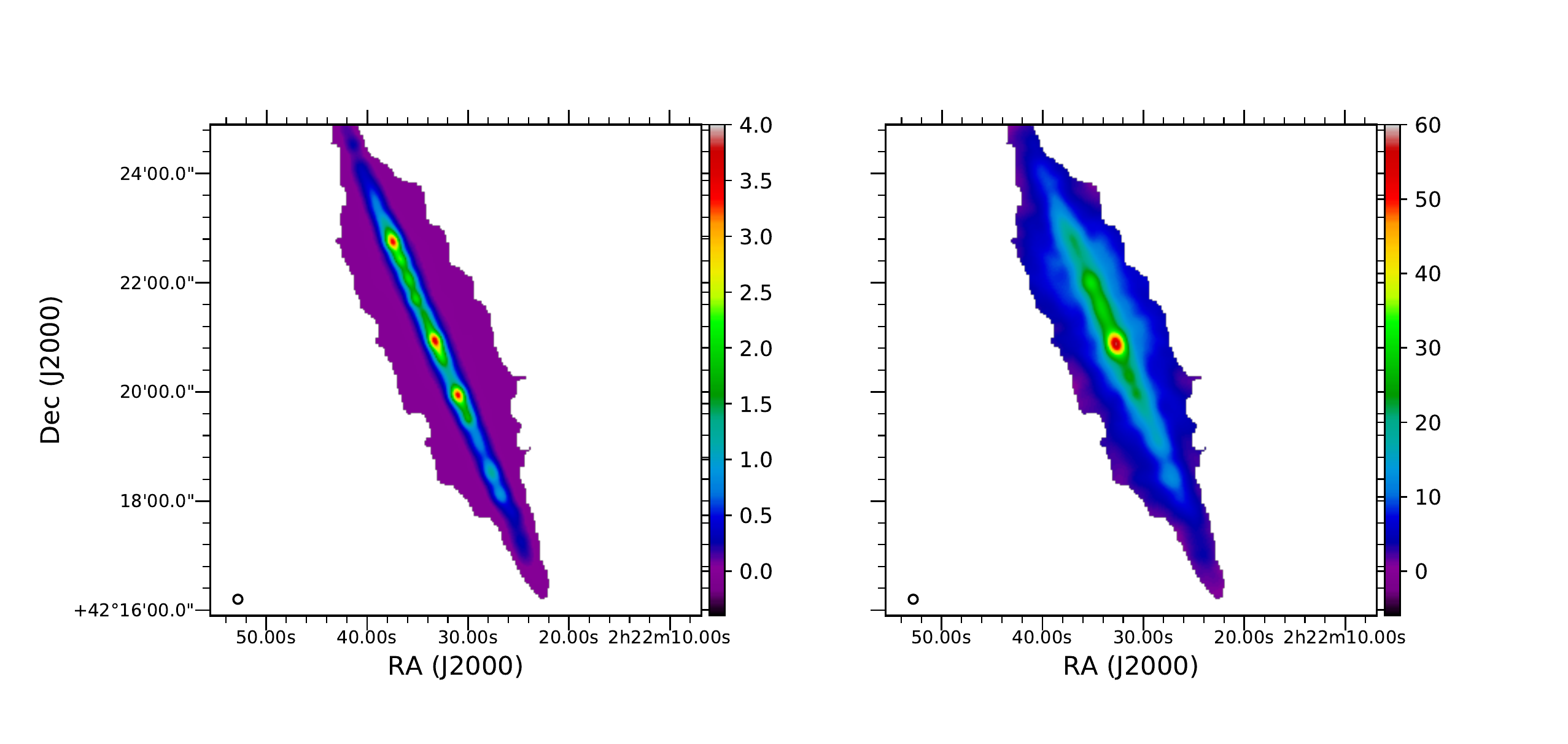}
 \caption{Thermal map (left) and nonthermal map (right) of NGC\,891 at 146\,MHz at a resolution of 12$\arcsec$, shown by the circle in the bottom left corner. The colour scale is in mJy/beam.}
\label{fig:NGC891thermnonthermmap} 
\end{center}
\end{figure*}

\begin{figure}
\includegraphics[height=6.8cm]{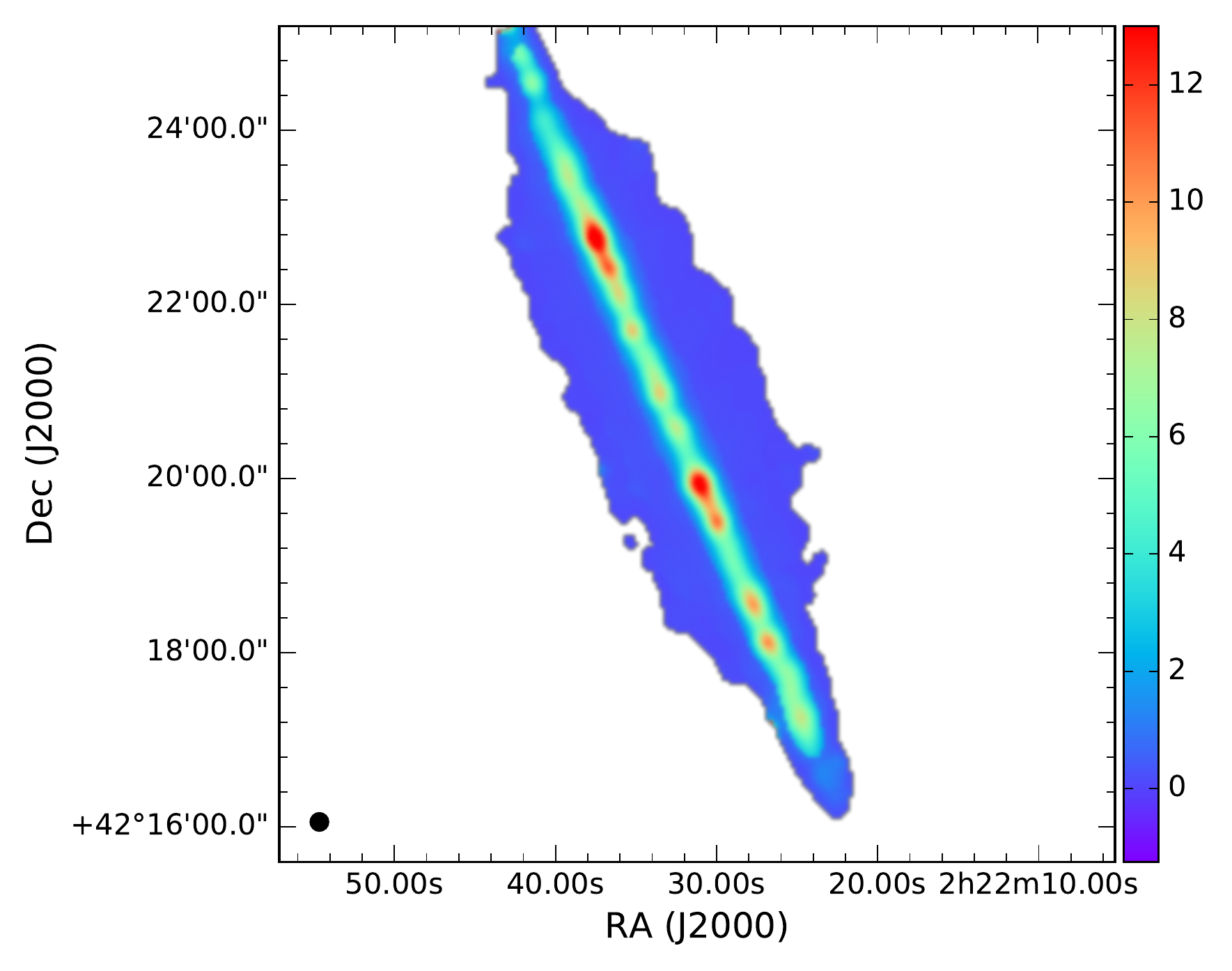}
\caption{Thermal fraction map of NGC\,891 at 146\,MHz at a resolution of 12$\arcsec$. The colour scale is in percent. The beam is shown in the bottom left corner. 
}
\label{fig:thermalfrac}
\end{figure}

\subsection{X-rays}
\label{subsec:comparexray}

X-ray observations performed by \citet{Bregman1994} were able to detect a considerable amount of diffuse X-ray emission from the halo of NGC\,891. The vertical profile is Gaussian with a vertical scale height of 3.5\,kpc, corresponding to a full width at half maximum of 5.8\,kpc \citep{Bregman1997}. \citet{Temple2005} observed X-ray emission protruding from the disk in the north-western direction up to approximately 6\,kpc which showed a sharp cut-off, suggesting that this is the maximum extent that the outflowing hot gas has reached. The authors also concluded that NGC\,891 has a larger star-formation rate than a normal spiral, but not as extreme as the starburst galaxy NGC\,253. Cosmic rays and magnetic fields will also be transported by the outflow, but radio emission cannot be detected at such large heights due to energy losses of the cosmic-ray electrons and the limited sensitivity of present-day radio observations.

\section{NGC\,891's spectral properties}
\label{sec:spectrum}

\subsection{Thermal and nonthermal emission}
\label{subsec:thermalprop}

To measure the nonthermal intensity $I_\mathrm{n}$ and the magnetic field strength, subtracting the free-free (thermal) emission is essential. Furthermore, in order to investigate the energy losses of the CREs, the nonthermal spectral index $\alpha_\mathrm{n}$ ($I_\mathrm{n} \propto \nu^{\alpha_\mathrm{n}}$) needs to be known.

Our AMI LA image cannot be used as a tracer of thermal emission because the angular resolution is too coarse and the emission is still dominated by nonthermal emission (Section~\ref{subsec:integratedspectrum}). A 1.5\,GHz thermal map at a resolution of 12$\arcsec$ was derived from the CHANG-ES VLA data \citep{Schmidt2016,Schmidt2018}, based on an H$\alpha$ image corrected for internal extinction with help of 24\,$\mu$m dust emission. In the inner disk of edge-on galaxies, the 24\,$\mu$m dust emission may become optically thick, so that the extinction correction may be insufficient and hence may lead to an underestimate of the thermal emission \citep{Vargas2018}. This effect was taken into account when estimating the thermal emission the 1.5\,GHz, but leads to a large relative uncertainty of about 40\%.

The integrated thermal flux density of $53\pm23$\,mJy corresponds to a thermal fraction of $7\pm3$\% at 1.5\,GHz. This thermal map was scaled to 146\,MHz, using the spectral index of optically thin thermal emission of $-0.1$. If the assumption of optically thin emission is not valid at 146\,MHz in dense regions of the disk, the thermal emission will be overestimated. Our high resolution LOFAR image was convolved to the same resolution and re-gridded to the same grid size as the thermal map. The scaled thermal map was subtracted from the LOFAR image to produce the nonthermal image of NGC\,891 at 146\,MHz.

The thermal and nonthermal maps at 146\,MHz and at a resolution of 12$\arcsec$ are shown in Fig.~\ref{fig:NGC891thermnonthermmap}. The thermal image displays a thin disk plus a weak diffuse halo, similar to the maps of thermal emission of cold dust observed at 850\,$\mu$m \citep{Alton1998,Israel1999} and of warm dust observed at 24\,$\mu$m \citep{Whaley2009}. In contrast to the thermal emission, the nonthermal disk is not ``thin'' and reveals a smooth transition to the halo.

The thermal fraction image at 146\,MHz was computed from the thermal image (Fig.~\ref{fig:NGC891thermnonthermmap} left) and a total intensity image at the same resolution of 12$\arcsec$ and is shown in Fig.~\ref{fig:thermalfrac}.

The thermal fractions in the disk are between 5\% and 10\% at 146\,MHz compared to 10\% to 20\% at 1.5\,GHz. The largest thermal fraction of approximately 16\% at 146\,MHz and 30\% at 1.5\,GHz occurs in the northern disk of the galaxy. The halo reveals very small thermal fractions at 146\,MHz of 0.1--0.2\% compared to 0.7--0.9\% at 1.5\,GHz, meaning that the radio emission observed in the halo is almost entirely nonthermal.

Free-free absorption by ionized gas in the disk lowers the radio synchrotron intensity at 146\,MHz (Sec.~\ref{subsec:nonthermspectral}) and causes a further overestimation of the thermal fraction. Therefore, the thermal fractions at 146\,MHz estimated for the disk should be regarded as upper limits. Further observations at frequencies lower than 146\,MHz are needed to measure thermal absorption and correct the synchrotron intensity (see Section~\ref{subsec:discussionthermal}).

\subsection{Integrated spectrum of NGC\,891}
\label{subsec:integratedspectrum}

We obtain an integrated flux density of NGC\,891 with LOFAR at 146\,MHz of 2.85$\pm$0.28\,Jy. The flux density of SN1986J was subtracted. The largest cause of uncertainty (about 10\%) is the limited accuracy of the beam model of LOFAR affecting the transfer of gains.
The integrated flux density with AMI at 15.5\,GHz is 120$\pm$12\,mJy, assuming a 10\% uncertainty.

A whole range of flux density measurements from the literature was found with many of the flux density measurements taken from \citet{Hummel1991}.
Several flux density measurements from the literature were found to have either no uncertainty quoted or seriously underestimated their uncertainties by only including the rms noise (of typically a few \%) and not including any calibration uncertainty. In these cases we have inserted a 10\% uncertainty to these flux density values. A full list of the flux density measurements is given in Table~\ref{tab:integratedflux}.

\begin{figure}[h!]
 \begin{center}
 \includegraphics[scale=0.35]{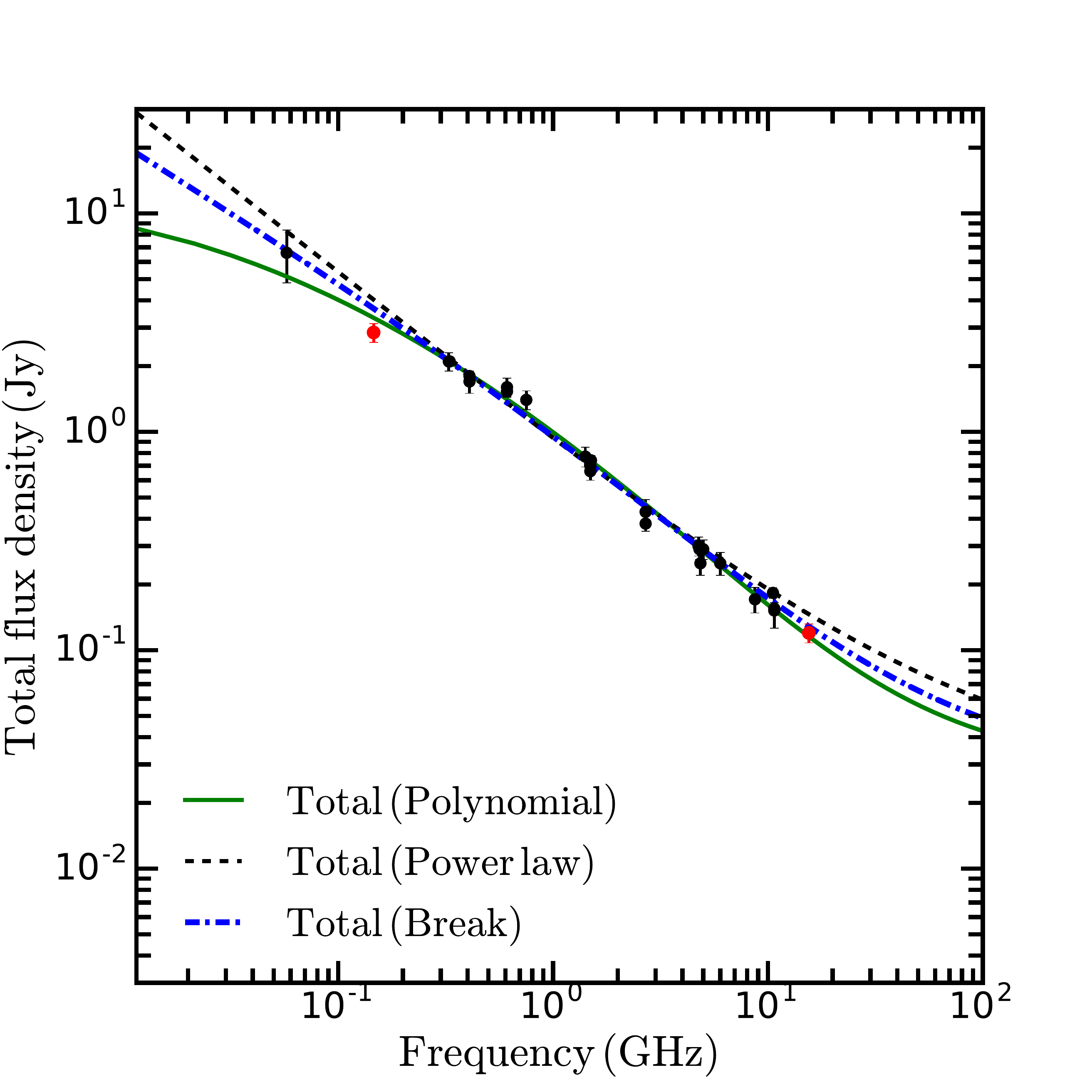}
 \includegraphics[scale=0.35]{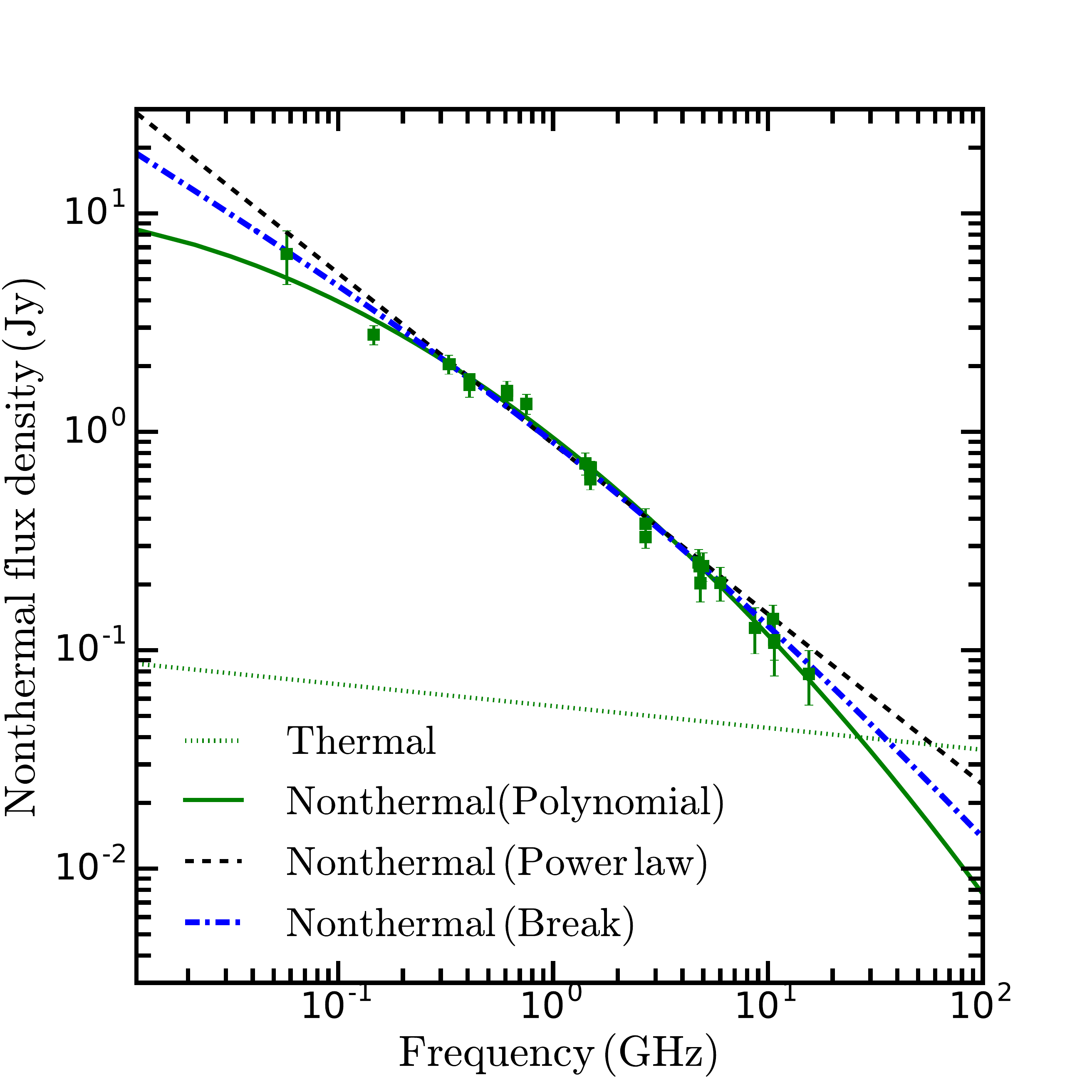}
 \caption{Top: Integrated total flux densities of NGC\,891 as listed in Table~\ref{tab:integratedflux}. The two new total flux densities from this paper are marked in red. Bottom: Integrated nonthermal flux densities of NGC\,891 after subtracting an estimate of the optically thin thermal free--free emission (shown as the green dotted line) from the total flux densities. The black dashed, solid green, and blue dashed--dotted lines in the bottom panel show the best-fit power law, second-order polynomial model, and spectral break model to the nonthermal emission, respectively. The corresponding lines in the top panel show the expected total flux density spectrum after adding the thermal free--free emission to the different models of the nonthermal spectrum.}
\label{fig:NGC891integratedflux} 
\end{center}
\end{figure}

We subtracted the thermal emission at all frequencies, based on the value of $53 \pm 23$\,mJy at 1.5\,GHz \citep{Schmidt2016,Schmidt2018} and using the spectral index of optically thin thermal emission of $-0.1$. The thermal fraction of the integrated flux density is $\le2\%$ at 146\,MHz, $7 \pm 3$\% at 1.5\,GHz, and $35 \pm 15$\% at 15.5\,GHz. The spectrum of integrated nonthermal emission is plotted in Fig.~\ref{fig:NGC891integratedflux}.

The thermal fraction of NGC\,891 at 1.5\,GHz fits well to the average thermal fractions at 1.4\,GHz obtained for the samples of spiral galaxies by \citet{Niklas1997} ($8\pm1\%$), \citet{Marvil2015} ($9\pm3\%$), and \citet{Tabatabaei2017} ($10\pm9\%$). \footnote{The average thermal fraction derived for a sample of star-forming galaxies by \citet{Klein2017} is more than twice larger, but these galaxies show indications for a break or an exponential decline in their nonthermal radio spectra which makes the estimate of the thermal fraction difficult.} This result indicates that the 24\,$\mu$m intensities of edge-on galaxies indeed needs to be corrected for extinction, as proposed by \citet{Vargas2018}.







The shape of the spectrum of total radio continuum emission is determined by the relative contribution of the thermal free--free emission and the shape of the nonthermal (synchrotron) spectrum. As the disk dominates the radio emission from NGC\,891, spectral effects in the disk are more important for the integrated emission than those in the halo. For example, at high frequencies, typically above 5\,GHz, the increasing thermal fraction in the disk could lead to a spectral flattening. At lower frequencies, depending on the level of free--free absorption by ionized gas in the disk and the nature of the nonthermal spectrum, the total radio continuum spectrum could develop a spectral flattening, typically at $\lesssim 300$\,MHz. The shape of the nonthermal spectrum depends on the dominant energy loss/gain mechanisms which the synchrotron-emitting CREs undergo (see Sec.~\ref{subsec:lossprocesses}). A transition from dominating bremsstrahlung and adiabatic losses to dominating synchrotron and/or inverse-Compton (IC) losses leads to a spectral steepening by $-0.5$ beyond a certain frequency.
Dominating ionization losses of low-energy CREs ($\lesssim 1.5$\,GeV) could also lead to a flattening of the nonthermal spectrum at low radio frequencies by $+0.5$ in regions of high gas density \citep[e.g.][]{Basu2015}. We note that throughout the disk and halo spatially varying magnetic fields and gas densities (both neutral and ionized) lead to locally varying breaks in the CRE energy spectrum, such that the corresponding breaks in the galaxy's integrated radio continuum spectrum are smoothed out \citep{Basu2015}.

\begin{table}[h!]
\caption{Integrated flux densities of NGC\,891. 
Uncertainties marked in bold were increased from their original values, as explained in the text. 
} 
\centering 
\begin{tabular}{c l l} 
\hline\hline 
$\nu$ (GHz) & Flux density (Jy) & Ref. \\ 
\hline 
15.5 & 0.120$\pm$0.012 & This work \\
10.7 & 0.152$\pm$0.026 & \cite{Gioia1982} \\
10.7 & 0.155$\pm$0.010 & \cite{Israel1983} \\
10.55 & 0.183$\pm$0.010 & \cite{Dumke1995} \\
8.7 & 0.171$\pm$0.023 & \cite{Beck1979} \\
6.0 & 0.25$\pm$0.03 & \cite{Schmidt2018} \\
4.995 & 0.29$\pm${\bf 0.03} & \cite{Allen1978} \\
4.85 & 0.25$\pm${\bf 0.03} & \cite{Gregory1991}\\
4.8 & 0.29$\pm${\bf 0.03} & \cite{Stil2009} \\
4.75 & 0.30$\pm${\bf 0.03} & \cite{Gioia1982} \\
2.695 & 0.43$\pm$0.06 & \cite{Kazes1970}\\
2.695 & 0.38$\pm$0.03 & \cite{deJong1967} \\
1.5 & 0.74$\pm$0.04 & \cite{Schmidt2018} \\
1.49 &0.74$\pm$0.02 & \cite{Hummel1991} \\
1.49 & 0.66$\pm$0.06 & \cite{Gioia1987} \\
1.49 &0.70$\pm${\bf 0.07} & \cite{Condon1987} \\
1.412 & 0.77$\pm${\bf 0.08} & \cite{Allen1978} \\
0.75 & 1.4$\pm${\bf 0.14} & \cite{Heeschen1964} \\
0.61 & 1.53$\pm$0.08 & \cite{Hummel1991} \\
0.61 & 1.6$\pm${\bf 0.16} & \cite{Allen1978} \\
0.408 & 1.8$\pm$0.1 & \cite{Gioia1980} \\
0.408 & 1.7$\pm${\bf 0.2} & \cite{Baldwin1973} \\
0.327 & 2.1$\pm${\bf 0.2} & \cite{Hummel1991} \\
0.330 & 2.1$\pm$0.1 & \cite{Rengelink1997} \\
0.146 & 2.85$\pm$0.28 & This work\\
0.0575 & 6.6$\pm$1.8 & \cite{Israel1990} \\
\hline
\end{tabular}
\label{tab:integratedflux}
\end{table}

To study the nonthermal spectrum of NGC\,891
we first model it using a simple power law of the form
$S_{\rm n}(\nu) = a_0\,\nu^{\alpha_{\rm n}}$.
Here, $S_{\rm n}$ is the nonthermal flux density, $\alpha_{\rm n}$ is the
nonthermal spectral index and $a_0$ is the normalization at 1\,GHz. We find the
best fit $\alpha_{\rm n} = -0.78 \pm 0.02$, with a reduced $\chi^2 =
1.87$. The best-fit power-law spectrum is shown as the black dashed line in
Fig.~\ref{fig:NGC891integratedflux} (bottom panel). The expected total intensity spectrum after adding the thermal spectrum to the nonthermal power-law spectrum is shown as the black dashed line in the top panel of Fig.~\ref{fig:NGC891integratedflux}. This total flux density spectrum has a
reduced $\chi^2 = 2.62$. Clearly, a simple power law does not represent the
integrated spectrum well.

In order to assess any curvature in the nonthermal spectrum of NGC\,891, we
empirically model it with a second-order polynomial of the form $\log\,S_{\rm
n}(\nu) = \log\,a_0 + \alpha\,\log\,\nu + \beta\,(\log\,\nu)^2$. Here,
$\alpha$ is the spectral index and $\beta$ is the curvature parameter. The best
fit values are found to be $a_0 = 0.94 \pm 0.02$, $\alpha = -0.76 \pm 0.02$, and
$\beta = -0.14 \pm 0.03$. The best-fit model is shown as the green solid line
in Fig.~\ref{fig:NGC891integratedflux}. The reduced $\chi^2$ for the fit is 0.86 and that for the total emission is 1.32, suggesting that the nonthermal spectrum significantly deviates from that of a simple power law.

To understand the physical origin of the curvature, we also performed modelling of the nonthermal spectrum of NGC\,891 with a spectral break given by $S_{\rm n}(\nu) = a_0\,\nu^{\alpha_{\rm inj}}/[(\nu/\nu_{\rm br})^{0.5} + 1]$. Here,
$\alpha_{\rm inj}$ is the injection spectral index of the CREs and $\nu_{\rm
br}$ is the break frequency beyond which the spectrum steepens by $-0.5$.
Unfortunately, with the current data the parameters for this model cannot be
well constrained. We therefore fixed $\alpha_{\rm inj} = -0.6$ \citep[see Fig.~3 in][]{Caprioli2011}. This yields a break at $\nu_{\rm br} = 3.7 \pm 1.3$\,GHz which corresponds to a synchrotron age (Eq.~\ref{eq:sync}) of $\lesssim 2.5 \times 10^7$\,yrs for the CREs in the disk emitting in a magnetic field of $\gtrsim10\,\mu$G (Section~\ref{sec:magneticfield}). The best fit is shown as the blue dashed--dotted line in Fig.~\ref{fig:NGC891integratedflux}. It gives a reduced $\chi^2 = 1.19$ for the nonthermal emission and a reduced $\chi^2 = 1.57$ for the total emission, better than the values for a simple power law.

As discussed above, the curvature in the nonthermal spectrum could also arise as the result of synchrotron--free absorption and/or ionization losses at frequencies below about 150\,MHz \citep[see Fig.~3 in][]{Basu2015}. To investigate the first scenario, a rigorous modelling of the synchrotron radiative transfer in an inclined disk is necessary, while for the later case additional information on the $\gamma$--ray spectrum is needed. Given the scarcity of low-frequency observations and the large uncertainty in the estimated thermal emission, the current data are insufficient to constrain the origin of the curvature in the nonthermal spectrum of NGC\,891.

To better constrain the radio continuum spectrum, additional radio frequency observations at low frequencies would be helpful. The 146\,MHz flux density derived in this paper seems to indicate a spectral flattening at low frequencies by free-free absorption. However, this is inconsistent with the value at 57.5\,MHz. The latter value is based on observations with the Clark Lake Radio Observatory synthesis radio telescope with a synthesized beam size of about $7\arcmin$ \citep{Israel1990}. As the frequency range below 100\,MHz is crucial to detect spectral flattening by free-free absorption, new observations with higher resolution are needed, e.g. with the LOFAR Low Band Antenna (see Sec.~\ref{subsec:discussionthermal}).

\subsection{Spectral index map of NGC\,891}
\label{subsec:spectralindexmaps}

\begin{figure*}[h!]
\begin{center}
\includegraphics[width=\textwidth]{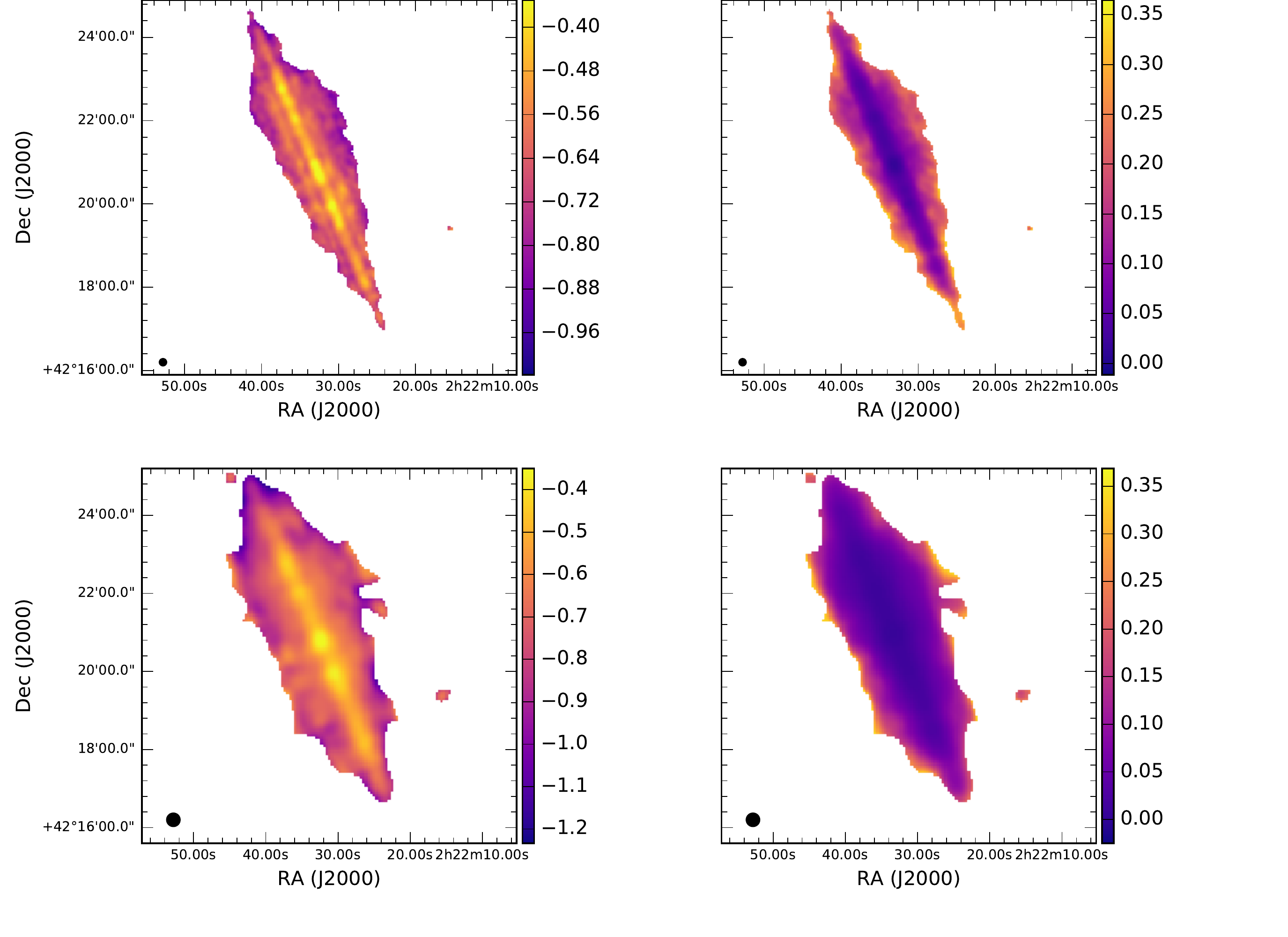}
\caption{Spectral index (left) and error map (right) of NGC\,891 between 146\,MHz and 1.5\,GHz at resolutions of 10$\arcsec$ (top) and 20$\arcsec$ (bottom).}
\end{center}
\label{fig:NGC891totalspecindexmap} 
\end{figure*}

To derive spectral index maps of high accuracy, the frequency span should be as large as possible. However, our new AMI image at 15.5\,GHz does not provide sufficient angular resolution to be combined with our new LOFAR image at 146\,MHz. Hence, spectral index maps were created from the CHANG-ES VLA image of NGC\,891 at 1.5\,GHz \citep{Schmidt2016,Schmidt2018}~\footnote{We did not use the 6\,GHz CHANG-ES image because the signal-to-noise ratios are smaller than those at 1.5\,GHz.} and our LOFAR image. In order to be sensitive to the same angular scales, both images were generated with the same minimum and maximum \textit{uv} distance and the same weighting scheme, smoothed to two different resolutions, $10\arcsec$ and $20\arcsec$, and placed onto the same grid via the {\tt AIPS} task {\tt OHGEO}. Only pixels with flux densities above 8x the rms ($\sigma$) level in the four input images were used.

The minimum projected baseline for the VLA is about 27\,m, corresponding in units of wavelength to a baseline about 270\,m for LOFAR at 146\,MHz. This \textit{uv$_{min}$} value corresponds to an angular scale of about $25\arcmin$ which is much larger than the angular scale of NGC\,891. Hence, we do not expect the spectral index distribution to be affected by systematic errors due to missing angular scales.

Before determining the spectral index, both images were set on the same flux density scale. The 146\,MHz LOFAR image was calibrated on the Scaife \& Heald flux density scale \citep{Scaife2012}, while the 1.5\,GHz image was calibrated on the Perley \& Butler (2010) flux density scale, a preliminary version of \cite{Perley2013}. We first converted the 1.5\,GHz image to the Baars scale \citep{Baars1977A&A} by a factor of 1.021 taken from Table~13 of \cite{Perley2013}. We then converted this flux density to the KPW scale \citep{Kellermann1969} which is identical to the Scaife \& Heald flux density scale at frequencies above 325\,MHz, using a factor of 1.029 taken from Table~7 of \cite{Baars1977A&A}.

The spectral index between 146\,MHz and 1.5\,GHz was computed pixel by pixel and is shown in Fig.~\ref{fig:NGC891totalspecindexmap} along with the image of uncertainties (errors) due to rms noise for both resolutions. The spectral index map at 10$\arcsec$ resolution (Fig.~\ref{fig:NGC891totalspecindexmap} top) reveals great detail on the disk's spectral features. We observe very flat spectra in the central region ($\alpha \approx -0.3$), coincident with prominent $\HII$ regions. In other regions in the disk we observe spectra with $\alpha \approx-0.5$. This is flatter than what is observed at higher frequencies. \citet{Schmidt2018} found the spectral index between 1.5\,GHz and 6\,GHz to be $\alpha \approx-0.7$ in the disk.
Immediately away from the disk, we observe spectral indices of $\approx -0.6$ to $-0.7$.

The spectral index map at 20$\arcsec$ resolution (Fig.~\ref{fig:NGC891totalspecindexmap} bottom) shows a similar steepening of the spectral index from the disk to the halo. Due to the higher sensitivity with respect to weak extended emission, we can trace the spectral index further away from the disk into the halo.

\subsection{Nonthermal spectral index of NGC\,891}
\label{subsec:nonthermspectral}

\begin{figure*}[h!]
\begin{center}
\includegraphics[height=6.8cm]{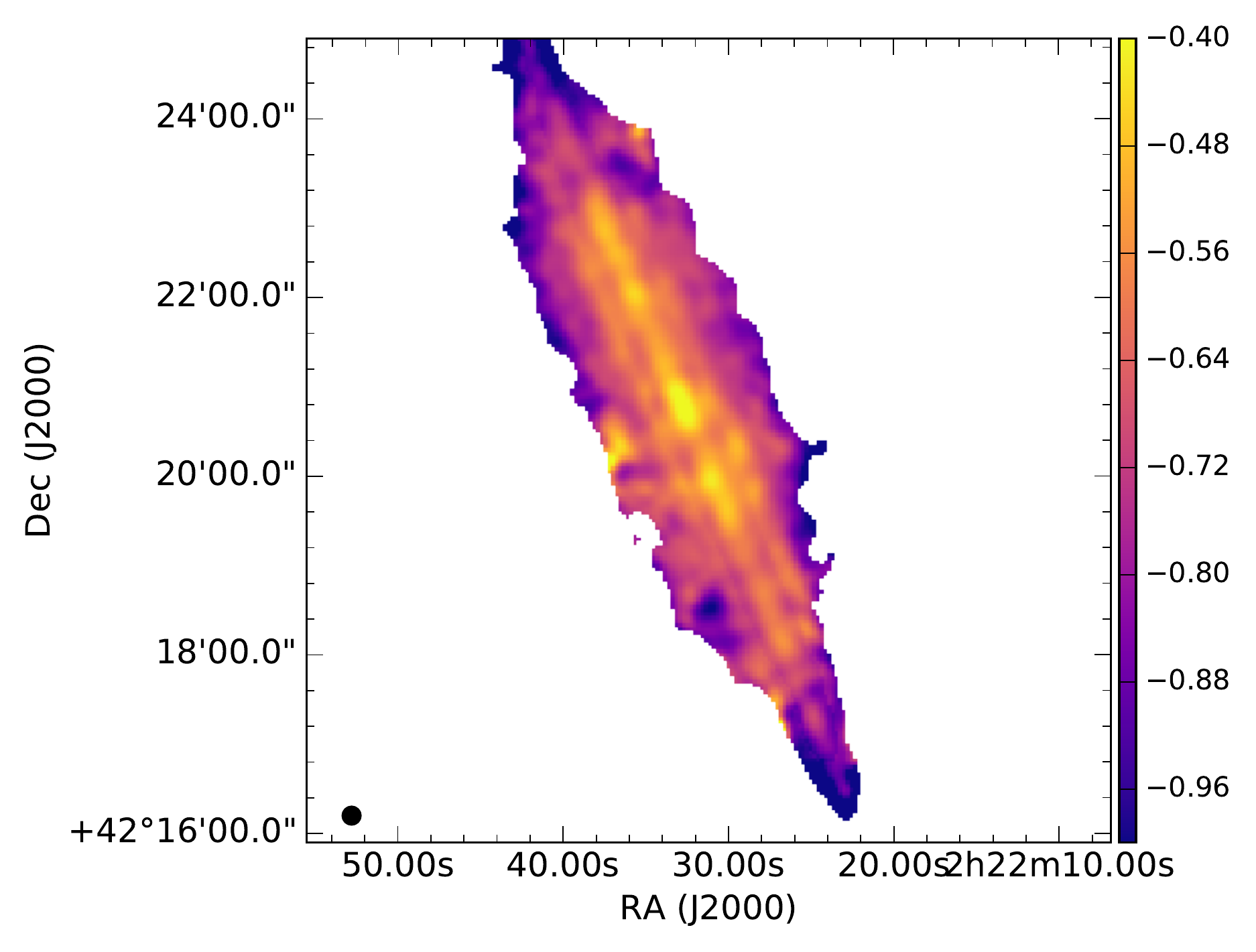}
\includegraphics[height=6.8cm]{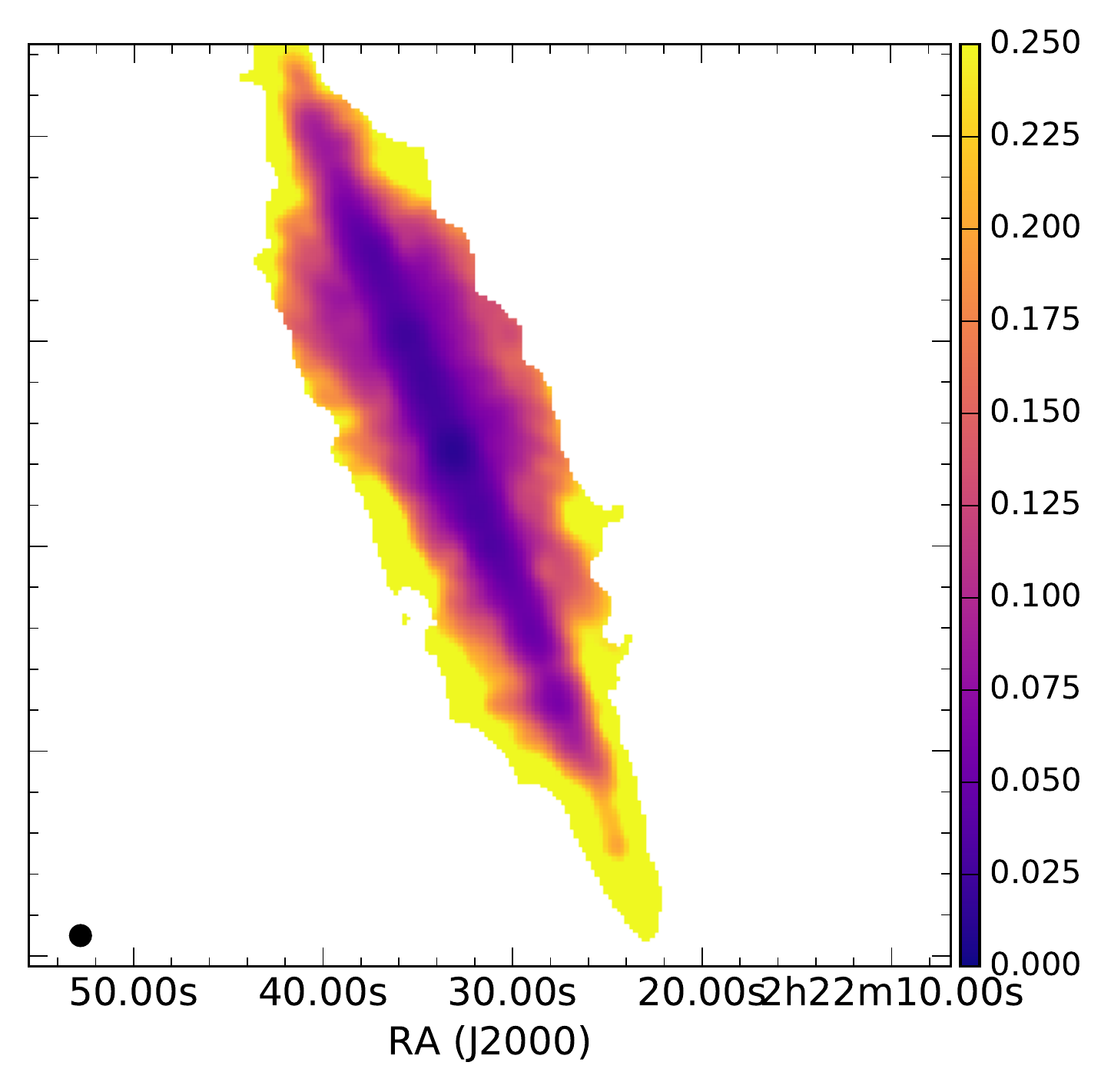}
\end{center}
\caption{Maps of nonthermal spectral index (left) and corresponding error map (right) between 146\,MHz and 1.5\,GHz at 12$\arcsec$ resolution.
The beam is shown in the bottom left corner.}
\label{fig:NGC891nonthermspecindexmap} 
\end{figure*}


The spectral index is contaminated by thermal emission that is unrelated to CREs and magnetic fields. Therefore, we computed a map of the nonthermal spectral index $\alpha_\mathrm{n}$, using the nonthermal maps at 146\,MHz at 12$\arcsec$ resolution (Fig.~\ref{fig:NGC891thermnonthermmap} right) and at 1.5\,GHz at the same resolution.

The nonthermal spectral indices in the disk of the galaxy show significant signs of flattening. In the very centre of the galaxy we measure a spectral index of $-0.37$. We measure spectral indices of $-0.43$ to $-0.48$ in the northern and southern star-forming regions of the disk, respectively. Other regions in the disk show spectral indices of $-0.5$ to $-0.6$.

In the disk the observed CRE population is a superposition of various spectral ages with young CREs located in the major star-forming regions, while older CREs would exist in the inter-arm regions \citep{Tabatabaei2007} and in the halo. Therefore, we expect to observe the injection spectral index in star-forming regions and a steeper spectral index in inter-arm regions and in the halo. Indeed, a significant arm--interarm contrast was observed for M\,51 \citep{mulcahy2014}, with $\alpha_\mathrm{n}=-0.8$ between 151\,MHz and 1.4\,GHz in the inter-arm regions.

CREs are accelerated in the shock fronts of supernova remnants (SNRs). Observations of the $\gamma$-ray emission from bright SNRs yield an average energy spectral index of CRs in the energy range 100\,MeV--50 GeV of $\gamma_\mathrm{inj} -2.2\pm 0.2$ \citep{Caprioli2011} where the CRE number density spectrum is a power law as a function of energy, $N \propto E^{\gamma_\mathrm{inj}}$. This is equivalent to an average injection spectral index in the radio range of $\alpha_\mathrm{inj} -0.6 \pm 0.1$, consistent with the average spectral index of the radio emission from SNRs of $-0.5$, though with a large dispersion \citep{Green2014}.

Nonthermal spectral indices of $\alpha_\mathrm{n} \approx -0.7$ between 1.4\,GHz and 4.86\,GHz are observed in the disk of NGC\,891 \citep{Heesen2018a}.
CREs in the disk emitting at frequencies of a few GHz are still young and have almost maintained their injection spectrum. At lower frequencies we are observing flatter spectra with $\alpha_\mathrm{n, 146MHz-1.5GHz} > \alpha_\mathrm{inj}$ which indicates that we are observing significant thermal (free-free) absorption of synchrotron emission in star-forming regions in the disk (see Section~\ref{subsec:discussionthermal}). 

In the halo we find $\alpha_\mathrm{n} \leq -0.8$, revealing CRE ageing through synchrotron and/or IC emission. These spectra are still significantly flatter compared to what is observed at higher frequencies \citep{Heesen2018a}.
The nonthermal spectra in the halo steepen with increasing frequency, as expected from energy losses of CREs by synchrotron or IC emission (see Section~\ref{subsec:lossprocesses}).

\section{Magnetic field strength in the halo of NGC\,891}
\label{sec:magneticfield}

The strength of the magnetic field (including turbulent and ordered components) can be determined from the nonthermal emission by assuming equipartition between the energy densities of total cosmic rays and magnetic fields, using the revised formula of \cite{Beck2005}. The equipartition magnetic field strength $B$ scales with the synchrotron intensity $I_\mathrm{syn}$ as:
\begin{equation}
B \, \propto \, \left( 
\frac{I_\mathrm{syn}}{(K+1) \,\, L}
\right)^{1/(3-\alpha_\mathrm{n})}
\end{equation}
where $\alpha_\mathrm{n}$ is the nonthermal spectral index. 
The adopted ratio of CR proton to electron number densities of $K = 100$ is a reasonable assumption in the star-forming regions in the disk \citep{Bell1978}. Strong energy losses of CREs due to synchrotron and/or IC emission (Sec.~\ref{subsec:lossprocesses}) are signified by $\alpha_\mathrm{n} \leq -1.1$ in Fig.~\ref{fig:NGC891nonthermspecindexmap} and occur at the outskirts of NGC\,891. These lead to an increase of $K$ by a factor of a few and hence to an underestimate of $B$ in these regions by about 10--30\%.

The effective path length through NGC\,891, $L$, is assumed to be 0.8\,times the radio diameter, resulting in $L = 19 $\,kpc. Depending on the shape and extent of the magnetic fields in the halo, $L$ decreases with distance from the major axis and also with distance from the galaxy's centre along the major axis. We assumed a constant value of $L$ which results in an underestimate of $B$ by another 10--30\%. An improved model of the distribution of magnetic field strengths will be presented in a forthcoming paper \citep{Schmidt2018}.


\begin{figure}
\includegraphics[height=6.8cm]{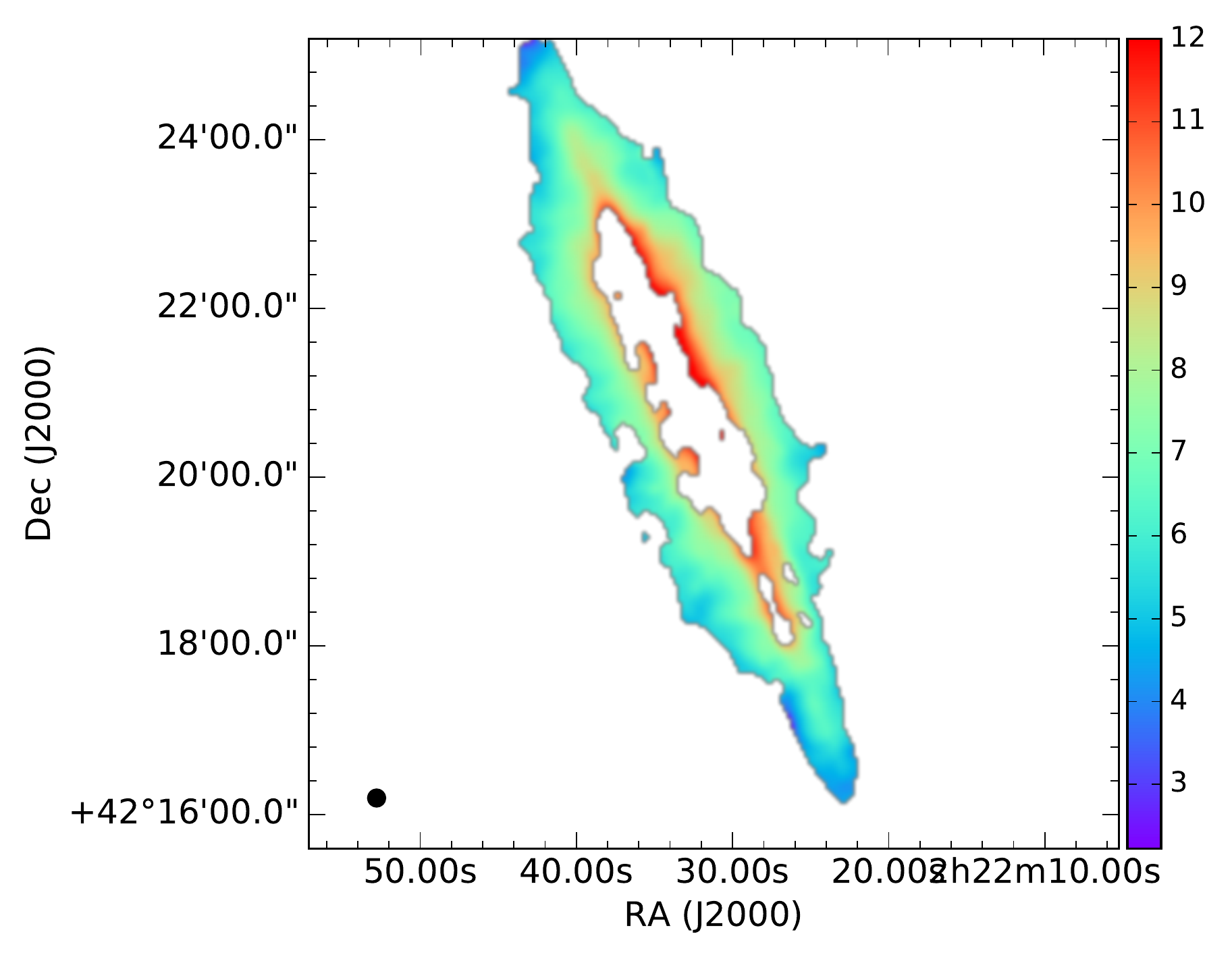}
\caption{Equipartition strength of the magnetic field $B$ at 12$\arcsec$ resolution, in units of $\mu$G, determined by assuming energy equipartition between cosmic rays and magnetic fields. Pixels with $\alpha_\mathrm{n} > -0.6$ are blanked.
The beam is shown in the bottom left corner.
}
\label{fig:bfieldmap}
\end{figure}

We created an image of the equipartition field strength $B$ in NGC\,891 from the nonthermal total intensity (Fig.~\ref{fig:NGC891thermnonthermmap} right) and the nonthermal spectral index (Fig.~\ref{fig:NGC891nonthermspecindexmap} left), presented in Fig.~\ref{fig:bfieldmap}. We blanked pixels with $\alpha_\mathrm{n} > -0.6$ as such regions are likely suffering from free-free absorption and would significantly underestimate the magnetic field strength. Blanking regions where $\alpha_\mathrm{n} > -0.6$ results in a large fraction of the inner disk without a magnetic field strength estimate and highlights the difficulty posed by free-free absorption in determining the magnetic field strength.

The field strength is found to be highest near to the plane with a field strength of 10--11\,$\mu$G, similar to the average field strength in the disks of mildly inclined galaxies \citep{Beck2015} and of edge-on galaxies \citep{Krause2017}. The average equipartition field strength in the halo, between 1\,kpc and 3\,kpc height, is 7\,$\mu$G with a standard deviation of 2\,$\mu$G. 

\section{Radio emission profiles and scale heights}
\label{sec:radioemissionprofile}

The vertical distributions of radio emission and their scale heights in the halos of star-forming edge-on galaxies have been studied in numerous works \citep[e.g.][]{Dumke1995,Heesen2009,Krause2017} and is a vital tool for the analysis of cosmic ray propagation and CRE energy losses. Data below 300\,MHz have never been used before, because sensitivity and angular resolution at low frequencies were insufficient so far. The MWA images of NGC\,253 between 88 and 215\,MHz have too low angular resolutions (between $5\arcmin$ and $2\arcmin$) for this purpose.

\begin{figure}
 \begin{center}
 \includegraphics[scale=0.35]{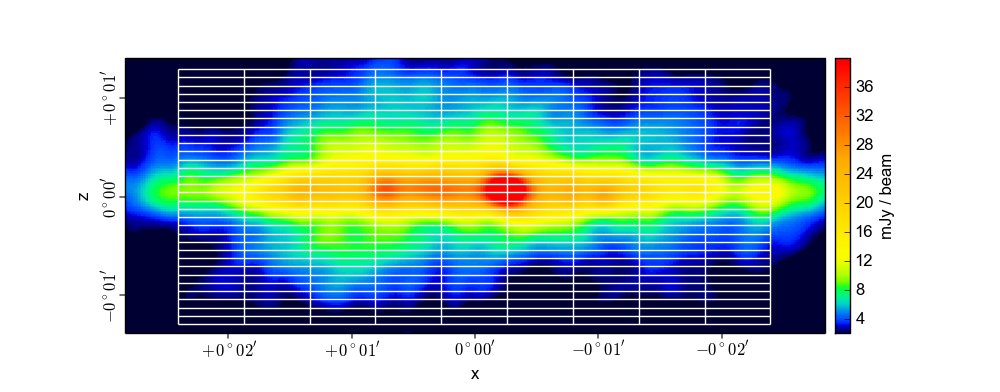}
 \caption{Image of NGC\,891 with indication of the boxes used for computing the average intensities, at different distances from the major axis of the galaxy's projected disk, and in 9 different strips along the major axis.}
\label{fig:ngc891boxes}
\end{center}
\end{figure}

\begin{table*}[h!]
\caption{Scale heights $h$ of the fits to the total intensity at 146\,MHz at $12\arcsec$ (0.55\,kpc) resolution for the 9 strips at different distances, $x$, from the galaxy's centre along the major axis, and to the nonthermal intensity at 1.5\,GHz, taken from \citet{Schmidt2018}. $x$ is positive on the north-eastern side and negative on the south-western side. $q$ is the ratio of the scale heights at 146\,MHz and 1.5\,GHz.} 
\centering 
\begin{tabular}{c c c c c | c c c | c c} 
\hline\hline 
$x$ ($\arcsec$) & $x$ (kpc) &  $h_\mathrm{disk}$ (kpc) & $h_\mathrm{halo}$ (kpc) &  $\chi_\mathrm{red}^2$ & $h_\mathrm{disk}$ (kpc)  & $h_\mathrm{halo}$ (kpc) & $\chi_\mathrm{red}^2$ & $q_\mathrm{disk}$ & $q_\mathrm{halo}$\\ 
  &     &      146\,MHz     &       146\,MHz   &          &  1.5\,GHz & 1.5\,GHz & & & \\
\hline 
+160 & +7.37 & 0.47$\pm$0.11   & 3.6$\pm$2.0 & 2.5 & 0.25$\pm$0.03 & 1.66$\pm$0.07 & 0.31 & 1.9$\pm$0.5  & 2.1$\pm$1.2 \\
+120 & +5.53 & 0.23$\pm$0.04   & 2.0$\pm$0.2 & 2.4 & 0.15$\pm$0.01 & 1.40$\pm$0.03 & 0.37 & 1.5$\pm$0.3  & 1.5$\pm$0.2 \\
+80  & +3.68 & 0.21$\pm$0.07    & 2.0$\pm$0.2 & 4.4 & 0.14$\pm$0.03 & 1.38$\pm$0.06 & 4.8  & 1.5$\pm$0.6  & 1.4$\pm$0.2 \\
+40  & +1.84 & 0.34$\pm$0.09    & 2.1$\pm$0.4 & 6.1 & 0.29$\pm$0.06 & 1.48$\pm$0.19 & 13   & 1.2$\pm$0.4  & 1.4$\pm$0.3 \\
0    & 0.00  & 0.26$\pm$0.06     & 1.8$\pm$0.3 & 12 & 0.13$\pm$0.02 & 1.13$\pm$0.09 & 30   & 2.0$\pm$0.6  & 1.6$\pm$0.3 \\
--40 & --1.84 & 0.40$\pm$0.19  & 1.6$\pm$0.6 & 9.0 & 0.21$\pm$0.06 & 1.07$\pm$0.08 & 7.0  & 1.9$\pm$1.1  & 1.5$\pm$0.6 \\
--80 & --3.68 & 0.30$\pm$0.06  & 1.7$\pm$0.2 & 2.8 & 0.21$\pm$0.04 & 1.14$\pm$0.08 & 4.6  & 1.4$\pm$0.4  & 1.5$\pm$0.2 \\
--120 & --5.53 & 0.30$\pm$0.06 & 2.5$\pm$0.4 & 1.3 & 0.22$\pm$0.03 & 1.45$\pm$0.06 & 1.1  & 1.4$\pm$0.3  & 1.7$\pm$0.3 \\
--160 & --7.37 & 0.39$\pm$0.08 & 3.5$\pm$1.2 & 3.9 & 0.21$\pm$0.02 & 1.51$\pm$0.05 & 0.26 & 1.9$\pm$0.4  & 2.3$\pm$0.8 \\
\hline 
\end{tabular}
\label{tab:scaleheights}
\end{table*}

\begin{figure*}
 \begin{center}
 \includegraphics[width=16cm]{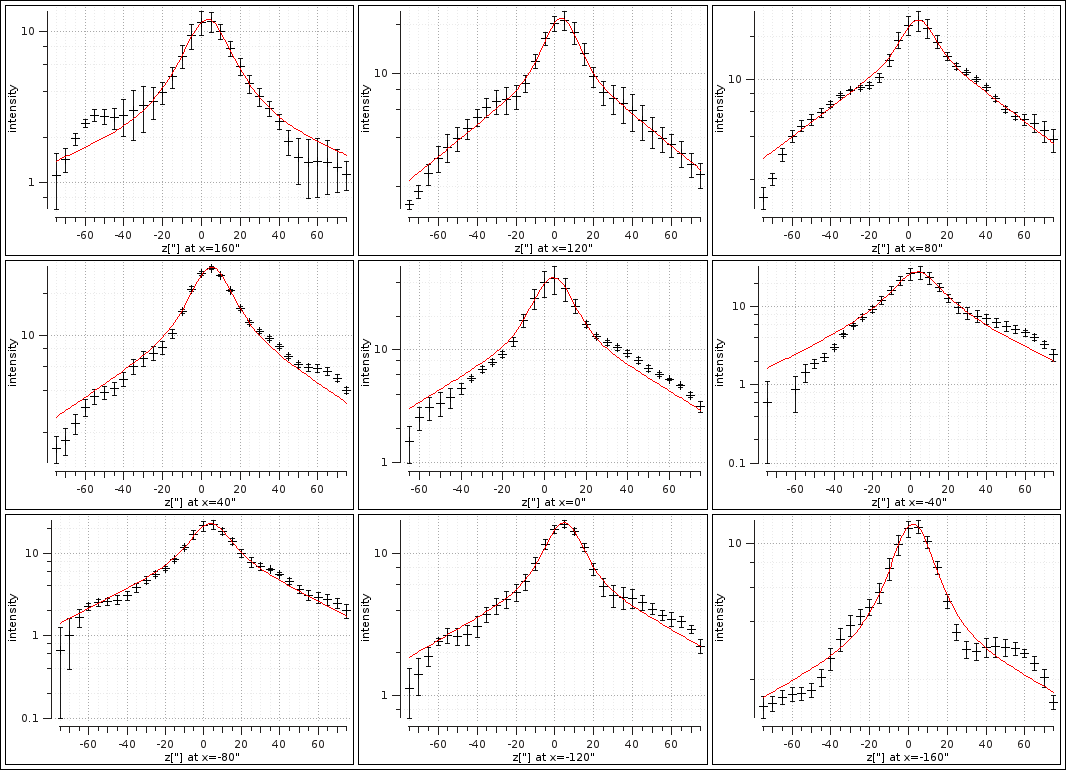}
 \caption{Profiles of the average intensity at 146\,MHz at $12\arcsec$ resolution (in mJy/beam) as a function of distance $z$ from the major axis (in arcseconds, $10\arcsec \simeq 0.46~\rm kpc$), for the 9 strips of NGC\,891 at different distances, $x$, from the galaxy's centre along the major axis, corresponding to Fig.~\ref{fig:ngc891boxes}. The first panel is at x=+7.37\,kpc (northern side), the last one at x=--7.37\,kpc (southern side). $z$ is negative (positive) on the eastern (western) side of the halo. The red lines show the two-component exponential functions fitted to the data.}
\label{fig:scaleheights}
\end{center}
\end{figure*}

The sensitivity of the observations limits the detectability of an extended halo. Therefore, the total observed extent is not suited as a physical parameter describing the halo emission, and a more objective parameter such as the scale height is required \citep{Krause2017}.

\begin{figure}[h!]
 \begin{center}
\includegraphics[angle=90,width=8.5cm]{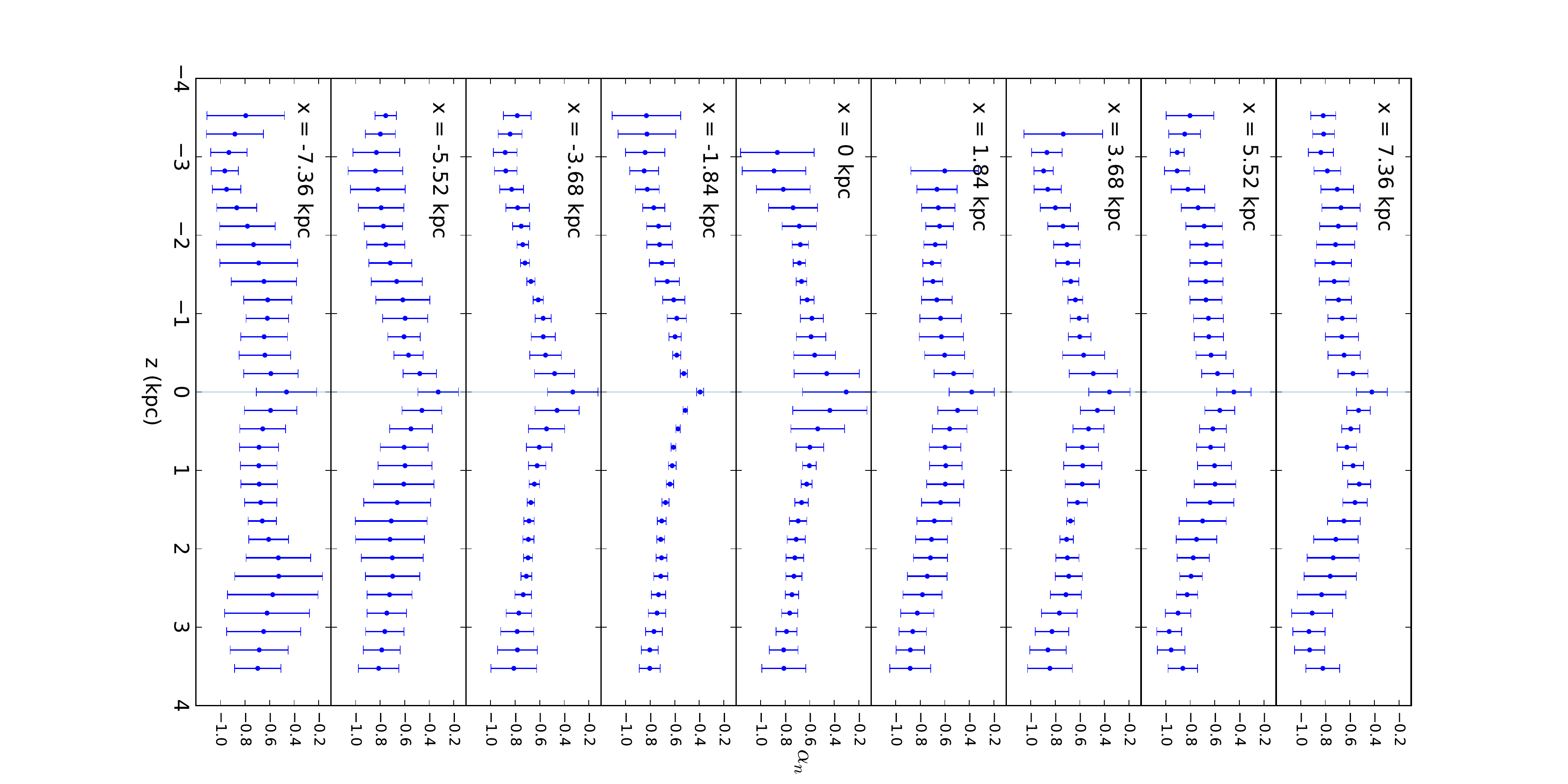}
\caption{Profiles of the nonthermal spectral index, $\alpha_n$, as a function of height $z$ above the major axis (in kpc) for the 9 strips at different distances, $x$, from the galaxy's centre along the major axis. The top four panels ($x>0$) correspond to the northern side of the galaxy and the bottom four panels ($x<0$) to the southern side; the middle panel is centred on the minor axis. $z$ is negative (positive) on the eastern (western) side of the halo.}
\label{fig:spectralz}
\end{center}
\end{figure}

To allow for testing CRE propagation models (see Sec.~\ref{subsec:discussioncreprop}), we need to measure the scale heights of the nonthermal emission at 146\,MHz and 1.5\,GHz. However, free-free absorption in the disk hampers a reliable estimate of thermal and nonthermal emission at 146\,MHz, so that we use instead the total emission at this frequency. The difference between these two quantities can be neglected because the thermal fractions are small (Sec.~\ref{subsec:thermalprop}). At 1.5\,GHz we use the nonthermal map from \citet{Schmidt2018}.

The disk scale heights of the nonthermal emission at 1.5\,GHz are on average 12\% larger than those of the total emission, indicating that the nonthermal disk is broader than the thermal disk. The halo scale heights of the nonthermal and total emission at 1.5\,GHz are the same within the errors.

Following a similar approach as in \citet{Krause2017}, we fitted two intrinsically exponential distributions to the vertical profiles of the disk and the halo. The vertical profile of the observed radio emission in an edge-on galaxy is a superposition of the vertical profile of the disk + halo emission, the radial profile of the disk emission projected onto the sky plane, and broadening by the telescope beam (see \citet{Mueller2017} and \citet{Krause2017} for full details).
For an almost perfectly edge-on galaxy such as NGC\,891, the radial profile of the disk emission does not contribute to the vertical profile.

To obtain the profiles of radio emission at 146\,MHz with distance from the major axis for both the total intensity map and the nonthermal map (Fig.~\ref{fig:NGC891thermnonthermmap} right) at $12\arcsec$ resolution, we performed averaging within strips of $40\arcsec$ ($\simeq 1.84$\,kpc) width and $75\arcsec$ ($\simeq 3.45$\,kpc) maximum height above and below the major axis, applying the method described in \citet{Mueller2017}. Each data point is separated by $5\arcsec$, slightly less than half the FWHM of the beam. We also calculated profiles with different maximum heights above and below the major axis, without significantly different results. The profiles of the total intensity are presented in Fig.~\ref{fig:scaleheights}; those of the nonthermal intensity are very similar and are not shown. A visualisation of the distribution of strips is shown Fig.~\ref{fig:ngc891boxes}.
The same process was also performed on the VLA 1.5\,GHz total intensity map from \citet{Schmidt2018}. The distributions were fitted by a least-squares routine and are shown as red lines in Fig.~\ref{fig:scaleheights}. The corresponding scale heights, corrected for beam smearing, are summarized in Table~\ref{tab:scaleheights}.


The exponential scale heights of the total radio emission at 146\,MHz for both the disk and halo are consistently larger than those at 1.5\,GHz.
The mean scale height of the disk emission is $0.32\pm0.08$\,kpc at 146\,MHz, compared to $0.20\pm0.05$\,kpc at 1.5\,GHz. However, the scale heights of the disk at 146\,MHz should be regarded with caution. The 146\,MHz emission in the inner disk is attenuated by free-free absorption which artificially increases the scale heights. 
Hence, we regard the systematic difference in the disk scale heights between 146\,MHz and 1.5\,GHz not to be significant.

The mean scale height of the halo emission is $2.30\pm0.70$\,kpc at 146\,MHz compared to $1.36\pm0.19$\,kpc at 1.5\,GHz. The mean ratio $q_\mathrm{halo}$ of scale heights of the halo emission between 146\,MHz and 1.5\,GHz (Table~\ref{tab:scaleheights}) is $1.7 \pm 0.3$. These results will be further discussed in Sec.~\ref{subsec:discussioncreprop}.

The profiles of the nonthermal spectral index between 146\,MHz and 1.5\,GHz for different strips along the major axis (see Fig.~\ref{fig:ngc891boxes}) using the average intensities are shown in Fig.~\ref{fig:spectralz}. The profiles show a consistently flat spectrum ($\approx -0.4$) on the major axis of the galaxy and a roughly linear steepening with increasing distance from the major axis on both sides, reaching a spectral index of about $-0.8$ at 3\,kpc height.

The halo scale heights and spectral index profiles will be discussed in Sec.~\ref{subsec:discussioncreprop}, using simplified models for the CRE propagation. A more thorough treatment, fitting the spectral index profiles with numerical models of CRE propagation such as of \cite{Heesen2016,Heesen2018a} and \cite{mulcahy2016} is deferred to future work.

\section{Discussion}
\label{sec:discussion}

\subsection{Thermal absorption in the disk of NGC\,891}
\label{subsec:discussionthermal}

The total and nonthermal spectral index between 146\,MHz and 1.5\,GHz in the star-forming disk is observed to be significantly flatter than the injection spectrum ($\alpha_\mathrm{inj} \simeq -0.6$, see Sec.~\ref{subsec:nonthermspectral}) and far flatter than what is observed at higher frequencies by \citet{Schmidt2018}. Synchrotron self-absorption is unlikely to occur for such a diffuse source and is seen only for compact sources such as radio supernovae \citep{Chevalier1998}. Bremsstrahlung losses of CREs dominate at low frequencies, but cannot flatten the radio spectrum \citep[e.g.][]{Basu2015}. We propose that free-free absorption of nonthermal emission by ionized gas occurs in the disk. 

Observations with the VLA at 330\,MHz and 74\,MHz suggested a flattening of the spectrum along the disk of the Milky Way (Kassim, private communication). Spectral turnovers due to free-free absorption have been observed in the integrated spectra of nearby starburst galaxies such M\,82 at $\approx$ 500\,MHz \citep{Adebahr2013}, NGC\,253 at 230\,MHz \citep{Kapinska2017}, and IC\,10 at 322\,MHz \citep{Basu2017}.
Spatially resolved free-free absorption has been observed in M\,82 with Merlin at 408\,MHz \citep{Wills1997} and with LOFAR at 154\,MHz \citep{Varenius2015}. \citet{Lacki2010} showed that for M\,82 free-free absorption from $\HII$ regions is the most important absorption process at frequencies down to 10\,MHz.
Low-frequency observations of several starburst galaxies with the Murchison Widefield Array (MWA) showed clear signs of free-free absorption in the spectrum of integrated flux densities \citep{Galvin2018}. Modelling of thermal absorption in nearby galaxies observed with the LOFAR Multifrequency Snapshot Sky Survey (MSSS) shows that absorption effects in non-starburst galaxies are visible in the integrated spectra only much below 100\,MHz, but can be strong in the spectra of individual regions at higher frequencies, especially for edge-on galaxies (Chyży et al. in prep.).

While common in starburst galaxies, it is under debate if a flattening of the spectrum of integrated emission can be observed also in normal star-forming galaxies. From our NGC\,891 data, we are able to study the spatially resolved distribution of free-free absorption in a highly inclined spiral galaxy for the first time. The distribution of nonthermal spectral index (Fig.~\ref{fig:NGC891nonthermspecindexmap}) reveals regions in the disk with significant spectral flattening, while the spectrum of integrated emission (Fig.~\ref{fig:NGC891integratedflux}) shows a less obvious flattening towards lower frequencies, probably due to the contribution of steep-spectrum synchrotron emission from the galaxy's halo.

NGC\,891 has a lower surface density of the star-formation rate compared with M\,82 and the nuclear starburst of NGC\,253, so that the space density of $\HII$ regions is lower in NGC\,891; consequently, the emission measure (the square of the electron density integrated along the line of sight) and thus the spectral turnover frequency are lower as well. Furthermore, the emission measure and hence the turnover frequency vary for different lines of sight through the galaxy disk. As a result, any flattening of the integrated spectrum is smeared out over a large frequency range. This may explain why the integrated spectrum of NGC\,891 around 146\,MHz hardly deviates from a power law.


To measure the possible flattening of the integrated nonthermal spectrum with higher accuracy, we first should ascertain the nature of the total emission spectrum of NGC\,891. Since we are subtracting a simple optically thin free--free emission of NGC\,891 throughout the radio frequency range studied here, any curvature in the total flux density spectrum itself will naturally give rise to a curvature in the nonthermal spectrum. A cursory look at the total flux densities in Fig.~\ref{fig:NGC891integratedflux} (top panel) suggests that the measurements at the lowest frequencies, viz. 57.5 and 146\,MHz, play an important role in determining the shape of the total intensity spectrum of NGC\,891. New observations at similar frequencies, e.g. using the LOFAR Low Band Antenna (LBA) at around 55\,MHz, could help us to ascertain whether the total intensity spectrum is indeed curved and thereby help to understand whether the curved nonthermal spectrum of NGC\,891 is due to flattening at the lower frequencies or steepening at the higher frequencies.
We performed fits to the total flux density spectrum by a power law, assuming varying values of the expected flux density at 55\,MHz. If the LBA measures a flux density between $8\text{--}9$\,Jy with less than about 10\% uncertainty in flux density, one can confidently rule out a curvature in the total flux density spectrum of NGC\,891.



While spatially resolved observations of free-free absorption in the disks of external galaxies are now feasible, the same effect hampers the estimate of the magnetic field strength. As mentioned in Sec.~\ref{sec:magneticfield}, the presence of free-free absorption leads to an underestimate of the nonthermal intensity at low frequencies and hence of the magnetic field strength. On the other hand, low-frequency observations still play an essential role in estimating the magnetic field strength in the halo because they are less affected by spectral ageing of the CREs.


High-resolution observations with the Giant Meterwave Radio Telescope (GMRT) and the LOFAR LBA are currently in progress and will help to constrain the optical depth of the ISM in the disk of galaxies. With NGC\,891 being nearby and having an inclination angle of $\approx90^{\circ}$ and thus a large pathlength, it is an ideal target for further studies of free-free absorption with high spatial resolution.

\subsection{Loss processes of CREs in the halo of NGC\,891}
\label{subsec:lossprocesses}

The maximum contribution to the synchrotron emission at the frequency $\nu$ comes from CREs at the energy:

\begin{equation}
E_\nu = \sqrt{
\left( \frac{\nu}{16~\mathrm{MHz}} \right) \,
\left( \frac{B_\perp}{\mu\mathrm{G}} \right)^{-1}
}
\,\,\mathrm{GeV} \, .
\label{eq:nuEnergy}
\end{equation}

CREs emitting at 146\,MHz have an energy of $\lesssim 1.0$\,GeV in the disk of NGC\,891 ($B\gtrsim 10\,\mu$G, Section~\ref{sec:magneticfield}) and $\simeq 1.1$\,GeV in the halo ($B\simeq7\,\mu$G).

The half-power lifetime of the observable synchrotron-emitting CREs is:

\begin{equation}
t_\mathrm{syn} = 8.35 \times 10^{9}~\mathrm{yrs} \,\, 
\left( \frac{E_\nu}{\rm GeV} \right)^{-1}
\left( \frac{B_{\perp}}{\mu\mathrm{G}} \right)^{-2} \,
\label{eq:sync}
\end{equation}
where $B_{\perp}$ is the magnetic field strength perpendicular to the line of sight (assuming an isotropic pitch angle distribution). For an isotropic turbulent field, $B_{\perp} = \sqrt{2/3} \, B$. In NGC\,891, $t_\mathrm{syn}$ varies between $\lesssim 1.2\times10^{8}$\,yrs in the disk and $\simeq 1.9\times10^{8}$\,yrs in the halo at 146\,MHz.

CREs losing energy via inverse Compton (IC) radiation losses have a half-power lifetime of:

\begin{equation}
t_\mathrm{IC} = 3.55 \times 10^8~\mathrm{yrs} \,\, 
\left( \frac{E}{\mathrm{GeV}} \right)^{-1}  
\left( \frac{U_\mathrm{ph}}{10^{-12}~\mathrm{erg\,cm}^{-3}} 
\right)^{-1} 
\, 
\end{equation}
where $U_\mathrm{ph}$ is the total photon energy density. \cite{Heesen2014} found that the ratio between $t_\mathrm{syn}$ and $t_\mathrm{IC}$ (identical to the ratio between $U_\mathrm{ph}$ and the magnetic energy density) varies typically between 10\% and 80\% in normal star-forming galaxies, although it may be higher in galaxies undergoing a starburst. In NGC\,891, this ratio is about 16\%, estimated from the total infrared luminosity and assuming energy equipartition  \citep{Heesen2018a}, so that synchrotron losses dominate over IC losses.

The CRE half-power lifetime against bremsstrahlung losses is:
\begin{equation}
t_\mathrm{\mathrm{brems}} = 3.96 \times 10^7~\mathrm{yrs} \, 
\left( \frac{n}{\mathrm{cm}^{-3}} \right)^{-1} \,
\end{equation}
where $n$ is the density of the neutral gas. According to \citet{Basu2015}, bremsstrahlung losses of CREs emitting in the range 100--300\,MHz dominate in regions with gas column densities beyond about 5\,M$_\sun$/pc$^2$, as is the case in the disk of NGC\,891.

The CRE half-power lifetime against ionization losses is \citep{Longair2011}:
\begin{equation}
t_{\rm ion} = 4.1\times 10^9\,{yrs}\,\left(\frac{n}{cm^{-3}}\right)^{-1} \left( \frac{E}{GeV}\right) \left[3\ln\left(\frac{E}{GeV}\right)\,+\,42.5 \right]^{-1}.
\end{equation}

At frequencies below 300\,MHz, ionization losses are slightly less important than bremsstrahlung losses \citep{Basu2015}, except perhaps in highly dense gas clouds.

\subsection{CRE propagation in NGC\,891}
\label{subsec:discussioncreprop}

\begin{figure*}[h!]
 \includegraphics[width=18.8cm]{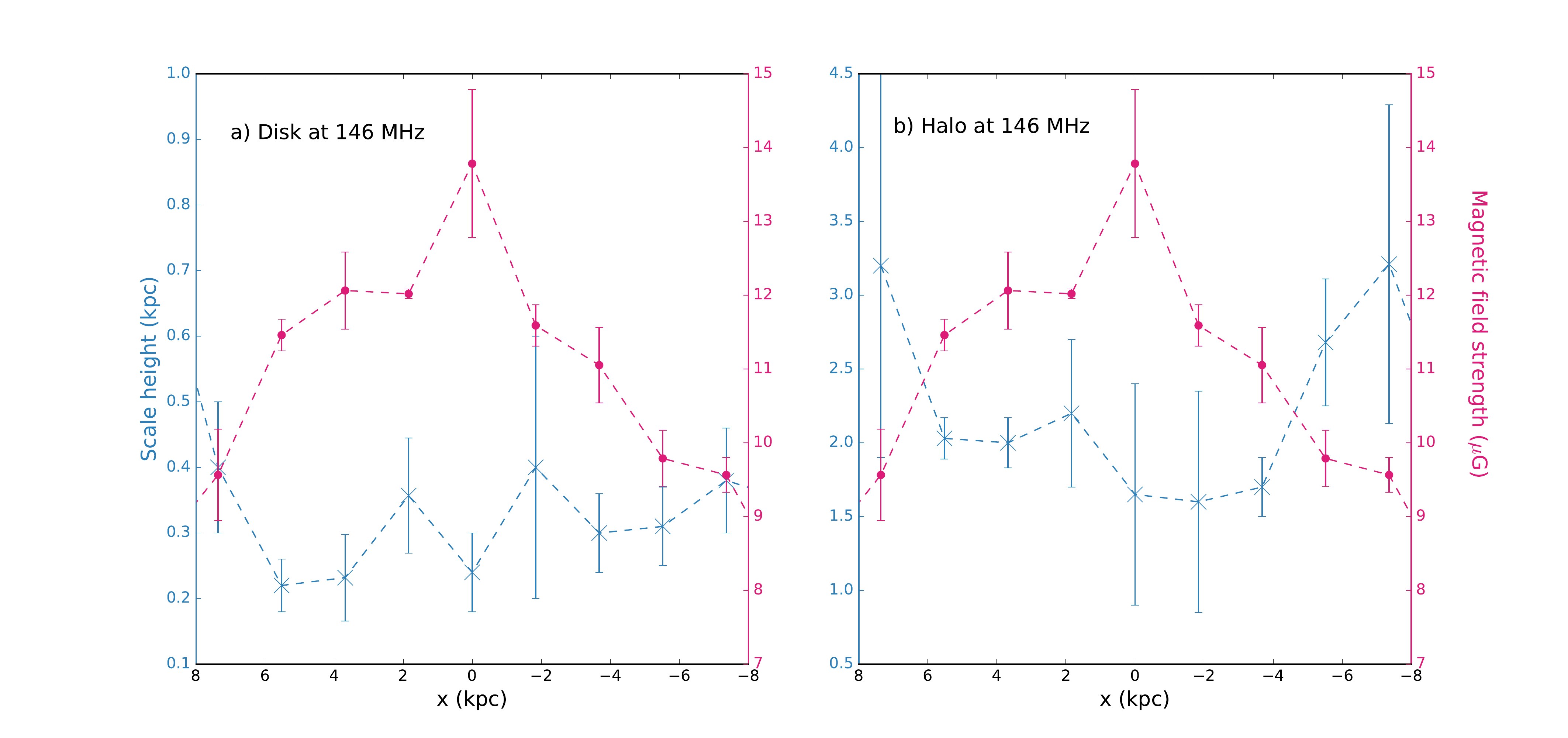}
 \caption{Plot of the scale heights in kpc (dark blue) at 146\,MHz
and of the magnetic field strengths in $\mu$G (red) at different distances $x$ from the galaxy's centre along the major axis (in kpc). $x>0$ corresponds to the north-eastern side of the galaxy and $x<0$ to the south-western side. The disk scale heights are shown in the left panel, the halo scale heights in the right panel. The magnetic field strengths are measured from the 1.5\,GHz data on the galaxy's major axis and are identical in all four panels.}
\label{fig:NGC891Bfieldpic} 
\end{figure*}

The scale heights of the nonthermal halo emission at 146\,MHz are consistently larger than those at 1.5\,GHz (Table~\ref{tab:scaleheights}). The probable reason is that CRE energy loss processes are weaker at lower frequencies, so that the CRE lifetime and hence the propagation height of CREs are larger.

We investigated two basic models of vertical CRE propagation, namely diffusion or advection (outflow) \citep{Tabatabaei2013,mulcahy2014,mulcahy2016,Krause2017}. \citet{mulcahy2016} showed via numerical modelling that diffusion is the dominant CRE propagation process in the disk of M\,51. As the observed synchrotron spectral indices in the disk are too flat to be explained by dominating synchrotron losses, the CREs of M\,51 must escape from the disk before losing their energy, e.g. by diffusion perpendicular to the disk.

We assume that the lifetime of CREs in the halo of NGC\,891 is dominated by synchrotron loss (Eq.~\ref{eq:sync}), i.e.\ the halo is a calorimeter for CREs. In the case that radiation losses are important, a galaxy is referred to as an ``electron calorimeter'', where the CREs are losing their energy through radiation and ionization losses before they escape. Further assuming that the diffusion coefficient $D$ is constant (no dependencies on height $z$, nor on CRE energy $E$, nor on field strength $B$), the diffusive propagation height is:
\begin{equation}
h_\mathrm{diff} = \sqrt{D \,\, t_\mathrm{CRE}} \propto B^{\,\,-3/4} \, \nu^{\,\,-1/4}
\label{eq:diffusion}
\end{equation}
where $t_\mathrm{CRE}$ is the lifetime of the CREs (see Sec.~\ref{subsec:lossprocesses}) and $B$ is the field strength. With the frequency ratio between 146\,MHz and 1.5\,GHz of 1/10.3, the expected ratio of propagation heights is 1.8. In case of dominating IC losses of the CREs, the frequency dependence of the diffusive propagation height $h_\mathrm{diff}$ remains the same, while there is only a weak positive dependence on field strength ($h_\mathrm{diff} \propto B^{1/4}$).

For a constant advection speed $v$ and a lifetime of CRE dominated by synchrotron loss, the advective propagation height is:
\begin{equation}
h_\mathrm{adv} = \mathrm{v} \,\, t_\mathrm{CRE} \propto B^{\,\,-3/2} \, \nu^{\,\,-1/2} \, . 
\label{eq:adv}
\end{equation}
With the same frequency ratio as above, the expected ratio of propagation heights is 3.2. In the case of dominating IC losses of the CREs, the frequency dependence of the advective propagation height $h_\mathrm{adv}$ remains the same, while the dependence on field strength becomes positive ($h_\mathrm{adv} \propto B^{1/2}$).

Several spiral galaxies in the CHANG-ES sample show indications for advective CRE propagation and CRE timescales dominated by escape from the halo \citep{Krause2017}. In this case, the propagation height does neither depend on frequency nor on magnetic field strength.

The mean observed ratio $q_\mathrm{halo}$ of scale heights of the halo emission between 146\,MHz and 1.5\,GHz (Table~\ref{tab:scaleheights}) of $1.7 \pm 0.3$ agrees well with the diffusion model and dominating synchrotron and/or IC losses.
It seems that the CRE propagation from the disk to the halo of NGC\,891 is slower than in other spiral edge-on galaxies in the CHANG-ES sample, so that the CRE synchrotron lifetime is smaller than the escape time.

The halo scale heights from Table~\ref{tab:scaleheights} may suffer from various systematic bias effects. Firstly, the halo scale heights at 146\,MHz can be affected by free-free absorption in the disk. As a test, we artificially increased the nonthermal intensity in the inner disk by a factor of 3. This leads to about $2\times$ smaller scale heights of the inner disk, while the halo scale heights do not change within the limits of uncertainty.

{NGC\,891 was observed in L-band with the VLA D-array for $3\times10$\,min. According to the VLA website, the maximum visible structure of a single snapshot observation at 1.5\,GHz is about $8\arcmin$ which is similar to the angular extent of NGC\,891. The maximum visible structure of a multiple snapshot observation is somewhat larger but hard to estimate. We cannot exclude that some diffuse halo emission is missing in the L-band image by \citet{Schmidt2018}. The effect on the total flux density must be small because Fig.~\ref{fig:NGC891integratedflux} does not indicate an obvious deficit, but the emission in the outer halo may still be underestimated, so that the measured halo scale heights could be too small. As a test, we added an artificial halo component which is constructed by smoothing the galaxy's image with a Gaussian function of $5\arcmin$ full-width-at-half-maximum and decreasing the resulting intensities by a factor of 10. Hence, this component contributes 10\% to the total flux density which is still consistent with Fig.~\ref{fig:NGC891integratedflux}. It leads to an increase of the halo scale heights by only 5--10\% and to an insignificant decrease of the mean ratio of scale heights of the halo emission between 146\,MHz and 1.5\,GHz to 1.6.

Both CRE propagation models discussed above are rather simplistic.
A more detailed analysis needs a proper description of CRE sources, magnetic field distribution, CRE loss processes, and CRE propagation mechanisms. Such a study, based on the CHANG-ES data, is forthcoming \citep{Schmidt2018}.

Nevertheless, we can explore the above models further, taking advantage of the fact that synchrotron emission is the only CRE loss process related to the magnetic field strength. Hence, the local scale height, as measured in the strips, should depend on the local magnetic field strength in the same area. We expect an increase of scale height with decreasing field strength which is different for diffusive and advective CRE propagation (Eqs.~\ref{eq:diffusion} and \ref{eq:adv}).

The scale heights at 146\,MHz (Table~\ref{tab:scaleheights}) are plotted in Fig.~\ref{fig:NGC891Bfieldpic}. A variation of the disk scale height (left panel) with distance from the galaxy centre is not obvious, while the scale height of the halo (right panel) increases with increasing distance (``flaring''). This effect is similar to the one seen in the halo of the edge-on spiral galaxy NGC\,253 \citep{Heesen2009}.

To investigate whether the halo flaring in NGC\,891 is caused by a decrease of the synchrotron losses of CREs, we also show in Fig.~\ref{fig:NGC891Bfieldpic} the magnetic field strengths on the major axis which are a measure of the level of synchrotron losses in the different strips (Eq.~\ref{eq:sync}). As we were unable to obtain reliable magnetic field strength estimates in the disk from our 146\,MHz data due to free-free absorption effects (see above), we instead use the magnetic field strength estimates from \citet{Schmidt2018} derived from 1.5\,GHz data. Indeed, the scale heights at both frequencies are smallest near to the centre where the field strengths are largest.

\begin{figure}[h!]
 \begin{center}
 \includegraphics[scale=0.25]{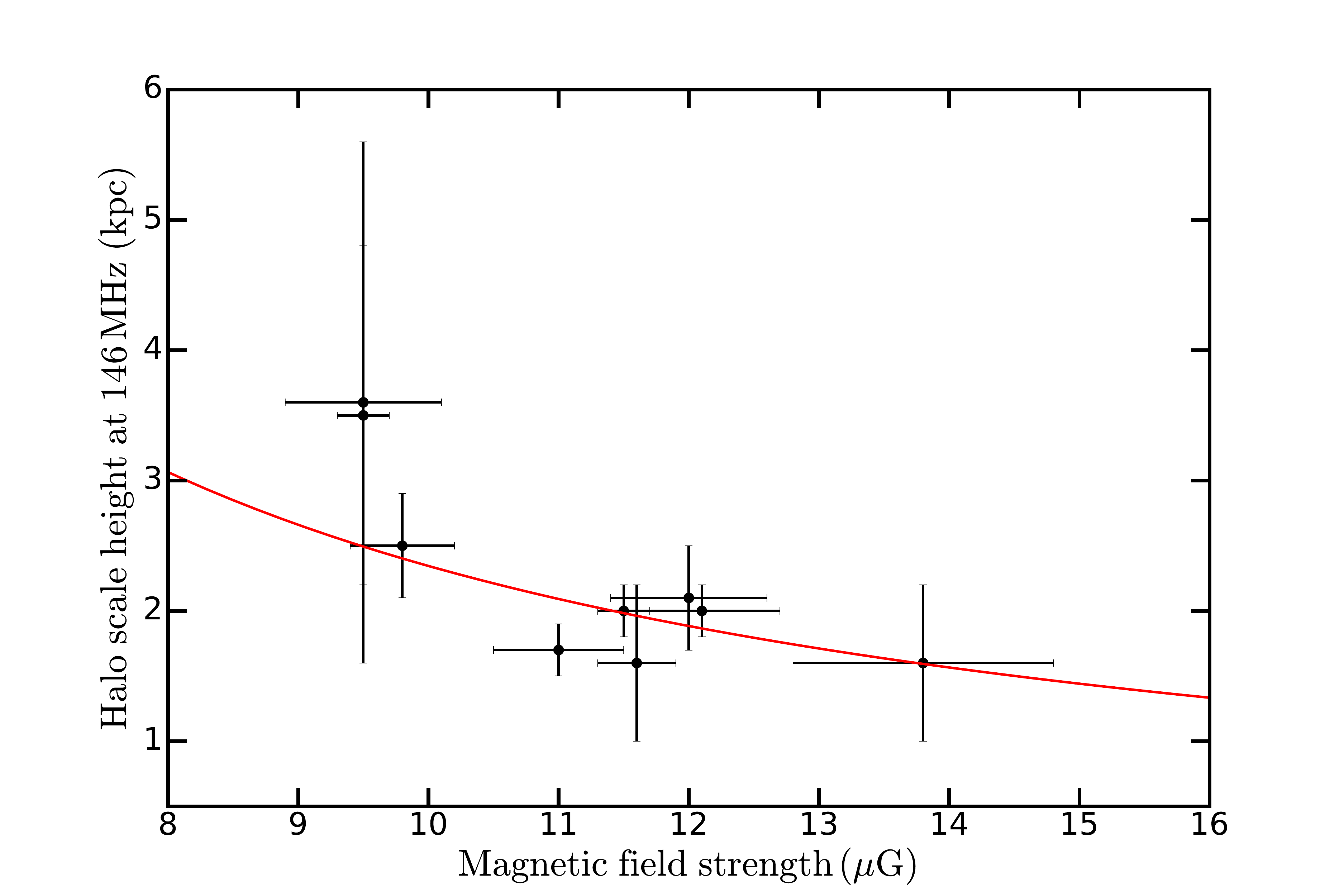}
 \caption{Halo scale heights at 146\,MHz
in the 9 strips along the major axis of NGC\,891 as a function of magnetic field strengths on the major axis. The red line shows the best-fit exponential function.}
\label{fig:bfieldagainstscaleheight} 
\end{center}
\end{figure}

The relation between the scale heights at 146\,MHz
and the magnetic field strength is shown in Fig.~\ref{fig:bfieldagainstscaleheight}.
The exponent of $-1.2\pm0.6$
is consistent with diffusive or advective propagation (Eqs.~\ref{eq:diffusion} and \ref{eq:adv}) within the uncertainty. The negative sign of the exponents at both frequencies supports our above result that synchrotron radiation is the dominating loss process of CREs in the halo of NGC\,891 and that IC losses are less important, which is generally the case for late-type spiral and dwarf irregular galaxies \citep{Heesen2014}.

The disk of NGC\,891 seen with LOFAR is not uniform everywhere, but bulges out in the north where the halo scale height is large (Fig.~\ref{fig:NGC891Bfieldpic}, at $x=7.36$\,kpc), even larger than expected for the weak magnetic field in this area, as indicated by the points above the fit line in the top left corners of Fig.~\ref{fig:bfieldagainstscaleheight}. Possible reasons could be a larger diffusion coefficient in the northern star-forming region or an increase of scale height by an enhanced, faster outflow.

According to \citet{Strong1978}, the vertical profile of the nonthermal spectral index is similar for the two types of CRE propagation, diffusion and advection. On the other hand, \citet{Heesen2016,Heesen2018a} showed that the vertical profile of the radio spectral index can help to distinguish between the two basic mechanisms of CRE transport. Advection of CREs with a constant outflow velocity
should lead to linear profiles of the nonthermal spectral index, where the spectral index steepens gradually with increasing distance from the disk plane. Conversely, diffusion should lead to parabolic-shaped spectral index profiles where the spectral index remains about constant until a few kpc height, from where on the synchrotron losses become strong and the nonthermal spectral index steepens rapidly. In NGC\,891, the vertical profiles of the radio spectral index between 146\,MHz and 1.5\,GHz, as presented in Fig.~\ref{fig:spectralz}, are falling into the first category, indicating that advection in an outflow may be the dominating CRE transport process. In non-calorimetric halos where CRE escape is important, the nonthermal scale height and CRE lifetime
are not related, and Eqs.~(\ref{eq:diffusion}) and (\ref{eq:adv}) do not apply.



In order to perform a more quantitative investigation, we need an estimate of the escape timescale for the cosmic rays in the halo of NGC\,891 and hence a model of the outflow velocity as a function of distance from the disk. Based on previous radio data, \citet{Heesen2018a} could not find a well-constrained advection speed for a possible outflow in NGC\,891.
We plan to combine our new LOFAR 146\,MHz data of NGC\,891 and the CHANG-ES data at $1.5$\ and 6\,GHz in a forthcoming work. An analysis of the profiles of synchrotron emission and spectral index will be performed, where the 1D diffusion-loss equation will be solved, such as has been done for the radial diffusion within the disk of the almost face-on galaxy M\,51 \citep{mulcahy2016} and for the vertical transport by advection and diffusion in the halos of many edge-on galaxies \citep{Heesen2016,Heesen2018a}. This will allow us to measure diffusion coefficients or advection speeds of CREs in the halo of NGC\,891.


\section{Conclusions}
\label{sec:conclusion}

In this work, we performed radio continuum observations of the edge-on spiral galaxy NGC\,891 with the LOFAR High Band Antenna (HBA) Array with a central frequency of 146\,MHz and with the Arcminute Microkelvin Imager (AMI) at 15.5\,GHz. Using the facet calibration scheme detailed in \cite{vanWeeren2016}, we achieved LOFAR images with a high quality near to thermal noise which provide a sensitive view of the extended halo of NGC\,891 and its physical origin. This is the first time that low-frequency, high-resolution observations are presented of this well-studied edge-on galaxy.

We used the AMI Small and Large Array to observe NGC\,891 at a central frequency of 15.5\,GHz. For the first time, a nearby galaxy has been studied using the upgraded correlator of AMI. Nearby galaxies have rarely been observed at radio frequencies larger than 10\,GHz. This work paves the way for an AMI nearby galaxy survey at 15.5\,GHz.
\\

The main findings of this work are:

\begin{itemize}
\item With our new measurements and assuming realistic uncertainties of previous measurements, we derived the spectrum of the integrated nonthermal flux density. The spectrum is probably not a simple power law. It can be fitted by a power law with a steepening in spectral index by $-0.5$ towards higher frequencies, or by a curved polynomial, indicating a flattening towards lower frequencies by free-free absorption and a steepening towards higher frequencies. Which physical mechanism produces the curvature in the spectrum cannot be constrained with the current measurements. A clear detection of a spectral curvature in the integrated spectrum needs observations with LOFAR LBA at around 55\,MHz with $<10\%$ uncertainty.
\item We find no significant flattening of the spectrum by thermal emission towards higher frequencies of up to 15.5\,GHz, suggesting that the synchrotron component is still dominant in this frequency range. Further observations at higher frequencies are necessary to confirm this result.
\item In the star-forming disk, our map of nonthermal spectral index between 146\,MHz and the CHANG-ES data at 1.5\,GHz reveals areas with significant spectral flattening towards lower frequencies, with values significantly flatter than $-0.5$. This is likely caused by free--free absorption in the ionized gas.
\item In the halo, we observe nonthermal spectral indices between 146\,MHz and 1.5\,GHz within a range of $-0.6$ to $-0.8$, significantly flatter than the spectral indices at higher frequencies. This supports the expectation that cosmic-ray electrons (CREs) emitting at low frequencies suffer less from energy losses. Consequently, low-frequency observations are better suited to estimate the magnetic field in the halo if one assumes energy equipartition between cosmic rays and magnetic fields.
\item The mean magnetic field strength in the halo is $7 \pm 2\,\mu$G. Due to the significant free-free absorption, we cannot confidently estimate magnetic field strengths in the disk of NGC\,891. High-frequency observations are more suited for this task, while low-frequency observations are more suitable to estimate magnetic field strengths in the halo.
\item Radio emission from the halo at 146\,MHz is detected out to a maximum of 7.3\,kpc distance from the major axis. The similar extension compared to the new CHANG-ES image at 1.5\,GHz is most likely due to the lower sensitivity at 146\,MHz.
\item The scale heights of the nonthermal halo emission at 146\,MHz are consistently larger at all radii than those at 1.5\,GHz, with a mean ratio of $1.7 \pm 0.3$, as predicted by diffusive CRE propagation. This ratio also suggests that spectral ageing of the CREs, caused by radiation losses, is important and hence that the halo is at least a partial calorimeter for CREs, i.e. they are losing a significant fraction of their energy before they can escape.
\item The scale height of the nonthermal halo emission at 146\,MHz correlates with the magnetic field strength with an exponent of $-1.2\pm0.6$,
consistent with diffusive or advective CRE propagation. The negative exponent is a signature of dominating synchrotron losses in the halo. IC losses appear to be less important.
\item The low-frequency radio halo in NGC\,891 seems to be different from the ones observed in the CHANG-ES sample at GHz frequencies that are escape-dominated (i.e.\ non-calorimetric). More low-frequency observations of edge-on spiral galaxies are needed in order to decide whether this is a specific property of the low-frequency halo in NGC\,891 or holds true for spiral galaxies in general.
\item The linearly steepening spectral index profiles of the nonthermal emission in the halo seem to favour advection of CREs in a galactic wind \citep{Heesen2016}.
Further studies are forthcoming, using refined modelling of CRE propagation and exploiting the combined LOFAR and CHANG-ES data at 1.5 and 6\,GHz.
\item To measure energy losses and propagation of CREs and magnetic field strengths in the disk and halo with higher accuracy, further observations of NGC\,891 and other edge-on spiral galaxies over a wide frequency coverage and with high spatial resolution are needed.
\end{itemize}

\begin{figure}[h!]
 \begin{center}
\includegraphics[width=8.5cm]{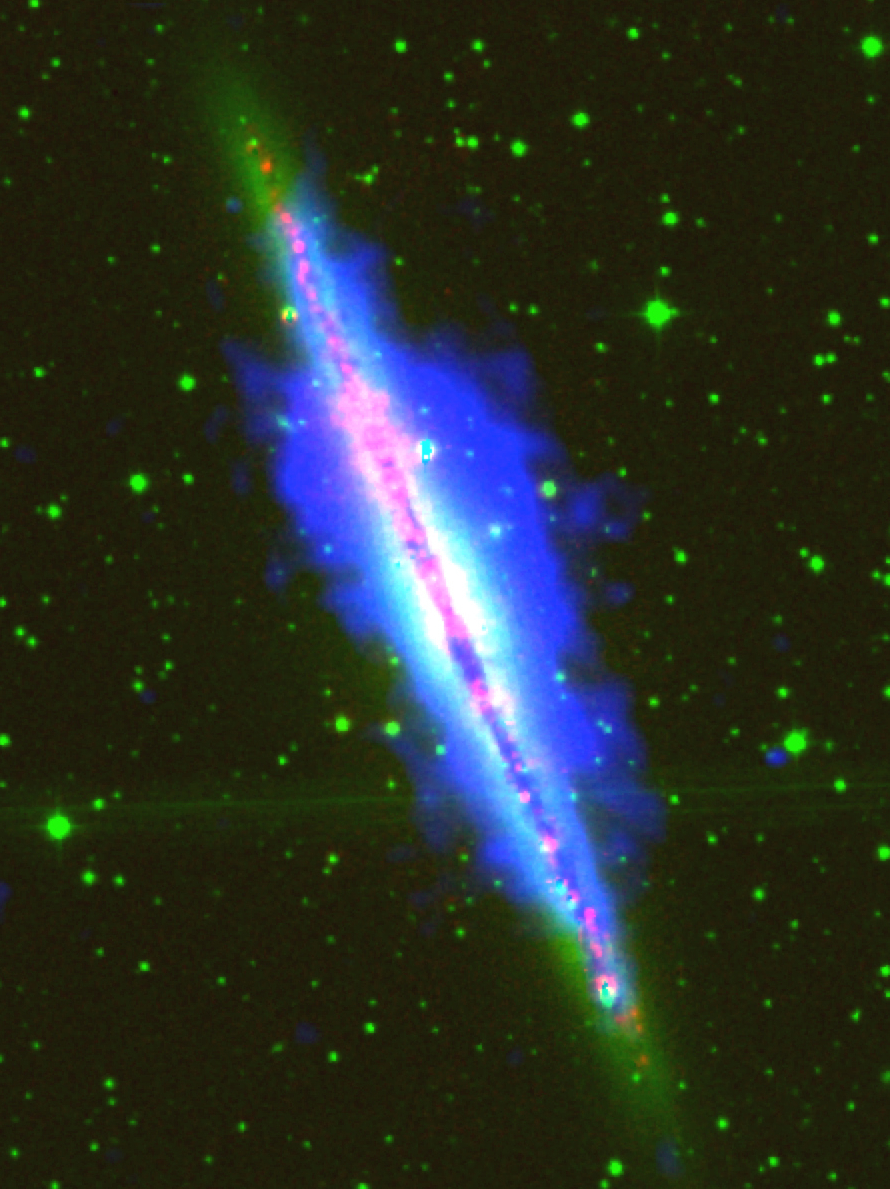}
\caption{Composite of NGC\,891: LOFAR 146\,MHz at $12\arcsec \times 12\arcsec$ resolution (blue), optical from DSS2 (green), and H$\alpha$ from \citet{Rand1990} (red). The size of the image is $9\arcmin \times 12\arcmin$.}
\label{fig:composite}
\end{center}
\end{figure}


\begin{acknowledgements}
This research was performed in the framework of the DFG Research Unit 1254 ``Magnetisation of Interstellar and Intergalactic Media: The Prospects of Low-Frequency Radio Observations''. --
LOFAR, designed and constructed by ASTRON, has facilities in several countries that are owned by various parties (each with their own funding sources) and that are collectively operated by the International LOFAR Telescope (ILT) foundation under a joint scientific policy. We acknowledge support by the FZ J\"ulich under Jureca computing grant HTB00. --
We thank the staff of the Mullard Radio Astronomy Observatory for their invaluable assistance in the commissioning and operation of AMI which is supported by Cambridge and Oxford Universities. We acknowledge support from the European Research Council under grant ERC-2012-StG-307215 LODESTONE. --
This research made use of the NASA/IPAC Extragalactic Database (NED) which is operated by the Jet Propulsion Laboratory, California Institute of Technology, under contract with the National Aeronautics and Space Administration. --
A.B. acknowledges financial support by the German Federal Ministry of Education and Research (BMBF) under grant 05A17PB1 (Verbundprojekt D-MeerKAT). F.S.T. acknowledges financial support from the Spanish Ministry of Economy and Competitiveness (MINECO) under grant AYA2016-76219-P. --
We thank Frank Israel (Sterrewacht Leiden) and Olaf Wucknitz (MPIfR Bonn) for carefully reading of the manuscript and our anonymous referee for many useful comments.
\end{acknowledgements}

\bibliographystyle{aa} 
\bibliography{n891} 

\begin{thebibliography}{119}
\expandafter\ifx\csname natexlab\endcsname\relax\def\natexlab#1{#1}\fi

\bibitem[{{Adebahr} {et~al.}(2013){Adebahr}, {Krause}, {Klein},
  {We{\.z}gowiec}, {Bomans}, \& {Dettmar}}]{Adebahr2013}
{Adebahr}, B., {Krause}, M., {Klein}, U., {et~al.} 2013, \aap, 555, A23

\bibitem[{{Allen} {et~al.}(1978){Allen}, {Sancisi}, \& {Baldwin}}]{Allen1978}
{Allen}, R.~J., {Sancisi}, R., \& {Baldwin}, J.~E. 1978, \aap, 62, 397

\bibitem[{{Alton} {et~al.}(1998){Alton}, {Bianchi}, {Rand}, {Xilouris},
  {Davies}, \& {Trewhella}}]{Alton1998}
{Alton}, P.~B., {Bianchi}, S., {Rand}, R.~J., {et~al.} 1998, \apjl, 507, L125

\bibitem[{{Arshakian} {et~al.}(2011){Arshakian}, {Stepanov}, {Beck}, {Krause},
  \& {Sokoloff}}]{Arshakian2011}
{Arshakian}, T.~G., {Stepanov}, R., {Beck}, R., {Krause}, M., \& {Sokoloff}, D.
  2011, Astronomische Nachrichten, 332, 524

\bibitem[{{Baars} {et~al.}(1977){Baars}, {Genzel}, {Pauliny-Toth}, \&
  {Witzel}}]{Baars1977A&A}
{Baars}, J.~W.~M., {Genzel}, R., {Pauliny-Toth}, I.~I.~K., \& {Witzel}, A.
  1977, \aap, 61, 99

\bibitem[{{Baldwin} \& {Pooley}(1973)}]{Baldwin1973}
{Baldwin}, J.~E. \& {Pooley}, G.~G. 1973, \mnras, 161, 127

\bibitem[{{Basu} {et~al.}(2015){Basu}, {Beck}, {Schmidt}, \& {Roy}}]{Basu2015}
{Basu}, A., {Beck}, R., {Schmidt}, P., \& {Roy}, S. 2015, \mnras, 449, 3879

\bibitem[{{Basu} {et~al.}(2017){Basu}, {Roychowdhury}, {Heesen}, {Beck},
  {Brinks}, {Westcott}, \& {Hindson}}]{Basu2017}
{Basu}, A., {Roychowdhury}, S., {Heesen}, V., {et~al.} 2017, \mnras, 471, 337

\bibitem[{{Beck}(2015)}]{Beck2015}
{Beck}, R. 2015, \aapr, 24, 4

\bibitem[{{Beck} {et~al.}(1979){Beck}, {Biermann}, {Emerson}, \&
  {Wielebinski}}]{Beck1979}
{Beck}, R., {Biermann}, P., {Emerson}, D.~T., \& {Wielebinski}, R. 1979, \aap,
  77, 25

\bibitem[{{Beck} {et~al.}(1996){Beck}, {Brandenburg}, {Moss}, {Shukurov}, \&
  {Sokoloff}}]{Beck1996}
{Beck}, R., {Brandenburg}, A., {Moss}, D., {Shukurov}, A., \& {Sokoloff}, D.
  1996, \araa, 34, 155

\bibitem[{{Beck} \& {Krause}(2005)}]{Beck2005}
{Beck}, R. \& {Krause}, M. 2005, Astronomische Nachrichten, 326, 414

\bibitem[{{Bell}(1978)}]{Bell1978}
{Bell}, A.~R. 1978, \mnras, 182, 443

\bibitem[{{Bietenholz} {et~al.}(2002){Bietenholz}, {Bartel}, \&
  {Rupen}}]{Bietenholz2002}
{Bietenholz}, M.~F., {Bartel}, N., \& {Rupen}, M.~P. 2002, \apj, 581, 1132

\bibitem[{{Bietenholz} {et~al.}(2010){Bietenholz}, {Bartel}, \&
  {Rupen}}]{Bietenholz2010}
{Bietenholz}, M.~F., {Bartel}, N., \& {Rupen}, M.~P. 2010, \apj, 712, 1057

\bibitem[{{Bregman} \& {Houck}(1997)}]{Bregman1997}
{Bregman}, J.~N. \& {Houck}, J.~C. 1997, \apj, 485, 159

\bibitem[{{Bregman} \& {Pildis}(1994)}]{Bregman1994}
{Bregman}, J.~N. \& {Pildis}, R.~A. 1994, \apj, 420, 570

\bibitem[{{Buffie} {et~al.}(2013){Buffie}, {Heesen}, \& {Shalchi}}]{Buffie2013}
{Buffie}, K., {Heesen}, V., \& {Shalchi}, A. 2013, \apj, 764, 37

\bibitem[{{Caprioli}(2011)}]{Caprioli2011}
{Caprioli}, D. 2011, \jcap, 5, 026

\bibitem[{{Chevalier}(1998)}]{Chevalier1998}
{Chevalier}, R.~A. 1998, \apj, 499, 810

\bibitem[{{Condon}(1987)}]{Condon1987}
{Condon}, J.~J. 1987, \apjs, 65, 485

\bibitem[{{Cornwell}(2008)}]{Cornwell2008}
{Cornwell}, T.~J. 2008, IEEE Journal of Selected Topics in Signal Processing,
  2, 793

\bibitem[{{Dahlem} {et~al.}(1994){Dahlem}, {Dettmar}, \& {Hummel}}]{Dahlem1994}
{Dahlem}, M., {Dettmar}, R.-J., \& {Hummel}, E. 1994, \aap, 290, 384

\bibitem[{{de Jong}(1967)}]{deJong1967}
{de Jong}, M.~L. 1967, \apj, 150, 1

\bibitem[{{de Vaucouleurs} {et~al.}(1991){de Vaucouleurs}, {de Vaucouleurs},
  {Corwin}, {Buta}, {Paturel}, \& {Fouqu{\'e}}}]{deVaucouleurs1991}
{de Vaucouleurs}, G., {de Vaucouleurs}, A., {Corwin}, Jr., H.~G., {et~al.}
  1991, {Third Reference Catalogue of Bright Galaxies} (New York: Springer)

\bibitem[{{Dettmar}(1990)}]{Dettmar1990}
{Dettmar}, R.-J. 1990, \aap, 232, L15

\bibitem[{{Dumke} {et~al.}(1995){Dumke}, {Krause}, {Wielebinski}, \&
  {Klein}}]{Dumke1995}
{Dumke}, M., {Krause}, M., {Wielebinski}, R., \& {Klein}, U. 1995, \aap, 302,
  691

\bibitem[{{Ekers} \& {Sancisi}(1977)}]{Ekers1977}
{Ekers}, R.~D. \& {Sancisi}, R. 1977, \aap, 54, 973

\bibitem[{{Elstner} {et~al.}(1995){Elstner}, {Golla}, {R\"udiger}, \&
  {Wielebinski}}]{Elstner1995}
{Elstner}, D., {Golla}, G., {R\"udiger}, G., \& {Wielebinski}, R. 1995, \aap,
  297, 77

\bibitem[{{Fraternali}(2017)}]{Fraternali2017}
{Fraternali}, F. 2017, in Astrophysics and Space Science Library, Vol. 430, Gas
  Accretion onto Galaxies, ed. A.~{Fox} \& R.~{Dav{\'e}}, 323

\bibitem[{{Galvin} {et~al.}(2018){Galvin}, {Seymour}, {Marvil},
  {Filipovi{\'c}}, {Tothill}, {McDermid}, {Hurley-Walker}, {Hancock},
  {Callingham}, {Cook}, {Norris}, {Bell}, {Dwarakanath}, {For}, {Gaensler},
  {Hindson}, {Johnston-Hollitt}, {Kapi{\'n}ska}, {Lenc}, {McKinley}, {Morgan},
  {Offringa}, {Procopio}, {Staveley-Smith}, {Wayth}, {Wu}, \&
  {Zheng}}]{Galvin2018}
{Galvin}, T.~J., {Seymour}, N., {Marvil}, J., {et~al.} 2018, \mnras, 474, 779

\bibitem[{{Garcia-Burillo} {et~al.}(1992){Garcia-Burillo}, {Guelin},
  {Cernicharo}, \& {Dahlem}}]{Garcia1992}
{Garcia-Burillo}, S., {Guelin}, M., {Cernicharo}, J., \& {Dahlem}, M. 1992,
  \aap, 266, 21

\bibitem[{{Gioia} \& {Fabbiano}(1987)}]{Gioia1987}
{Gioia}, I.~M. \& {Fabbiano}, G. 1987, \apjs, 63, 771

\bibitem[{{Gioia} \& {Gregorini}(1980)}]{Gioia1980}
{Gioia}, I.~M. \& {Gregorini}, L. 1980, \aaps, 41, 329

\bibitem[{{Gioia} {et~al.}(1982){Gioia}, {Gregorini}, \& {Klein}}]{Gioia1982}
{Gioia}, I.~M., {Gregorini}, L., \& {Klein}, U. 1982, \aap, 116, 164

\bibitem[{{Green}(2014)}]{Green2014}
{Green}, D.~A. 2014, Bulletin of the Astronomical Society of India, 42, 47

\bibitem[{{Gregory} \& {Condon}(1991)}]{Gregory1991}
{Gregory}, P.~C. \& {Condon}, J.~J. 1991, \apjs, 75, 1011

\bibitem[{{Heeschen} \& {Wade}(1964)}]{Heeschen1964}
{Heeschen}, D.~S. \& {Wade}, C.~M. 1964, \aj, 69, 277

\bibitem[{{Heesen} {et~al.}(2009){Heesen}, {Beck}, {Krause}, \&
  {Dettmar}}]{Heesen2009}
{Heesen}, V., {Beck}, R., {Krause}, M., \& {Dettmar}, R.-J. 2009, \aap, 494,
  563

\bibitem[{{Heesen} {et~al.}(2014){Heesen}, {Brinks}, {Leroy}, {Heald}, {Braun},
  {Bigiel}, \& {Beck}}]{Heesen2014}
{Heesen}, V., {Brinks}, E., {Leroy}, A.~K., {et~al.} 2014, \aj, 147, 103

\bibitem[{{Heesen} {et~al.}(2016){Heesen}, {Dettmar}, {Krause}, {Beck}, \&
  {Stein}}]{Heesen2016}
{Heesen}, V., {Dettmar}, R.-J., {Krause}, M., {Beck}, R., \& {Stein}, Y. 2016,
  \mnras, 458, 332

\bibitem[{{Heesen} {et~al.}(2018{\natexlab{a}}){Heesen}, {Krause}, {Beck},
  {Adebahr}, {Bomans}, {Carretti}, {Dumke}, {Heald}, {Irwin}, {Koribalski},
  {Mulcahy}, {Westmeier}, \& {Dettmar}}]{Heesen2018a}
{Heesen}, V., {Krause}, M., {Beck}, R., {et~al.} 2018{\natexlab{a}}, \mnras,
  476, 158

\bibitem[{{Heesen} {et~al.}(2018{\natexlab{b}}){Heesen}, {Rafferty},
  {Horneffer}, {Beck}, {Basu}, {Westcott}, {Hindson}, {Brinks}, {Chy{\.Z}y},
  {Scaife}, {Br{\"u}ggen}, {Heald}, {Fletcher}, {Horellou}, {Tabatabaei},
  {Paladino}, {Nikiel-Wroczy{\'n}ski}, {Hoeft}, \& {Dettmar}}]{Heesen2018b}
{Heesen}, V., {Rafferty}, D.~A., {Horneffer}, A., {et~al.} 2018{\natexlab{b}},
  \mnras, 476, 1756

\bibitem[{{Henriksen} \& {Irwin}(2016)}]{Henriksen2016}
{Henriksen}, R.~N. \& {Irwin}, J.~A. 2016, \mnras, 458, 4210

\bibitem[{{Hickish} {et~al.}(2018){Hickish}, {Razavi-Ghods}, {Perrott},
  {Titterington}, {Carey}, {Scott}, {Grainge}, {Scaife}, {Alexander},
  {Saunders}, {Crofts}, {Javid}, {Rumsey}, {Jin}, {Ely}, {Shaw}, {Northrop},
  {Pooley}, {D'Alessandro}, {Doherty}, \& {Willatt}}]{Hickish2018}
{Hickish}, J., {Razavi-Ghods}, N., {Perrott}, Y.~C., {et~al.} 2018, \mnras,
  475, 5677

\bibitem[{{Hodges-Kluck} \& {Bregman}(2013)}]{Hodges2013}
{Hodges-Kluck}, E.~J. \& {Bregman}, J.~N. 2013, \apj, 762, 12

\bibitem[{{H{\"o}gbom}(1974)}]{Hogbom1974}
{H{\"o}gbom}, J.~A. 1974, \aaps, 15, 417

\bibitem[{{Howk} \& {Savage}(1997)}]{Howk1997}
{Howk}, J.~C. \& {Savage}, B.~D. 1997, \aj, 114, 2463

\bibitem[{{Hummel}(1991)}]{Hummel1991b}
{Hummel}, E. 1991, \aap, 251, 442

\bibitem[{{Hummel} {et~al.}(1991){Hummel}, {Dahlem}, {van der Hulst}, \&
  {Sukumar}}]{Hummel1991}
{Hummel}, E., {Dahlem}, M., {van der Hulst}, J.~M., \& {Sukumar}, S. 1991,
  \aap, 246, 10

\bibitem[{{Irwin} {et~al.}(2012){Irwin}, {Beck}, {Benjamin}, {Dettmar},
  {English}, {Heald}, {Henriksen}, {Johnson}, {Krause}, {Li}, {Miskolczi},
  {Mora}, {Murphy}, {Oosterloo}, {Porter}, {Rand}, {Saikia}, {Schmidt},
  {Strong}, {Walterbos}, {Wang}, \& {Wiegert}}]{Irwin2012}
{Irwin}, J., {Beck}, R., {Benjamin}, R.~A., {et~al.} 2012, \aj, 144, 43

\bibitem[{{Israel} \& {Mahoney}(1990)}]{Israel1990}
{Israel}, F.~P. \& {Mahoney}, M.~J. 1990, \apj, 352, 30

\bibitem[{{Israel} \& {van der Hulst}(1983)}]{Israel1983}
{Israel}, F.~P. \& {van der Hulst}, J.~M. 1983, \aj, 88, 1736

\bibitem[{{Israel} {et~al.}(1999){Israel}, {van der Werf}, \&
  {Tilanus}}]{Israel1999}
{Israel}, F.~P., {van der Werf}, P.~P., \& {Tilanus}, R.~P.~J. 1999, \aap, 344,
  L83

\bibitem[{{Kapi{\'n}ska} {et~al.}(2017){Kapi{\'n}ska}, {Staveley-Smith},
  {Crocker}, {Meurer}, {Bhandari}, {Hurley-Walker}, {Offringa}, {Hanish},
  {Seymour}, {Ekers}, {Bell}, {Callingham}, {Dwarakanath}, {For}, {Gaensler},
  {Hancock}, {Hindson}, {Johnston-Hollitt}, {Lenc}, {McKinley}, {Morgan},
  {Procopio}, {Wayth}, {Wu}, {Zheng}, {Barry}, {Beardsley}, {Bowman}, {Briggs},
  {Carroll}, {Dillon}, {Ewall-Wice}, {Feng}, {Greenhill}, {Hazelton}, {Hewitt},
  {Jacobs}, {Kim}, {Kittiwisit}, {Line}, {Loeb}, {Mitchell}, {Morales},
  {Neben}, {Paul}, {Pindor}, {Pober}, {Riding}, {Sethi}, {Udaya Shankar},
  {Subrahmanyan}, {Sullivan}, {Tegmark}, {Thyagarajan}, {Tingay}, {Trott},
  {Webster}, {Wyithe}, {Cappallo}, {Deshpande}, {Kaplan}, {Lonsdale},
  {McWhirter}, {Morgan}, {Oberoi}, {Ord}, {Prabu}, {Srivani}, {Williams}, \&
  {Williams}}]{Kapinska2017}
{Kapi{\'n}ska}, A.~D., {Staveley-Smith}, L., {Crocker}, R., {et~al.} 2017,
  \apj, 838, 68

\bibitem[{{Kaz{\`e}s} {et~al.}(1970){Kaz{\`e}s}, {Le Squeren}, \&
  {Nguyen-Quang-Rieu}}]{Kazes1970}
{Kaz{\`e}s}, I., {Le Squeren}, A.~M., \& {Nguyen-Quang-Rieu}. 1970, \aplett, 6,
  193

\bibitem[{{Kellermann} {et~al.}(1969){Kellermann}, {Pauliny-Toth}, \&
  {Williams}}]{Kellermann1969}
{Kellermann}, K.~I., {Pauliny-Toth}, I.~I.~K., \& {Williams}, P.~J.~S. 1969,
  \apj, 157, 1

\bibitem[{{Klein} {et~al.}(2017){Klein}, {Lisenfeld}, \& {Verley}}]{Klein2017}
{Klein}, U., {Lisenfeld}, U., \& {Verley}, S. 2017, ArXiv e-prints

\bibitem[{{Klein} {et~al.}(1984){Klein}, {Wielebinski}, \& {Beck}}]{Klein1984b}
{Klein}, U., {Wielebinski}, R., \& {Beck}, R. 1984, \aap, 133, 19

\bibitem[{{Krause}(2009)}]{Krause2009}
{Krause}, M. 2009, in Revista Mexicana de Astronomia y Astrofisica, Vol.~36,
  Revista Mexicana de Astronomia y Astrofisica Conference Series, 25

\bibitem[{{Krause}(2012)}]{Krause2012}
{Krause}, M. 2012, in Magnetic Fields in the Universe III - From Laboratory and
  Stars to Primordial Structures, ed. M.~{Soida}, K.~{Otmianowska-Mazur}, E.~M.
  {de Gouveia Dal Pino}, \& A.~{Lazarian}, 155

\bibitem[{{Krause} {et~al.}(2017){Krause}, {Irwin}, {Wiegert}, {Miskolczi},
  {Damas-Segovia}, {Beck}, {Li}, {Heald}, {M{\"u}ller}, {Stein}, {Rand},
  {Heesen}, {Walterbos}, {Dettmar}, {Vargas}, {English}, \&
  {Murphy}}]{Krause2017}
{Krause}, M., {Irwin}, J., {Wiegert}, T., {et~al.} 2017, ArXiv e-prints

\bibitem[{{Kulsrud} \& {Pearce}(1969)}]{Kulsrud1969}
{Kulsrud}, R. \& {Pearce}, W.~P. 1969, \apj, 156, 445

\bibitem[{{Lacki} \& {Thompson}(2010)}]{Lacki2010}
{Lacki}, B.~C. \& {Thompson}, T.~A. 2010, \apj, 717, 196

\bibitem[{{Lepage} \& {Billard}(1992)}]{Lepage1992}
{Lepage}, R. \& {Billard}, L. 1992, {Exploring the Limits of Bootstrap, Wiley
  Series in Probability and Mathematical Statistics, New York: Wiley, 1992 }

\bibitem[{{Lerche} \& {Schlickeiser}(1981)}]{Lerche1981}
{Lerche}, I. \& {Schlickeiser}, R. 1981, \apjs, 47, 33

\bibitem[{{Lerche} \& {Schlickeiser}(1982)}]{Lerche1982}
{Lerche}, I. \& {Schlickeiser}, R. 1982, \aap, 107, 148

\bibitem[{{Licquia} \& {Newman}(2014)}]{Licquia2014}
{Licquia}, T. \& {Newman}, J. 2014, in American Astronomical Society Meeting
  Abstracts, Vol. 223, 336.04

\bibitem[{Longair(2011)}]{Longair2011}
Longair, M. 2011, {High energy astrophysics, 3rd ed} (Cambridge: Cambridge
  University Press)

\bibitem[{{Marvil} {et~al.}(2015){Marvil}, {Owen}, \& {Eilek}}]{Marvil2015}
{Marvil}, J., {Owen}, F., \& {Eilek}, J. 2015, \aj, 149, 32

\bibitem[{{McMullin} {et~al.}(2007){McMullin}, {Waters}, {Schiebel}, {Young},
  \& {Golap}}]{MCMULLIN2007}
{McMullin}, J.~P., {Waters}, B., {Schiebel}, D., {Young}, W., \& {Golap}, K.
  2007, in Astronomical Society of the Pacific Conference Series, Vol. 376,
  Astronomical Data Analysis Software and Systems XVI, ed. R.~A. {Shaw},
  F.~{Hill}, \& D.~J. {Bell}, 127

\bibitem[{{Moss} \& {Sokoloff}(2017)}]{Moss2017}
{Moss}, D. \& {Sokoloff}, D. 2017, \aap, 598, A72

\bibitem[{{Moss} {et~al.}(2010){Moss}, {Sokoloff}, {Beck}, \&
  {Krause}}]{Moss2010}
{Moss}, D., {Sokoloff}, D., {Beck}, R., \& {Krause}, M. 2010, \aap, 512, A61

\bibitem[{{Mulcahy} {et~al.}(2016){Mulcahy}, {Fletcher}, {Beck}, {Mitra}, \&
  {Scaife}}]{mulcahy2016}
{Mulcahy}, D.~D., {Fletcher}, A., {Beck}, R., {Mitra}, D., \& {Scaife},
  A.~M.~M. 2016, \aap, 592, A123

\bibitem[{{Mulcahy} {et~al.}(2014){Mulcahy}, {Horneffer}, {Beck}, {Heald},
  {Fletcher}, {Scaife}, {Adebahr}, {Anderson}, {Bonafede}, {Br{\"u}ggen},
  {Brunetti}, {Chy{\.z}y}, {Conway}, {Dettmar}, {En{\ss}lin}, {Haverkorn},
  {Horellou}, {Iacobelli}, {Israel}, {Junklewitz}, {Jurusik}, {K{\"o}hler},
  {Kuniyoshi}, {Orr{\'u}}, {Paladino}, {Pizzo}, {Reich}, \&
  {R{\"o}ttgering}}]{mulcahy2014}
{Mulcahy}, D.~D., {Horneffer}, A., {Beck}, R., {et~al.} 2014, \aap, 568, A74

\bibitem[{{M{\"u}ller} {et~al.}(2017){M{\"u}ller}, {Krause}, {Beck}, \&
  {Schmidt}}]{Mueller2017}
{M{\"u}ller}, P., {Krause}, M., {Beck}, R., \& {Schmidt}, P. 2017, \aap, 606,
  A41

\bibitem[{{Niklas} {et~al.}(1997){Niklas}, {Klein}, \&
  {Wielebinski}}]{Niklas1997}
{Niklas}, S., {Klein}, U., \& {Wielebinski}, R. 1997, \aap, 322, 19

\bibitem[{{Nixon} {et~al.}(2018){Nixon}, {Hands}, {King}, \&
  {Pringle}}]{Nixon2018}
{Nixon}, C.~J., {Hands}, T.~O., {King}, A.~R., \& {Pringle}, J.~E. 2018, ArXiv
  e-prints

\bibitem[{{Norman} \& {Ikeuchi}(1989)}]{Norman1989}
{Norman}, C.~A. \& {Ikeuchi}, S. 1989, \apj, 345, 372

\bibitem[{{Offringa} {et~al.}(2010){Offringa}, {de Bruyn}, {Biehl}, {Zaroubi},
  {Bernardi}, \& {Pandey}}]{Offringa2010}
{Offringa}, A.~R., {de Bruyn}, A.~G., {Biehl}, M., {et~al.} 2010, \mnras, 405,
  155

\bibitem[{{Offringa} {et~al.}(2012){Offringa}, {van de Gronde}, \&
  {Roerdink}}]{Offringa2012}
{Offringa}, A.~R., {van de Gronde}, J.~J., \& {Roerdink}, J.~B.~T.~M. 2012,
  \aap, 539, A95

\bibitem[{{Okabe} {et~al.}(2000){Okabe}, {Boots}, {Sugihara}, \&
  {Chiu}}]{Okabe2000}
{Okabe}, A., {Boots}, B., {Sugihara}, K., \& {Chiu}, S. 2000, {Spatial
  tessellations. Concepts and Applications of Voronoi Diagrams : Wiley Series
  in Probability and Mathematical Statistics, Chichester, New York, 2nd Edition
  }

\bibitem[{{Oosterloo} {et~al.}(2007){Oosterloo}, {Fraternali}, \&
  {Sancisi}}]{Oosterloo2007}
{Oosterloo}, T., {Fraternali}, F., \& {Sancisi}, R. 2007, \aj, 134, 1019

\bibitem[{{Pakmor} {et~al.}(2014){Pakmor}, {Marinacci}, \&
  {Springel}}]{Pakmor2014}
{Pakmor}, R., {Marinacci}, F., \& {Springel}, V. 2014, \apjl, 783, L20

\bibitem[{{Perley} \& {Butler}(2013)}]{Perley2013}
{Perley}, R.~A. \& {Butler}, B.~J. 2013, \apjs, 204, 19

\bibitem[{{Pohl} \& {Schlickeiser}(1990)}]{Pohl1990}
{Pohl}, M. \& {Schlickeiser}, R. 1990, \aap, 234, 147

\bibitem[{{Rand} {et~al.}(1990){Rand}, {Kulkarni}, \& {Hester}}]{Rand1990}
{Rand}, R.~J., {Kulkarni}, S.~R., \& {Hester}, J.~J. 1990, \apjl, 352, L1

\bibitem[{{Rau} \& {Cornwell}(2011)}]{Rau2011}
{Rau}, U. \& {Cornwell}, T.~J. 2011, \aap, 532, A71

\bibitem[{{Rengelink} {et~al.}(1997){Rengelink}, {Tang}, {de Bruyn}, {Miley},
  {Bremer}, {Roettgering}, \& {Bremer}}]{Rengelink1997}
{Rengelink}, R.~B., {Tang}, Y., {de Bruyn}, A.~G., {et~al.} 1997, \aaps, 124,
  259

\bibitem[{{Rossa} {et~al.}(2004){Rossa}, {Dettmar}, {Walterbos}, \&
  {Norman}}]{Rossa2004}
{Rossa}, J., {Dettmar}, R.-J., {Walterbos}, R.~A.~M., \& {Norman}, C.~A. 2004,
  \aj, 128, 674

\bibitem[{{Rupen}(1991)}]{Rupen1991}
{Rupen}, M.~P. 1991, \aj, 102, 48

\bibitem[{{Scaife} \& {Heald}(2012)}]{Scaife2012}
{Scaife}, A.~M.~M. \& {Heald}, G.~H. 2012, \mnras, 423, L30

\bibitem[{{Schmidt}(2016)}]{Schmidt2016}
{Schmidt}, P. 2016, {The Radio Continuum Halos of the Edge-on Galaxies NGC 891
  and NGC 4565, PhD Thesis, University of Bonn}

\bibitem[{{Schmidt} {et~al.}(2018){Schmidt}, {Krause}, {Heesen}, {Basu},
  {Beck}, \& {et al.}}]{Schmidt2018}
{Schmidt}, P., {Krause}, M., {Heesen}, H., {et~al.} 2018, \aap, to be submitted

\bibitem[{{Scoville} {et~al.}(1993){Scoville}, {Thakkar}, {Carlstrom}, \&
  {Sargent}}]{Scoville1993}
{Scoville}, N.~Z., {Thakkar}, D., {Carlstrom}, J.~E., \& {Sargent}, A.~I. 1993,
  \apjl, 404, L59

\bibitem[{{Shapiro} \& {Field}(1976)}]{Shapiro1976}
{Shapiro}, P.~R. \& {Field}, G.~B. 1976, \apj, 205, 762

\bibitem[{{Sofue}(1987)}]{Sofue1987}
{Sofue}, Y. 1987, \pasj, 39, 547

\bibitem[{{Soida} {et~al.}(2011){Soida}, {Krause}, {Dettmar}, \&
  {Urbanik}}]{Soida2011}
{Soida}, M., {Krause}, M., {Dettmar}, R.-J., \& {Urbanik}, M. 2011, \aap, 531,
  A127

\bibitem[{{Stil} {et~al.}(2009){Stil}, {Krause}, {Beck}, \&
  {Taylor}}]{Stil2009}
{Stil}, J.~M., {Krause}, M., {Beck}, R., \& {Taylor}, A.~R. 2009, \apj, 693,
  1392

\bibitem[{{Strickland} {et~al.}(2004){Strickland}, {Heckman}, {Colbert},
  {Hoopes}, \& {Weaver}}]{Strickland2004}
{Strickland}, D.~K., {Heckman}, T.~M., {Colbert}, E.~J.~M., {Hoopes}, C.~G., \&
  {Weaver}, K.~A. 2004, \apjs, 151, 193

\bibitem[{{Strong}(1978)}]{Strong1978}
{Strong}, A.~W. 1978, \aap, 66, 205

\bibitem[{{Tabatabaei} {et~al.}(2007){Tabatabaei}, {Beck}, {Kr{\"u}gel},
  {Krause}, {Berkhuijsen}, {Gordon}, \& {Menten}}]{Tabatabaei2007}
{Tabatabaei}, F.~S., {Beck}, R., {Kr{\"u}gel}, E., {et~al.} 2007, \aap, 475,
  133

\bibitem[{{Tabatabaei} {et~al.}(2013){Tabatabaei}, {Berkhuijsen}, {Frick},
  {Beck}, \& {Schinnerer}}]{Tabatabaei2013}
{Tabatabaei}, F.~S., {Berkhuijsen}, E.~M., {Frick}, P., {Beck}, R., \&
  {Schinnerer}, E. 2013, \aap, 557, A129

\bibitem[{{Tabatabaei} {et~al.}(2017){Tabatabaei}, {Schinnerer}, {Krause},
  {Dumas}, {Meidt}, {Damas-Segovia}, {Beck}, {Murphy}, {Mulcahy}, {Groves},
  {Bolatto}, {Dale}, {Galametz}, {Sandstrom}, {Boquien}, {Calzetti},
  {Kennicutt}, {Hunt}, {De Looze}, \& {Pellegrini}}]{Tabatabaei2017}
{Tabatabaei}, F.~S., {Schinnerer}, E., {Krause}, M., {et~al.} 2017, \apj, 836,
  185

\bibitem[{{Temple} {et~al.}(2005){Temple}, {Raychaudhury}, \&
  {Stevens}}]{Temple2005}
{Temple}, R.~F., {Raychaudhury}, S., \& {Stevens}, I.~R. 2005, \mnras, 362, 581

\bibitem[{{Uhlig} {et~al.}(2012){Uhlig}, {Pfrommer}, {Sharma}, {Nath},
  {En{\ss}lin}, \& {Springel}}]{Uhlig2012}
{Uhlig}, M., {Pfrommer}, C., {Sharma}, M., {et~al.} 2012, \mnras, 423, 2374

\bibitem[{{van der Kruit} \& {Searle}(1981)}]{vanderKruit1981}
{van der Kruit}, P.~C. \& {Searle}, L. 1981, \aap, 95, 116

\bibitem[{{van Gorkom} {et~al.}(1986){van Gorkom}, {Rupen}, {Knapp}, {Gunn},
  {Neugebauer}, \& {Matthews}}]{vangorkom1986}
{van Gorkom}, J., {Rupen}, M., {Knapp}, G., {et~al.} 1986, \iaucirc, 4248

\bibitem[{{van Haarlem} {et~al.}(2013){van Haarlem}, {Wise}, {Gunst}, {Heald},
  {McKean}, {Hessels}, {de Bruyn}, {Nijboer}, {Swinbank}, {Fallows},
  {Brentjens}, {Nelles}, {Beck}, {Falcke}, {Fender}, {H{\"o}randel},
  {Koopmans}, {Mann}, {Miley}, {R{\"o}ttgering}, {Stappers}, {Wijers},
  {Zaroubi}, {van den Akker}, {Alexov}, {Anderson}, {Anderson}, {van Ardenne},
  {Arts}, {Asgekar}, {Avruch}, {Batejat}, {B{\"a}hren}, {Bell}, {Bell}, {van
  Bemmel}, {Bennema}, {Bentum}, {Bernardi}, {Best}, {B{\^i}rzan}, {Bonafede},
  {Boonstra}, {Braun}, {Bregman}, {Breitling}, {van de Brink}, {Broderick},
  {Broekema}, {Brouw}, {Br{\"u}ggen}, {Butcher}, {van Cappellen}, {Ciardi},
  {Coenen}, {Conway}, {Coolen}, {Corstanje}, {Damstra}, {Davies}, {Deller},
  {Dettmar}, {van Diepen}, {Dijkstra}, {Donker}, {Doorduin}, {Dromer}, {Drost},
  {van Duin}, {Eisl{\"o}ffel}, {van Enst}, {Ferrari}, {Frieswijk}, {Gankema},
  {Garrett}, {de Gasperin}, {Gerbers}, {de Geus}, {Grie{\ss}meier}, {Grit},
  {Gruppen}, {Hamaker}, {Hassall}, {Hoeft}, {Holties}, {Horneffer}, {van der
  Horst}, {van Houwelingen}, {Huijgen}, {Iacobelli}, {Intema}, {Jackson},
  {Jelic}, {de Jong}, {Juette}, {Kant}, {Karastergiou}, {Koers}, {Kollen},
  {Kondratiev}, {Kooistra}, {Koopman}, {Koster}, {Kuniyoshi}, {Kramer},
  {Kuper}, {Lambropoulos}, {Law}, {van Leeuwen}, {Lemaitre}, {Loose}, {Maat},
  {Macario}, {Markoff}, {Masters}, {McFadden}, {McKay-Bukowski}, {Meijering},
  {Meulman}, {Mevius}, {Middelberg}, {Millenaar}, {Miller-Jones}, {Mohan},
  {Mol}, {Morawietz}, {Morganti}, {Mulcahy}, {Mulder}, {Munk}, {Nieuwenhuis},
  {van Nieuwpoort}, {Noordam}, {Norden}, {Noutsos}, {Offringa}, {Olofsson},
  {Omar}, {Orr{\'u}}, {Overeem}, {Paas}, {Pandey-Pommier}, {Pandey}, {Pizzo},
  {Polatidis}, {Rafferty}, {Rawlings}, {Reich}, {de Reijer}, {Reitsma},
  {Renting}, {Riemers}, {Rol}, {Romein}, {Roosjen}, {Ruiter}, {Scaife}, {van
  der Schaaf}, {Scheers}, {Schellart}, {Schoenmakers}, {Schoonderbeek},
  {Serylak}, {Shulevski}, {Sluman}, {Smirnov}, {Sobey}, {Spreeuw}, {Steinmetz},
  {Sterks}, {Stiepel}, {Stuurwold}, {Tagger}, {Tang}, {Tasse}, {Thomas},
  {Thoudam}, {Toribio}, {van der Tol}, {Usov}, {van Veelen}, {van der Veen},
  {ter Veen}, {Verbiest}, {Vermeulen}, {Vermaas}, {Vocks}, {Vogt}, {de Vos},
  {van der Wal}, {van Weeren}, {Weggemans}, {Weltevrede}, {White}, {Wijnholds},
  {Wilhelmsson}, {Wucknitz}, {Yatawatta}, {Zarka}, {Zensus}, \& {van
  Zwieten}}]{vanhaarlem2013}
{van Haarlem}, M.~P., {Wise}, M.~W., {Gunst}, A.~W., {et~al.} 2013, \aap, 556,
  A2

\bibitem[{{van Weeren} {et~al.}(2016){van Weeren}, {Williams}, {Hardcastle},
  {Shimwell}, {Rafferty}, {Sabater}, {Heald}, {Sridhar}, {Dijkema}, {Brunetti},
  {Br{\"u}ggen}, {Andrade-Santos}, {Ogrean}, {R{\"o}ttgering}, {Dawson},
  {Forman}, {de Gasperin}, {Jones}, {Miley}, {Rudnick}, {Sarazin}, {Bonafede},
  {Best}, {B{\^i}rzan}, {Cassano}, {Chy{\.z}y}, {Croston}, {Ensslin},
  {Ferrari}, {Hoeft}, {Horellou}, {Jarvis}, {Kraft}, {Mevius}, {Intema},
  {Murray}, {Orr{\'u}}, {Pizzo}, {Simionescu}, {Stroe}, {van der Tol}, \&
  {White}}]{vanWeeren2016}
{van Weeren}, R.~J., {Williams}, W.~L., {Hardcastle}, M.~J., {et~al.} 2016,
  \apjs, 223, 2

\bibitem[{{Varenius} {et~al.}(2015){Varenius}, {Conway}, {Mart{\'{\i}}-Vidal},
  {Beswick}, {Deller}, {Wucknitz}, {Jackson}, {Adebahr}, {P{\'e}rez-Torres},
  {Chy{\.z}y}, {Carozzi}, {Mold{\'o}n}, {Aalto}, {Beck}, {Best}, {Dettmar},
  {van Driel}, {Brunetti}, {Br{\"u}ggen}, {Haverkorn}, {Heald}, {Horellou},
  {Jarvis}, {Morabito}, {Miley}, {R{\"o}ttgering}, {Toribio}, \&
  {White}}]{Varenius2015}
{Varenius}, E., {Conway}, J.~E., {Mart{\'{\i}}-Vidal}, I., {et~al.} 2015, \aap,
  574, A114

\bibitem[{{Vargas} {et~al.}(2018){Vargas}, {Mora-Partiarroyo}, {Schmidt},
  {Rand}, {Stein}, {Walterbos}, {Wang}, {Basu}, {Patterson}, {Kepley}, {Beck},
  {Irwin}, {Heald}, {Li}, \& {Wiegert}}]{Vargas2018}
{Vargas}, C.~J., {Mora-Partiarroyo}, S.~C., {Schmidt}, P., {et~al.} 2018, \apj,
  853, 128

\bibitem[{{Vigotti} {et~al.}(1989){Vigotti}, {Grueff}, {Perley}, {Clark}, \&
  {Bridle}}]{Vigotti1989}
{Vigotti}, M., {Grueff}, G., {Perley}, R., {Clark}, B.~G., \& {Bridle}, A.~H.
  1989, \aj, 98, 419

\bibitem[{{Whaley} {et~al.}(2009){Whaley}, {Irwin}, {Madden}, {Galliano}, \&
  {Bendo}}]{Whaley2009}
{Whaley}, C.~H., {Irwin}, J.~A., {Madden}, S.~C., {Galliano}, F., \& {Bendo},
  G.~J. 2009, \mnras, 395, 97

\bibitem[{{Wiegert} {et~al.}(2015){Wiegert}, {Irwin}, {Miskolczi}, {Schmidt},
  {Mora}, {Damas-Segovia}, {Stein}, {English}, {Rand}, {Santistevan},
  {Walterbos}, {Krause}, {Beck}, {Dettmar}, {Kepley}, {Wezgowiec}, {Wang},
  {Heald}, {Li}, {MacGregor}, {Johnson}, {Strong}, {DeSouza}, \&
  {Porter}}]{Wiegert2015}
{Wiegert}, T., {Irwin}, J., {Miskolczi}, A., {et~al.} 2015, \aj, 150, 81

\bibitem[{{Williams} {et~al.}(2016){Williams}, {van Weeren}, {R{\"o}ttgering},
  {Best}, {Dijkema}, {de Gasperin}, {Hardcastle}, {Heald}, {Prandoni},
  {Sabater}, {Shimwell}, {Tasse}, {van Bemmel}, {Br{\"u}ggen}, {Brunetti},
  {Conway}, {En{\ss}lin}, {Engels}, {Falcke}, {Ferrari}, {Haverkorn},
  {Jackson}, {Jarvis}, {Kapi{\'n}ska}, {Mahony}, {Miley}, {Morabito},
  {Morganti}, {Orr{\'u}}, {Retana-Montenegro}, {Sridhar}, {Toribio}, {White},
  {Wise}, \& {Zwart}}]{Williams2016}
{Williams}, W.~L., {van Weeren}, R.~J., {R{\"o}ttgering}, H.~J.~A., {et~al.}
  2016, \mnras, 460, 2385

\bibitem[{{Wills} {et~al.}(1997){Wills}, {Pedlar}, {Muxlow}, \&
  {Wilkinson}}]{Wills1997}
{Wills}, K.~A., {Pedlar}, A., {Muxlow}, T.~W.~B., \& {Wilkinson}, P.~N. 1997,
  \mnras, 291, 517

\bibitem[{{Xilouris} {et~al.}(1999){Xilouris}, {Byun}, {Kylafis}, {Paleologou},
  \& {Papamastorakis}}]{Xilouris1999}
{Xilouris}, E.~M., {Byun}, Y.~I., {Kylafis}, N.~D., {Paleologou}, E.~V., \&
  {Papamastorakis}, J. 1999, \aap, 344, 868

\bibitem[{{Zwart} {et~al.}(2008){Zwart}, {Barker}, {Biddulph}, {Bly}, {Boysen},
  {Brown}, {Clementson}, {Crofts}, {Culverhouse}, {Czeres}, {Dace}, {Davies},
  {D'Alessandro}, {Doherty}, {Duggan}, {Ely}, {Felvus}, {Feroz}, {Flynn},
  {Franzen}, {Geisb{\"u}sch}, {G{\'e}nova-Santos}, {Grainge}, {Grainger},
  {Hammett}, {Hills}, {Hobson}, {Holler}, {Hurley-Walker}, {Jilley}, {Jones},
  {Kaneko}, {Kneissl}, {Lancaster}, {Lasenby}, {Marshall}, {Newton}, {Norris},
  {Northrop}, {Odell}, {Petencin}, {Pober}, {Pooley}, {Pospieszalski}, {Quy},
  {Rodr{\'{\i}}guez-Gonz{\'a}lvez}, {Saunders}, {Scaife}, {Schofield}, {Scott},
  {Shaw}, {Shimwell}, {Smith}, {Taylor}, {Titterington}, {Veli{\'c}},
  {Waldram}, {West}, {Wood}, {Yassin}, \& {AMI Consortium}}]{Zwart2008}
{Zwart}, J.~T.~L., {Barker}, R.~W., {Biddulph}, P., {et~al.} 2008, \mnras, 391,
  1545

\end{thebibliography}
\end{document}